\documentclass{article}

\usepackage{amsmath,amssymb,amsfonts,amsthm}
\usepackage[all]{xy}
\usepackage{color}

\usepackage{hyperref}%To get clickable links to papers

\usepackage{multirow}
\usepackage{rotating}
\usepackage{fullpage}

\def\endofproof {\hfill{$\Box$}\\}

\renewcommand{\(}{\begin{equation}}
\renewcommand{\)}{\end{equation}}
\newcommand{\bea}{\begin{eqnarray}}
\newcommand{\eea}{\end{eqnarray}}

\theoremstyle{plain}
\newtheorem{theorem}{Theorem}[subsection]

\newtheorem{proposition}[theorem]{Proposition}
\newtheorem{corollary}[theorem]{Corollary}

\theoremstyle{definition}
\newtheorem{definition}[theorem]{Definition}
\newtheorem{example}[theorem]{Example}

\newtheorem{remark}[theorem]{Remark}

\title{Higher $U(1)$-gerbe connections in geometric prequantization}

\author{Domenico Fiorenza\thanks{University La Sapienza, Rome},
  Christopher L. Rogers\thanks{University of Louisiana, Lafayette}, Urs Schreiber\thanks{CAS Prague and MPI Bonn}}

\begin{document}

\maketitle

\begin{abstract}
  We promote geometric prequantization to
  higher geometry (higher stacks), where a prequantization is given by a
  higher principal connection (a higher gerbe with connection).
  We show fairly generally how there is canonically
  a tower of higher gauge groupoids and Courant groupoids assigned
  to a higher prequantization, and establish the corresponding
  Atiyah sequence as an integrated Kostant-Souriau $\infty$-group extension of higher
  Hamiltonian symplectomorphisms by higher quantomorphisms.
  We also exhibit the $\infty$-group cocycle
  which classifies this extension and discuss how its restrictions along
  Hamiltonian $\infty$-actions yield higher Heisenberg cocycles.
  In the special case of higher differential geometry over smooth manifolds
  we find the $L_\infty$-algebra extension of Hamiltonian vector fields -- which is
  the higher Poisson bracket of local observables --
  and show that it is equivalent to the
  construction proposed by the second author in $n$-plectic geometry. Finally we
  indicate a list of examples of applications of higher prequantization in the extended geometric
  quantization of local quantum field theories and specifically in string geometry.
\end{abstract}

\newpage

\tableofcontents

\newpage

\section{Motivation}

\subsection{Geometric prequantization via slicing}
\label{TraditionalPrequantumGeometryViaSlicing}

In {geometric quantization} (see \cite{BrylinskiLoop, BatesWeinstein, Bongers} for review),
a \emph{prequantization} of a symplectic manifold $(X,\omega)$ is a lift of the symplectic
form from cocycles in de Rham cohomology to cocycles in {differential cohomology}. In more geometrical terms, this may be seen as the datum of a $U(1)$-principal bundle with connection $(P,\nabla)$  on $X$ whose curvature 2-form is the symplectic form $\omega$   \cite{GQ}.\footnote{More precisey, this is an \emph{integral} prequantization. The whole theory goes through without changes to prequantizations for any discrete subgroup of the real
numbers, if only one allows diffeological structure groups. However here we will always assume our prequantizations to be integral, for ease of exposition.}
The bundle $(P,\nabla)$ is called a \emph{prequantum bundle} for $\omega$ and the space of sections of the associated line bundle is the space of \emph{prequantum states}. The choice of a prequantum bundle is an intermediate
step in the genuine {geometric quantization} of symplectic manifolds,
which is obtained by ``dividing this data in half'' by a choice of polarization.
While polarizations do play a central role in geometric quantum theory,
for instance in the orbit method in geometric representation theory
\cite{Kirillov}, to name just one example, the study of
geometric prequantizations is of interest in its own right.
%the automorphisms of prequantum bundles covering diffeomorphisms of the base
%-- the {prequantum operators} or {contactomorphisms} -- and the action of these on the space of sections
%of the associated line bundle -- the {prequantum states}.
For instance the quantomorphism group \cite{GQ, Kostant, Viz2}, i.e., the group of the automorphisms of the prequantum bundle covering diffeomorphisms of the base, naturally provides a %non-simply connected
Lie integration
of the {Poisson bracket} Lie algebra of the underlying symplectic
manifold. As the Poisson bracket Lie algebra is a central extension of the Lie algebra of Hamiltonian vector fields,  pulling back the quantomorphism groups along Hamiltonian group actions induces central extensions of (possibly infinite-dimensional) Lie groups, see for instance \cite{RS,Viz, Viz2}.
Moreover, as we discuss below in \ref{HigherCourantQuantomorphimsGroupoids}, the quantomorphism group of $(P,\nabla)$ comes equipped with a canonical
injection into the group of bisections of the groupoid integrating the
{Atiyah Lie algebroid} of $P$.

We now observe that all of this has a simple natural reformulation in terms of the maps into
the {smooth moduli stacks} \cite{FScSt, NikolausSchreiberStevensonII} that classify
-- or better: that {modulate} -- principal bundles with connections.
This reformulation %exhibits an abstract characterization of prequantum geometry which
immediately generalizes
to higher geometric contexts richer than traditional differential geometry.
In \ref{TheNeedForHigherPrequantumGeometryFromExtendedLocalGeometricQuantization} below we say why
such a generalization is indeed desireable and in \ref{Survey} we survey
constructions and results in higher geometric prequantum theory, which we will need in order to present the examples and applications in the final part
\ref{Examples} of the article. We refer the reader to \cite{dcct} for a comprehensive treatment of the higher geometrical constructions of which we will make use.

\medskip

To start with, if we write $\Omega^2_{\mathrm{cl}}$
for the sheaf of smooth closed differential 2-forms (on the site of all smooth manifolds), then
by the Yoneda lemma a closed 2-form $\omega$ on a smooth manifold $X$ is
equivalently a map of sheaves
$
  \omega\colon
    X \to \Omega^2_{\mathrm{cl}}
$.
It is useful to think of this as a simple first instance of moduli stacks:
$\Omega^2_{\mathrm{cl}}$ is the {universal moduli stack of smooth closed 2-forms}.\footnote{Notice that symplectic 2-forms do not form a stack on the site of smooth manifolds: the pullback of a symplectic form along a smooth map is closed but not necessarily non-degenerate.
Symplectic forms form a sheaf after slicing over the moduli stack $\mathbf{B}\mathrm{GL}$ \cite{FiorenzaSatiSchreiberCS}, but
for the purposes of \emph{pre}-quantization we do not need to consider this.} Similarly but more interestingly, there is a smooth moduli stack
of $U(1)$-principal connections, which is a stack of groupoids on the site of smooth manifolds \cite{FScSt}.
It is denoted by
$\mathbf{B}U(1)_{\mathrm{conn}}$ and has a natural ``forget the connection'' morphism to the universal moduli stack $\mathbf{B}U(1)$ of $U(1)$-principal bundles,
which in turn is a smooth refinement of the classifying space
$B U(1) \simeq K(\mathbb{Z},2)$ of equivalence classes of such bundles.
Hence $\mathbf{B}U(1)_{\mathrm{conn}}$ is the  %``smooth homotopy 1-type''
smooth stack which is uniquely characterized by the fact that maps $X \to \mathbf{B}U(1)_{\mathrm{conn}}$
from a smooth manifold $X$ to $\mathbf{B}U(1)_{\mathrm{conn}}$ are precisely $U(1)$-principal connections on $X$, and that
homotopies between such maps are precisely smooth gauge transformations between the corresponding
connections. The stack $\mathbf{B}U(1)_{\mathrm{conn}}$ may be seen as a refinement of $\Omega^2_{\mathrm{cl}}$:
the natural map which sends a $U(1)$-principal connection to its curvature 2-form constitutes
a map of moduli stacks
$F_{(-)} \colon \mathbf{B}U(1)_{\mathrm{conn}} \to \Omega^2_{\mathrm{cl}}$, which in turn may be regarded
as a universal closed 2-form on $\mathbf{B}U(1)_{\mathrm{conn}}$.
This {universal curvature form} is at the heart of traditional prequantization:
for $\omega \in \Omega^2_{\mathrm{cl}}(X)$
a (pre-)symplectic\footnote{In the context of geometric quantisation, closed 2-forms that are not of maximal but
still of constant rank are often called presymplectic forms. Here we do not even need to require constant rank.
Several constructions in symplectic geometry can be adapted to the context of presymplectic manifolds, see \cite{nextArticle}.} form as above, a \emph{prequantization} of $(X,\omega)$ is equivalently
a lift $\nabla$ in the diagram
$$
  \raisebox{20pt}{
    \xymatrix{
      X \ar@{-->}[rr]^\nabla \ar[dr]_{\omega} && \mathbf{B}U(1)_{\mathrm{conn}} \ar[dl]^{F_{(-)}}
  	  \\
	  & \Omega^2_{\mathrm{cl}}
    }}
  \,,
$$
where the commutativity of the diagram expresses the traditional prequantization condition
$\omega = F_{\nabla}$.

A triangular diagram as above may naturally be interpreted as exhibiting a map
\emph{from $\omega$ to $F_{(-)}$ in the slice topos over $\Omega^2_{\mathrm{cl}}$}.
This means that the map $F_{(-)}$ is itself a moduli stack, namely,
 the \emph{moduli stack of prequantizations}, not over the site of smooth manifolds, but over the site of smooth manifolds equipped with a closed 2-form (i.e., the site of presymplectic manifolds). Once the slice point of view has been adopted, there is a natural notion of automorphisms of a given prequantization $\nabla$: by looking at $\nabla$ as an object in the slice over $\mathbf{B}U(1)_{\mathrm{conn}}$ one immediately sees that an automorphism
of $\nabla$ is a homotopy commutative diagram of the form
\[
 \raisebox{20pt}{
    \xymatrix{
	  X \ar[rr]^\phi|\simeq_{\ }="s" \ar[dr]_\nabla^{\ }="t" && X \ar[dl]^\nabla
	  \\
	  & \mathbf{B}U(1)_{\mathrm{conn}}
	  \ar@{=>}^\eta "s"; "t"
	}
	}
\]
i.e., it is
a pair $(\phi, \eta)$, consisting of a
diffeomorphism $\phi \colon X \xrightarrow{\simeq}X$ together with an equivalence
of prequantum connections $\eta\colon \phi^* \nabla \xrightarrow{\simeq} \nabla$. This is precisely what is classically known as a \emph{quantomorphism}  of $\nabla$.
%
%A moment of reflection
%shows that such a pair is equivalently again a triangular diagram, now as on the right of
%$$
%  \mathbf{QuantMorph}(\nabla)
%  =
%  \left\{
%    {\phi \in \mathrm{Diff}(X)}\,, \atop {\eta : \phi^* \nabla \stackrel{\simeq}{\to} \nabla}
%  \right\}
%  \simeq
%  \left\{
%    \raisebox{20pt}{
%    \xymatrix{
%	  X \ar[rr]^\phi|\simeq_{\ }="s" \ar[dr]_\nabla^{\ }="t" && X \ar[dl]^\nabla
%	  \\
%	  & \mathbf{B}U(1)_{\mathrm{conn}}
%	  %
%	  \ar@{=>}^\eta "s"; "t"
%	}
%	}
%  \right\}
%  \,.
%$$
The slice point of view also makes the group structure on quantomorphisms manifest: it is given by the evident pasting of
triangular diagrams.
In this form, the \emph{quantomorphism group} of $\nabla$ is realized as an example of a very general
construction that directly makes sense also in higher geometry: it is the
{automorphism group} of a morphism regarded as an object in the slice
over its target: a \emph{relative} automorphism group.
Moreover, in this form one of the crucial properties of the quantomorphism group, namely, the
fact that over a connected manifold the quantomorphism group is a $U(1)$-central extension of the group of Hamiltonian
symplectomorphisms, is expressed by the schematic diagrams below:
\[
  \begin{array}{ccccc}
  U(1) &\to& \mathbf{QuantMorph}(\nabla) &\to& \mathbf{HamSympl}(\omega)
  \\
  \\
  \left\{
    \raisebox{20pt}{
    \xymatrix{
	  X
	  \ar@/^1pc/[d]^-\nabla_{\ }="s"
	  \ar@/_1pc/[d]_-\nabla^{\ }="t"
	  \\
	  \mathbf{B}U(1)_{\mathrm{conn}}
	  \ar@{=>} "s"; "t"
	}
	}
  \right\}
  &\to&
  \left\{
    \raisebox{20pt}{
    \xymatrix{
	  X \ar[dr]_{\nabla}^{\ }="t" \ar[rr]^\simeq_{\ }="s" && X \ar[dl]^{\nabla}
	  \\
	  & \mathbf{B}U(1)_{\mathrm{conn}}
	  \ar@{=>} "s"; "t"
	}
	}
  \right\}
  &\to&
  \left\{
    \raisebox{20pt}{
    \xymatrix@R=1pt{
	  \\
	  \\
	  X  \ar[rr]^\simeq_{\ }="s" && X
	  \\
	}
	}
  \right\}
  \end{array}
  \,,
\]
which is just a special case of a very general
extension phenomenon, see Section \ref{TheCentralTheorems}.

The above $U(1)$-extension is the hallmark of quantization: under Lie differentiation the
above sequence of (infinite-dimensional) Lie groups turns into the central extension of Lie algebras
$$
  \begin{array}{ccccc}
    \mathbb{R} &\to& \mathfrak{Poisson}(X, \omega) & \to &  \mathfrak{X}_{\mathrm{Ham}}(X, \omega)
  \end{array}
$$
that exhibits the Poisson bracket Lie algebra of a symplectic manifold $(X,\omega)$ as an
$\mathbb{R}$-central extension of the Lie algebra of Hamiltonian vector fields
on $X$: the \emph{Kostant-Souriau extension} (see, e.g \cite[section 2.3]{BrylinskiLoop}). Notice that  we wrote $\mathfrak{Poisson}(X, \omega)$ instead of $\mathfrak{Poisson}(X, \nabla)$: the Poisson bracket Lie algebra is independent of the prequantization chosen.\footnote{Actually it does not even require that an integral prequantization exists. It only requires a \emph{real} prequantization to exists, a condition which is always fulfilled.}
If we write $i\hbar\colon \mathbb{R}\to i\mathbb{R}$ for a choice of an $\mathbb{R}$-linear isomorphism between $\mathbb{R}$ and $i\mathbb{R}$, then
 the above central extension expresses the
quantum deformation of ``classical commutators'' in $\mathfrak{X}_{\mathrm{Ham}}(X,\omega)$ by scalar multiples of
the central term $i \hbar$ (where $\hbar$ plays the role of ``Planck's constant'').

More widely known than the quantomorphism groups of all prequantum operators
are certain classes of subgroups of them,
the \emph{Heisenberg groups}. The prototypical example is obtained from the symplectic vector space $(\mathbb{R}^{2n},\omega_{\mathbb{R}^{2n}})$ seen as a symplectic manifold endowed with a translation invariant symplectic form. Namely, the translation
group $\mathbb{R}^{2n}$ acts on $(\mathbb{R}^{2n},\omega_{\mathbb{R}^{2n}})$ by Hamiltonian symplectomorphisms,
and so defines a group homomorphism $\mathbb{R}^{2n} \to \mathbf{HamSympl}(\omega_{\mathbb{R}^{2n}})$.
The pullback of the above quantomorphism group extension along this map yields a
$U(1)$-extension of $\mathbb{R}^{2n}$, whose universal cover is the traditional Heisenberg group
$H(n,\mathbb{R})$. More generally, for $(X,\omega)$ any (prequantized) symplectic manifold
and $G$ any Lie group, one considers \emph{Hamiltonian $G$-actions}: smooth group
homomorphisms $\phi : G \to \mathbf{HamSympl}(\omega)$. Pulling back the quantomorphism
group extension now yields a $U(1)$-extension of $G$ and this we may call, more generally,
the Heisenberg-type group extension induced by the Hamiltonian $G$-action:
$$
  U(1) \to \mathbf{Heis}_\phi(\nabla) \to G.
$$

The crucial property of the
quantomorphism group and any of its Heisenberg subgroups,
at least for the purposes of geometric quantization,
is that these are canonically equipped with an action on the space
of prequantum states, i.e., on the space of sections of the complex line bundle which is associated to the
prequantum bundle. It is customary to call  a quantomorphism realised
this way, as an operator on the space of prequantum states, a \emph{prequantum operator}.
Notice that, by composition, an \emph{integrated moment map}, i.e.,
a group homomorphism $G \to \mathbf{QuantMorph}(\nabla)$ lifting a Hamiltonian $G$-action,
induces a representation of $G$ on the space of prequantum states.
After a choice of polarization this is the
construction that makes geometric quantization a valuable tool in geometric representation theory.

Also the action of prequantum operators on prequantum states admits a natural interpretation in
terms of slicings.  Indeed, the quotient stack $\mathbb{C}/\!/U(1)$ associated with the defining complex representation of $U(1)$
 comes equipped with a canonical map
$\rho\colon \mathbb{C}/\!/U(1) \to  \ast/\!/U(1) \simeq \mathbf{B}U(1) $
to the moduli stack of $U(1)$-principal bundles. This morphism is the {universal}
complex line bundle over the moduli stack of $U(1)$-principal bundles (see \cite{NikolausSchreiberStevensonI}
for this statement and its higher generalization, we recall this below in Section \ref{Survey}).
%, and it comes with a natural differential refinement $\rho_{\mathrm{conn}}\colon \mathbb{C}/\!/U(1)_{\mathrm{conn}} \to \mathbf{B}U(1)_{\mathrm{conn}} $.
Let $(P,\nabla)$ be a principal $U(1)$-bundle with connection on a smooth manifold $X$.
One sees that a map $\psi : P \to \rho$ in the slice over
$\mathbf{B}U(1)$, i.e., a diagram with specified homotopy of the form
 $$
   \raisebox{20pt}{
  \xymatrix{
    X \ar[dr]_{P}^{\  }="t" \ar[rr]^\psi_{\ }="s" && \mathbb{C}/\!/U(1) \ar[dl]^{\rho}
	\\
	& \mathbf{B}U(1)
	\ar@{=>} "s"; "t"
   }
   }
$$
is equivalent to the datum of a section of
the complex line bundle associated to $P$, i.e., to a prequantum state. With this identification, the action of quantomorphisms on prequantum states
%$$
%  (O_h, \psi) \mapsto O_h(\psi)
%$$
is
simply the precomposition action in the slice over $\mathbf{B}U(1)$: %$\mathbf{H}_{/\mathbf{B}U(1)}$,
$$
  \hspace{-1cm}
  \left(
   \raisebox{20pt}{
  \xymatrix{
    X \ar[dr]_{\nabla}^{\  }="t" \ar[rr]^\phi|\simeq_{\ }="s" && X \ar[dl]^{\nabla}
	\\
	& \mathbf{B}U(1)_{\mathrm{conn}}
	\ar@{=>}^{%O_h
	} "s"; "t"
   }
   }
   \;\,,\;
   \raisebox{20pt}{
  \xymatrix{
    X \ar[dr]_{P}^{\  }="t" \ar[rr]^\psi_{\ }="s" && \mathbb{C}/\!/U(1)
	\ar[dl]^{\rho}
	\\
	& \mathbf{B}U(1)
	\ar@{=>} "s"; "t"
   }
   }
  \right)
  \mapsto
    \raisebox{20pt}{
    \xymatrix{
	  X \ar[rdd]_{P}\ar[r]^\phi|\simeq_>>>>{\ }="s1" \ar[dr]|{\nabla}^{\ }="t1"
	   &
	   X \ar[d]|{\nabla}^{\ }="t2" \ar[r]^\psi_{\ }="s2" & \mathbb{C}/\!/U(1)
	   \ar[ddl]^{\rho}
	  \\
	  & \mathbf{B}U(1)_{\mathrm{conn}}\ar[d]
	  \ar@{=>} "s1"; "t1"
	  \ar@{=>} "s2"; "t2"\\
	  &\mathbf{B}U(1)
	}
	}
$$
where the vertical morphism $\mathbf{B}U(1)_{}\mathrm{conn}\to \mathbf{B}U(1)$ is the obvious forgetful morphism.
Differentiating this action, one obtains a canonical action of the Poisson bracket Lie algebra of
a (prequantized) symplectic manifold on the vector space of its prequantum states.
We show below in Section \ref{TheHigherPoissonBracketExtension} that this construction
 sends a Hamiltonian  function $H \in C^\infty(X)$ to the associated prequantum operator $O_H$ acting on a prequantum state $\psi$ as
$$
  O_H \colon\psi \mapsto \frac{i}{2\pi} \nabla_{v_H}\psi + H \psi
  \,,
$$
where the first term is the covariant derivative of the prequantum state $\psi$ along the Hamiltonian vector field $v_H$ corresponding to $H$. This is the traditional formula for the action of the Poisson algebra on prequantum states, see for instance {\cite[p. 94]{BatesWeinstein}}.

Once  prequantum geometry is formulated this way in terms of the slice over
the smooth stack of $U(1)$-principal connections, it is clear how it may naturally be promoted to {higher differential geometry}, simply by replacing the stack $\mathbf{B}U(1)_{\mathrm{conn}}$ by its higher versions $\mathbf{B}^nU(1)_{\mathrm{conn}}$, for $n\geq 1$ \cite{FScSt, NikolausSchreiberStevensonI, NikolausSchreiberStevensonII}. For instance, $\mathbf{B}^2U(1)_{\mathrm{conn}}$ is the higher moduli stack of $U(1)$-bundle gerbes with connections, and the Wess-Zumino-Witten gerbe is naturally interpreted as a map $G\to \mathbf{B}^2U(1)_{\mathrm{conn}}$ giving a prequantization of the canonical closed 3-form on a compact simple and simply connected Lie group $G$ in this context.
We discuss this example below in Section \ref{ExtendedWZW}.
%
%Morever,
%by carefully abstracting the minimum number of axioms on the ambient toposes actually needed
%in order to express the relevant
%constructions (this we discuss in \ref{HigherDifferentialGeometryInIntroduction}) one
%obtains generalizations to various other flavors of higher/derived geometry, such as
%higher/derived supergeometry.

Just as traditional prequantum geometry %and contact geometry
is of interest in itself,  this higher prequantum geometry is of interest in itself, and this is one of the motivations for
studying higher prequantum geometry.  But the strongest motivation for studying prequantum geometry is, as the name indicates,
as a means in quantum mechanics and quantum field theory. In
the next section \ref{TheNeedForHigherPrequantumGeometryFromExtendedLocalGeometricQuantization}
we discuss how the generalization of traditional geometric quantization to
local ``extended'' quantum field theory involves higher geometric prequantum theory. As an example,
in Section \ref{Examples} we indicate how various higher central extensions
of interest in {string geometry} can be constructed as higher Heisenberg-group
extensions in higher prequantum geometry.

\subsection{The need for higher prequantum theory}
 \label{TheNeedForHigherPrequantumGeometryFromExtendedLocalGeometricQuantization}

Important examples of prequantum bundles turn out to be
{transgressions} of {higher geometric bundles} to mapping spaces or more
generally to mapping stacks  \cite{FiorenzaSatiSchreiberCup, FiorenzaSatiSchreiberCS, SchreiberMPI}.

A classical example is the canonical prequantum bundle
over the loop group $LG$ of a compact simply connected group $G$. This prequantum bundle, whose
geometric quantization induces the positive energy representations of $LG$
\cite{Segal},  is the transgression of the WZW bundle gerbe on $G$ to the loop space of $G$.

Another example is the  Chern-Simons theory  prequantum
bundle on the space of $G$-principal connections over a oriented surface $\Sigma_2$,
which is the transgression to the mapping stack $[\Sigma_2,\mathbf{B}G_\mathrm{conn}]$ of the
morphism $\mathbf{B}G_{\mathrm{conn}}\to \mathbf{B}^3 U(1)_{\mathrm{conn}}$ induced by the canonical 3-cocycle on the Lie algebra of $G$, where $\mathbf{B}G_{\mathrm{conn}}$ denotes the stack of
 $G$-principal connections  \cite{FiorenzaSatiSchreiberCup, FiorenzaSatiSchreiberCS}. The curvature 2-form of this canonical prequantum bundle on $[\Sigma_2,\mathbf{B}G_\mathrm{conn}]$ is non-degenerate when restricted to the substack of flat $G$-connections on $\Sigma_2$, and this gives the canonical symplectic structure on the moduli space of flat $G$-connections on a Riemann surface as in { \cite{Witten89}}.
Since the WZW gerbe itself can be expressed in terms of the transgression of the canonical morphism $\mathbf{B}G_{\mathrm{conn}}\to \mathbf{B}^3 U(1)_{\mathrm{conn}}$ to the moduli stack of maps from $S^1$ to $\mathbf{B}G_{\mathrm{conn}}$, one sees that the first example is actually a particular instance of the second one, see the discussion in Section \ref{ExtendedWZW} below.

A motivation behind the study of higher geometry is that transgression in general loses
information, so that it is better to study the higher geometric pre-images before
transgression. An archetypical example of this phenomenon, that has motivated
many of the developments we are building on, is the relationship between
\emph{string structures} on a space $X$ and their transgression to
spin structures on the loop space $LX$.  A review and careful analysis of this case is in \cite{Waldorf12}.
This example suggests that when faced with a
prequantum bundle that is the transgression of a higher bundle, one should
de-transgress the traditional prequantum structure to a
higher prequantum structure.

For instance,  for the geomeric quantization of loop groups by the orbit method
\cite{Kirillov, Segal} such a detransgression
has been suggested some twenty years ago in \cite[p.\ 249]{BrylinskiLoop}:
``The main open question seems to be to obtain the representation theory of
$\hat LG$ from the canonical sheaf of groupoids on $G$ [\dots]
we ask whether some quantization method exists based on the sheaf of groupoids''.
%Here \emph{sheaf of groupoids} means \emph{stack of groupoids},
%and
Although there has been some progress (see \cite{Nuiten} and see our
discussion in  Section \ref{ExtendedWZW} below)
the question remains open.
This question was also a
motivation behind the investigation of the homotopy algebra aspects of multisymplectic geometry
developed in \cite{Rogers} and which led to a higher
  Bohr-Sommerfeld-like geometric quantization procedure for manifolds equipped
  with closed integral 3-forms \cite[Chap.\ 7]{RogersThesis}. This procedure
  includes not just higher prequantization, but a higher notion of real
  polarizations, namely, \emph{2-polarizations}, as well as of quantum 2-states. The output is
  a certain linear category, examples of which include the categories of
  representations of compact Lie groups. Some of the results which we present here may be seen as a natural development of constructions introduced in \cite{RogersThesis, Rogers}.
  In \cite{Nuiten}, using \cite{Bongers}, it is shown that prequantum 2-geometry may be geometrically quantized
  by pull-push in twisted K-theory. Here the role of choices of polarization is replaced by choices of
  orientations in K-theory. As discussed there, this suggests that higher prequantum geometry is naturally
  polarized and then quantized by choices of orientations for generalized cohomlogy theories of higher chromatic degree,
  but details remain to be worked out.

Finally, not unexpectedly, higher prequantization naturally makes its appearance when considering quantized field theories. Namely, the classification of local (or ``fully extended'')
topological quantum field theories (TQFTs) in \cite{LurieTQFT} shows that an $n$-dimensional
local quantum field theory assigns not just a vector space
of quantum states in codimension 1, but assigns some kind of
\emph{$k$-module of quantum  $k$-states}
in each codimension $k$ (see \cite{QFTIntroduction} for a survey of the formalization
of quantum field theory in higher category theory).
Moreover, all
these assignments are entirely determined by the assignment of a single $n$-module of
$n$-states over the point (i.e., in codimension $n$). This fully localized extension of
TQFTs is thought to reveal deep structures in quantum field theories of relevance in
low dimensional topology, such as Chern-Simons theories and their various siblings and higher
generalizations; see for instance \cite{Freed}. What is still missing is a refinement of the process of {geometric quantization}
that goes along with this local and homotopy-theoretic refinement of quantum field theory;
hence what is missing is a rigorous way to construct and control fully extended TQFTs from and by
prequantum geometric data.
Yet, the TQFT classification theorem suggests that to geometrically quantize a local
$n$-dimensional quantum field theory, one should have
a \emph{prequantum $n$-bundle}  over the higher moduli stack
of fields of the theory, and a notion of
(polarized) sections of a suitably associated higher vector bundle. The $n$-vector space of these sections would then be the
\emph{$n$-module of quantum $n$-states}
that, thanks to \cite{LurieTQFT}, would define the local TQFT, see \cite{FiorenzaSatiSchreiberCS, SchreiberSingapore}.
As in the traditional story, this process should proceed in two steps:
first higher prequantization, then higher polarization. Here we are concerned with
the first step.

%\begin{center}
%\begin{tabular}{|ccc|}
%  \hline
%  differential geometric (Lagrangian) data &$\xymatrix{\ar[rr]^{\mathrm{quantization}} &&}$& quantum field theory
%  \\
%  \hline\hline
%  traditional geometric prequantum theory &&  quantum mechanics
%  \\
%  \hline
%  higher geometric prequantum theory && local (extended) QFT
%  \\
%  \hline
%\end{tabular}
%\end{center}

At the level of the symplectic form data,
the physics literature describes this need of passing to higher codimensions as the
problem of  ``non-covariance of canonical quantization'', which refers
to the choice of spatial slices of spacetime involved in the construction
of the space of states of an $n$-dimensional quantum field theory in codimension 1.
There are at least two well established techniques for dealing with this by a ``covariant''
procedure that refines the symplectic form on the space
of fields over a chosen Cauchy surface to a differential $(n+1)$-form
on the jet bundle of the fields bundle; namely,
\emph{multisymplectic geometry} and the \emph{covariant phase space method}.
A review of both of these and a discussion of how they are related is in \cite{FR}.
This naturally leads to the following
question: what is to multisymplectic forms as
geometric prequantization is to symplectic forms?
For instance: what does replace the Poisson bracket Lie algebra as we pass from
global observables given by functions on phase space to local observables
given by differential forms (or currents) on the extended phase space?
In \cite{OtherAttempts,OtherAttempts2} it is observed that
the collection of differential form observables in such a context inherits the structure of a
(graded) Lie algebra only after exerting some force, i.e.,  after restricting to smaller subspaces
and/or after quotienting out terms that would
otherwise violate the Jacobi identity. The crucial insight of \cite{Rogers} is
that these terms that violate the Jacobi identity are coherent and hence instead of
being a nuisance are part of a natural but higher structure: Hamiltonian $(n-1)$-form observables
together with all lower degree form observables (i.e., not discarding or quotienting out any of them) constitute
not a Poisson bracket Lie algebra but its homotopy-theoretic refinement,
an \emph{$L_\infty$-algebra of local observables}.
This is exactly what one expects to see in a higher geometric version of geometric quantization
by the above reasoning, and one of purposes of the present article is to show how these
higher-Poisson bracket $L_\infty$-algebras of local observables are part of a general
theory of higher geometric prequantization. A comprehensive theory of such higher prequantized
covariant phase spaces in higher codimension is laid out in \cite{KhavkineSchreiber}.

\section{Smooth higher stacks}
 \label{Survey}

This section surveys some basics of the theory of smooth higher stacks, expanding on those items that we need below in
Section \ref{higherprequantumbundles}.
A comprehensive account is in \cite{dcct}, further exposition is in \cite{FiorenzaSatiSchreiberCS}.

 \subsection{Basic notions}
In this section we briefly recall a few basic constructions in smooth higher stacks that will be needed to present the main examples we are interested in. We point the reader to \cite{FScSt,FiorenzaSatiSchreiberCup} for an introduction to the language of smooth higher stacks and to \cite{dcct} for a comprehensive treatment. At the same time we reassure the reader unfamiliar with the language of stacks that everything works precisely as the intuition suggests and that all the examples and constructions in this section will be presented at a level of detail making them self-explanatory. Basically, all that will be needed is the following informal definition of what a (higher) smooth stack is: it is a rule $\mathbf{S}$ associating to a smooth manifold $X$ an $\infty$-groupoid $\mathbf{S}(X)$ in such a way that  $\mathbf{S}(X)$ can be pulled back along smooth maps of smooth manifolds and $\mathbf{S}(X)$ can be described by local data $\mathbf{S}(U_i)$, $\mathbf{S}(U_{ij}), \dots$ in terms of an open cover $\{U_i\}_{i\in I}$ of $X$. In fancier terms, this is expressed by saying that $\mathbf{S}$ is a contravariant functor on the site of smooth manifolds with values in Kan complexes\footnote{That is, fibrant simplicial sets, i.e., those particular simplicial sets which satisfy the ``horn fillings'' conditions, see \cite{GoerssJardine}} (i.e., it is what is called a smooth higher pre-stack) such that the descent data are effective. By the very definition above, one sees that essentially all constructions in differential geometry are smooth stacks: everything that can be described by local data and pulled back is a smooth stack. So, for instance, one has a stack $\Omega^n$ of smooth differential $n$-forms, and a stack $\Omega^n_{\mathrm{cl}}$ of closed $n$-forms. Note however that there is no stack $\Omega_{\mathrm{ex}}$ of exact $n$-forms since the pullback of an exact form is not necessarily exact.

If the $\infty$-groupoid $\mathbf{S}(X)$ has no non-identity $k$-morphisms for $k>n$, i.e., in more colloquial terms, if there are no local data beyond the $n$-fold intersections $U_{i_1,\dots,i_n}$ and no constrains on these local data beyond the $(n+1)$-fold intersections $U_{i_1,\dots,i_{n+1}}$, one says that $\mathbf{S}$ is an $n$-stack. In particular $0$-stacks are equivalently smooth sheaves of sets, while smooth 1-stacks are what is commonly called stack in the literature. However, to avoid repeating ``higher'' each time, we will always mean ``higher stack'' when we will write ``stack" in what follows, and we will explicitly write 1-stack when we want to emphasise that a certain stack is actually a 1-stack. For instance, both $\Omega^n$ and $\Omega^n_{\mathrm{cl}}$ are 0-stacks. Another classical example of a 0-stack is provided by a smooth manifold $M$, identified with the stack that maps a smooth manifold $X$ to the set $C^\infty(X,M)$ of smooth maps from $X$ to $M$. Note that, by Yoneda lemma, the smooth manifold $M$ is completely encoded in the 0-stack it represents. A basic example of a smooth 1-stack is the stack $\mathbf{B}G$ of principal $G$-bundles for a given Lie group $G$.

\begin{remark}
Since a smooth $m$-dimensional manifold $X$ always admits a good open cover, i.e., an open cover $\{U_i\}_{i\in I}$ such that all the nonempty $n$-fold intersections $U_{i_1,\dots,i_n}$ are diffeomorphic to the Cartesian space $\mathbb{R}^m$, smooth stacks can be equivalently defined on the site of all Cartesian spaces, with smooth maps between them as morphisms, and good open covers as coverings. For how simple this observation is, it provides an extremely powerful method to produce smooth stacks: one defines a prestack $\mathcal{S}$ on the site of Cartesian spaces and then stackifies it to get a stack on the site of smooth manifolds.
\end{remark}
\begin{example}
Given a Lie group $G$, consider the prestack on the site of Cartesian spaces given by
\[
\mathcal{S}_G\colon U\mapsto */\!/C^\infty(U,G)
\]
where on the right one has the action groupoid for the (trivial) action of the group $C^\infty(U,G)$ on the point.\footnote{More precisely, we should have written $N(*//C^\infty(U,G))$ on the right, where $N$ is the nerve functor from groupoids to Kan complexes. We will always identify a groupoid with its nerve in what follows.} The associated stack $\mathbf{S}_G$ works as follows: for $X$ a smooth manifold, let  $\{U_i\}_{i\in I}$ be a good open cover of $X$. Then an object in $\mathbf{S}_G(X)$ consists of the following local data:
\begin{itemize}
\item smooth maps $g_{ij}\colon U_{ij}\to G$ for every $i,j\in I$
\end{itemize}
such that
\begin{itemize}
\item  $g_{ij}g_{jk}g_{ki}=1_G$ on $U_{ijk}$ for every $i,j,k\in I$
\end{itemize}
But this are precisely the local data for a $G$-principal bundle on $X$. Similarly one sees that 1-morphisms in $\mathbf{S}_G(X)$ are precisely gauge transformations of $G$-principal bundles, and that there are no nontrivial higher morphism: $\mathbf{S}_G$ is precisely the stack $\mathbf{B}G$ of $G$-principal bundles.
\end{example}

\begin{example}
The above example has an immediate generalisation: to a Lie groupoid
  $\mathcal{G} = \left(\xymatrix{\mathcal{G}_1 \ar@<+4pt>[r]^t \ar@<-4pt>[r]_s\ar@{<-}@<+0pt>[r] & \mathcal{G}_0}\right)$
  is associated the prestack on Cartesian spaces
   $$
    \mathcal{S}_\mathcal{G} \colon U \mapsto
	\left(\xymatrix{C^\infty(U,\mathcal{G}_1) \ar@<+4pt>[rr]^-{C^\infty(U,t)} \ar@<-4pt>[rr]_-{C^\infty(U,s)}
	 \ar@{<-}@<+0pt>[rr] && C^\infty(U,\mathcal{G}_0})\right)\,.
  $$
Its stackification $\mathbf{S}_{\mathcal{G}}$ associates with a smooth manifold $X$ (with a given open cover $\{U_i\}_{i\in I}$) the groupoid whose objects are
\begin{itemize}
\item smooth maps $\gamma_i\colon U_{i}\to \mathcal{G}_0$ for every $i\in I$;
\item smooth maps $g_{ij}\colon U_{ij}\to \mathcal{G}_1$ for every $i,j\in I$
\end{itemize}
such that
\begin{itemize}
\item $\gamma_i\bigr\vert_{U_{ij}}=s\circ g_{ij}$ and $\gamma_{j}\bigr\vert_{U_{ij}}=t\circ g_{ij}$  for every $i,j\in I$;
\item  $g_{ij}\circ g_{jk}\circ g_{ki}=\iota\circ \gamma_{i}\bigr\vert_{U_{ijk}}$ on $U_{ijk}$ for every $i,j,k\in I$
\end{itemize}
where $\iota\colon \mathcal{G}_0\to \colon  \mathcal{G}_1$ is the identity section of the Lie groupoid $\mathcal{G}$. Similarly, one works out the explicit local data expression for the morphisms in $\mathbf{S}_{\mathcal{G}}$. From this explicit description one sees that $\mathbf{S}_{\mathcal{G}}(X)$ is precisely the groupoid of smooth maps from the manifold $X$ to the Lie groupoid $\mathcal{G}$.
%, see {\color{red}\cite{???}}.
See \cite{Orbifolds, Orbifolds2} for how orbifolds and foliations are special cases of Lie groupoids
  and therefore can naturally be seen as smooth stacks. Details on this example can be found in \cite{FScSt, NikolausSchreiberStevensonII,dcct}.
  \label{ToposOfSmoothInfinityGroupoids}
\end{example}

\begin{example}
Let $M$ be a smooth manifold and let $\mathcal{V}_M = \left(\xymatrix{\coprod_{\alpha,\beta}V_{\alpha\beta} \ar@<+4pt>[r]^{\partial_0} \ar@<-4pt>[r]_{\partial_1}\ar@{<-}@<+0pt>[r] & \coprod_\alpha V_\alpha}\right)$
be the Lie groupoid associated with a open cover $\mathcal{V}=\{V_\alpha\}_{\alpha\in A}$ of $M$. Then the stack $\mathbf{S}_{\mathcal{V}_M}$ is equivalent to $M$. Namely, given a smooth manifold $X$ endowed with a good open cover $\mathcal{U}=\{U_i\}_{i\in I}$,  local data for $\mathbf{S}_{\mathcal{V}_M}(X)$ reduce to
\begin{itemize}
\item a choice of an index $\alpha(i)$ in $A$ for any $i\in I$;
\item smooth maps $f_i\colon U_i\to V_{\alpha(i)}$, for any $i\in I$;
\end{itemize}
such that
\begin{itemize}
\item $f_i\bigr\vert_{U_{ij}}=f_j\bigr\vert_{U_{ij}}$, for any $i,j\in I$,
\end{itemize}
\end{example}
and these are precisely the local data for a smooth map from $X$ to $M$.
\subsection{The $\infty$-topos of smooth higher stacks}

Exactly as the category of sheaves of sets on a site is a topos, the $\infty$-category of higher stacks on a site is an $\infty$-topos \cite{Lurie}. So in particular we have the $\infty$-topos of smooth (higher) stacks, which we will denote by the symbol $\mathbf{H}$. Since $\mathbf{H}$ is an $\infty$-category, given two smooth stacks $X$ and $A$, we have a \emph{space}\footnote{i.e., equivalently, an $\infty$-groupoid.} of morphisms $\mathbf{H}(X,A)$, rather than just a set of morphism as we would have had in the case of an ordinary category. The set of path connected components of $\mathbf{H}(X,A)$ will be denoted by $H^0(X,A)$, i.e., we set
\[
H^0(X,A)=\pi_0\mathbf{H}(X,A).
\]
As this notation suggests, $H^0(X,A)$ is to be thought of as the 0-th cohomology group of $X$ with coefficients in $A$; this point of view will be supported by several examples below.
Also notice that, by the $\infty$-Yoneda lemma \cite[lemma 5.5.2.1]{Lurie}, if $X$ is a smooth manifold and $\mathbf{S}$ is a smooth stack, then one has a natural equivalence
\[
\mathbf{S}(X)\cong \mathbf{H}(X,\mathbf{S}),
\]
where on the right hand side $X$ stands for the smooth 0-stack it represents.

\begin{remark}
As in traditional homotopy theory, when we draw a commuting diagram of morphisms
in an $\infty$-category, it is always understood that they commute
up to a specified homotopy. We will often notationally suppress these homotopies that fill diagrams,
except if we want to give them explicit labels.
For instance, in the figure below, the diagram of morphisms in $\mathbf{H}$
on the left hand side always means the more explicit diagram displayed on the right hand side:
$$
  \raisebox{20pt}{
  \xymatrix{
    X \ar[rr] \ar[dr] \ar[rr] && Y \ar[dl]
	\\
	& Z
  }}
  \;\
  :=
  \raisebox{20pt}{
  \xymatrix{
    X \ar[rr] \ar[dr]^{\ }="t" \ar[rr]_{\ }="s" && Y \ar[dl]
	\\
	& Z
	\ar@{=>} "s"; "t"
  }}
  \,.
$$
In the same spirit all the universal constructions that we mention in the following refer to their
homotopy-correct version. For instance, ``fiber product'' will always mean ``homotopy fiber product''. With homotopies thus understood, most of the familiar basic
facts of category theory generalize verbatim to $\infty$-category theory.
For instance,
a basic fact that we make repeated use of is the pasting law for homotopy pullbacks:
if we have two adjacent square diagrams and the right (resp., the left) square is a homotopy pullback, then
the left (resp., the right) square is  a pullback if and only if the total rectangle is.
\end{remark}

{
\subsection{Internal homs and automorphism groups as objects in $\mathbf{H}$}
\label{InternalHoms}
\label{group-stacks}

An important feature of the $\infty$-category $\mathbf{H}$ of higher
smooth stacks is that it has internal homs. Namely, given two stacks
$X$ and $Y$ one can give a meaning to the notion of a \emph{family} of
morphisms from $X$ to $Y$ parametrized by a stack $\mathbf{S}$, simply
by declaring such a family to be a morphism of stacks $X\times
\mathbf{S}\to Y$. It turns out that the $\infty$-groupoid valued functor
\[
\mathbf{S}\mapsto \mathbf{H}(X\times \mathbf{S},Y)
\]
is representable, i.e., there exists a stack $[X,Y]$ in $\mathbf{H}$ (unique up to natural equivalence) such that
\[
\mathbf{H}(X\times \mathbf{S},Y)\cong \mathbf{H}(\mathbf{S},[X,Y]),
\]
naturally in all variables. (See \cite{Lurie,dcct}.)
The stack $[X,Y]$ is called the \emph{internal hom} of $X$ and $Y$. Notice how, by definition, a morphism $*\to [X,Y]$ is precisely a morphism between $X$ and $Y$. Also notice that, if $f\colon X\to Y$ is a morphism in $\mathbf{H}$ then one has morphisms of stacks
\[
f_*:=f\circ -\colon [Z,X] \to [Z,Y] ;\qquad \qquad f^*:=-\circ f\colon [Y,Z] \to [X,Z].
\]

If $X=Y$, then one has the stack $\mathbf{End}(X)=[X,X]$, which is the endomorphism stack of $X$. It is manifestly a monoid object in $\mathbf{H}$, with identity element $\mathbf{id}_X\colon *\to  \mathbf{End}(X)$ corresponding to the identity morphism of $X$.  We denote by $\mathbf{Aut}(X)$ the stack of automorphisms of $X$, i.e., the substack of $\mathbf{End}(X)$ consisting of invertible endomorphisms of $X$.

Now, let $f\colon X\to Z$ and $g\colon Y\to Z$ be fixed morphisms in $\mathbf{H}$. Then both $f$ and $g$ are objects in the slice category $\mathbf{H}_{/Z}$ and so we have a natural notion of morphisms between $f$ and $g$: they are homotopy commutative diagrams
 \[
 \raisebox{20pt}{
  \xymatrix{
    X \ar[rr]^{\varphi} \ar[dr]^{\ }="t"_{f} \ar[rr]_{\ }="s" && Y \ar[dl]^{g}
	\\
	& Z
	\ar@{=>}^\eta "s"; "t"
  }}
  \,.
\]
 We want to show that there exists also a natural notion of a stack of morphisms between $f$ and $g$. To do this, we must give a meaning to a family of morphisms between $f$ and $g$, parametrized by a stack $\mathbf{S}$. Let us start by considering the case when $\mathbf{S}$ is a smooth manifold, so that we can confidently talk of its points. Then, naively, a family of morphisms between $f$ and $g$ parametrized by $\mathbf{S}$ would be a collection of homotopy commutative diagrams
 \[
 \raisebox{20pt}{
  \xymatrix{
    X \ar[rr]^{\varphi_s} \ar[dr]^{\ }="t"_{f} \ar[rr]_{\ }="s" && Y \ar[dl]^{g}
	\\
	& Z
	\ar@{=>}^{\eta_s} "s"; "t"
  }}
  \,,
\]
with $s$ varying in $S$. Now, the family of morphisms $\{\varphi_s\}$ is the datum of a single morphism $\varphi\colon X\times \mathbf{S}\to Y$, and the family of homotopies $\{\eta_s\}$ is the datum of a single homotopy from $g\circ \varphi$ to the morphism
$f_{\mathbf{S}}^{}\colon X\times \mathbf{S}\xrightarrow{f\times  \pi_{\mathbf{S}}} Z\times \{*\}\cong Z$. This way we have obtained a definition of an $\mathbf{S}$-family of morphisms betwen $f$ and $g$, which is meaningful for an arbitrary stack $\mathbf{S}$: it is the datum of a homotopy commutative diagram
 \[
 \raisebox{20pt}{
  \xymatrix{
    X\times \mathbf{S} \ar[rr]^{\varphi} \ar[dr]^{\ }="t"_{f_\mathbf{S}^{}} \ar[rr]_{\ }="s" && Y \ar[dl]^{g}
	\\
	& Z
	\ar@{=>}^{\eta} "s"; "t"
  }}
  \,.
\]
In other words, the $\infty$-groupoid of $\mathbf{S}$-families of morphisms
betwen $f$ and $g$ is $\mathbf{H}_{/Z}(f_{\mathbf{S}}^{},g)$. Since $f_{\mathbf{S}}^{}$ is the product in $\mathbf{H}_{/Z}$ of $f\colon X\to Z$ with $(\mathrm{id}_{Z})_{\mathbf{S}}\colon Z\times \mathbf{S}\to Z$, this home space is $\mathbf{H}_{/Z}( (\mathrm{id}_Z)_{\mathbf{S}}^{}\times f,g)$ and so,  by definition of the internal hom in $\mathbf{H}_{/Z}$, it is equivalent to
\[
\mathbf{H}_{/Z}((\mathrm{id}_Z)_{\mathbf{S}}^{},[f,g]_{\mathbf{H}_{/\mathbf{S}}}).
\]
The morphism $(\mathrm{id}_Z)_{\mathbf{S}}^{}\colon Z\times \mathbf{S}\to Z$ is nothing but the projection on the first factor, and the functor ``multiply by $Z$''
\begin{align*}
\mathbf{H}&\xrightarrow{Z\times -} \mathbf{H}_{/Z}\\
\mathbf{S}&\mapsto (\mathrm{id}_Z)_{\mathbf{S}}^{}
\end{align*}
has a right adjoint
\[
\prod_Z\colon \mathbf{H}_{/Z}\to \mathbf{H},
\]
called the \emph{dependent product} over $Z$ \cite[Section 6.3.5]{Lurie}. Therefore the $\infty$-groupoid of $\mathbf{S}$-families of morphisms
betwen $f$ and $g$ is
\[
\mathbf{H}(\mathbf{S},\prod_Z[f,g]_{\mathbf{H}_{/\mathbf{S}}}),
\]
i.e. the stack of morphisms betwen $f$ and $g$ is $
[f,g]_{\mathbf{H}}:=\prod_Z[f,g]_{\mathbf{H}_{/\mathbf{S}}}$. Similarly, we denote by $\mathbf{End}_{\mathbf{H}}(f)$ the stack of endomorphisms of a morphism $f\colon X\to Z$ and by $\mathbf{Aut}_{\mathbf{H}}(f)$ the substack of automorphisms of $f$.

\begin{remark}
The ``multiply by $Z$'' functor has also a left adjoint, which is nothing but the forgetful functor
\begin{align*}
\mathbf{H}_{/Z}&\longrightarrow \mathbf{H}\\
\{f\colon X\to Z\}&\,\,\mapsto\,\, X
\end{align*}
keeping only the source of a morphism to $Z$. This functor is called the \emph{dependent sum} over $Z$ and denoted by the symbol $\sum_Z$.
\end{remark}

\begin{example}
\label{Bisections}
If
$\mathcal{G} = \left(\xymatrix{\mathcal{G}_1 \ar@<+4pt>[r]^t \ar@<-4pt>[r]_s\ar@{<-}@<+0pt>[r] & \mathcal{G}_0}\right)$ is a Lie groupoid,
a \emph{bisection} of $\mathcal{G}$ is defined to be a pair $(\phi,\sigma)$, where $\phi\colon \mathcal{G}_0\to \mathcal{G}_0$ is a diffeomorphism, and $\sigma\colon \mathcal{G}_0 \to \mathcal{G}_1$ is a smooth function such that at any point $x$ in $\mathcal{G}_0$,
\[
\sigma(x)\colon x\to \phi(x)
\]
is a morphism in $\mathcal{G}$. Note that, since $\phi$ is completely determined by $\sigma$, a bisection of $\mathcal{G}$
is equivalently a smooth map $\sigma\colon \mathcal{G}_0 \to \mathcal{G}_1$ such that $s \circ \sigma = \mathrm{id}_{\mathcal{G}_0}$
and such that $\phi := t \circ \sigma : \mathcal{G}_0 \to \mathcal{G}_0$
is a diffeomorphism, see e.g. \cite[p. 114]{MoerdijkMrcun}.
%\footnote{\color{red}I've found several references for this definition, unsure on which to chose: of course,
%the more classical the better. I'd like to have a reference here since there is also another notion of bisection,
%which is the one which Mackenzie uses and which is maybe more natural, in which one does not go from id
%to $\phi$ but from $\phi_0$ to $\phi_1$ for two arbitrary diffeomorphisms. With this definition one does not
%have a group of bisections but a groupoid of bisections.}
From the perspective of smooth stacks, the definition of a bisection of $\mathcal{G}$ admits an extremely simple rephrasing: it is the datum of an automorphism of $\mathcal{G}_0$ over $\mathcal{G}$, i.e., a homotopy commutative diagram of the form
$$
    \raisebox{20pt}{
    \xymatrix{
	  \mathcal{G}_0 \ar@{->>}[dr]_{\iota_{\mathcal{G}}}^{\ }="t" \ar[rr]^{\phi}|\simeq_{\ }="s" && \mathcal{G}_0
	  \ar@{->>}[dl]^{\iota_{\mathcal{G}}}
	  \\
	  & \mathcal{G}
	  \ar@{=>}^{\sigma^{-1}} "s"; "t"
	}
	}
  \,,
$$
where $\iota_{\mathcal{G}}\colon \mathcal{G}_0\to \mathcal{G}$ is the morphism of smooth stacks given by the identities in $\mathcal{G}$. From this description we do not only recover the group structure on the set of bisections, but we see we have a natural notion of the group stack of bisections:
\[
\mathbf{BiSect}(\mathcal{G}): = \mathbf{Aut}_{\mathbf{H}}(\iota_{\mathcal{G}}).
\]
More generally, one can define this way the (higher) group stack of bisections of a groupoid object in $\mathbf{H}$.
%In view of prop. \ref{1EpisAreEquivalentToGroupoidObjects} and
%remark \ref{HValuedHomsInSlice} we have the following natural
%generalization of the notion of groupoid bisections to higher geometry.
%
\end{example}
%\begin{definition}
% \begin{enumerate}
%% \item
%% For $G \in \mathrm{Grp}(\mathbf{H})$,
%% the \emph{$\mathbf{H}$-valued automorphism group} of a $G$-action $\rho$ is
%%$$
%%  \mathbf{Aut}_{\mathbf{H}}(\rho) := \underset{\mathbf{B}G}{\prod} \mathbf{Aut}(\rho)
%%  \,.
%%  \label{HValuedAutomorphismsGrouInIntroduction}
%%$$
%\item For $\mathcal{G}_\bullet \in \mathrm{Grpd}(\mathbf{H})$ a groupoid
% object, def. \ref{GroupoidObjects}, its \emph{$\infty$-group of bisections} is
% $$
%   \mathbf{BiSect}(\mathcal{G}_\bullet)
%   :=
%   \underset{\mathcal{G}}{\prod} \mathbf{Aut}(p_{\mathcal{G}})
%   \,,
% $$
%  where $p_\mathcal{G} : \xymatrix{\mathcal{G}_0 \ar@{->>}[r] & \mathcal{G}}$ is the
%  atlas of $\infty$-groupoids which corresponds to $\mathcal{G}_\bullet$ under
%  the equivalence of prop. \ref{EquivalenceOfGPrincipalInfinityBundlesWithCocycles}.
%\end{enumerate}
 \label{BisectionInIntroduction}
%\end{definition}

The following proposition is an exercise in the definitions.
\begin{proposition}\label{prop.aut-in-slice}
Let $f\colon X\to Y$ be a morphism in $\mathbf{H}$. Then one has a natural fiber sequence
\[
\xymatrix{
\mathbf{Aut}_{\mathbf{H}}(f)\ar[r]\ar[d] &\mathbf{Aut}(X)\ar[d]^{f_*}\\
{*}\ar[r]^{f} &[X,Y]
}\,.
\]
\end{proposition}
\begin{corollary}\label{cor.central-extension}
Let $f\colon X\to Y$ be a morphism in $\mathbf{H}$. Then one has a natural fiber sequence
\[
\xymatrix{
\mathbf{\Omega}_f[X,Y]\ar[r]\ar[d] &\mathbf{Aut}_{\mathbf{H}}(f)\ar[d]\\
{*}\ar[r] &\mathbf{Aut}(X)
}\,,
\]
where $\mathbf{\Omega}_f[X,Y]$ denotes the loop space of $[X,Y]$ based at the point $f$.
\end{corollary}

%\begin{proposition}
%
%  For $(\xymatrix{\underset{X}{\sum}E \ar[r]^E & X}) \in \mathbf{H}_{/X}$
%  an object in the slice $\infty$-topos of $\mathbf{H}$ over some $X \in \mathbf{H}$,
%  there is an $\infty$-fiber sequence in
%  $\mathbf{H}$ of the form
%  $$
%    \xymatrix{
%	  \Omega_{E}\left[
%	    \underset{X}{\sum}E, X
%	  \right]
%	  \ar[r]
%	  &
%      \mathbf{Aut}_{\mathbf{H}}\left(E\right)
%	  \ar[r]
%	  &
%	  \mathbf{Aut}\left(\underset{X}{\sum}E\right)
%	  \ar[r]^{E \circ (-)}
%	  &
%	  \left[
%	    \underset{X}{\sum}E, X
%	  \right]
%	}
%	\,.
%  $$
%  Here the object on the far right is regarded as pointed by $E$ and the object
%  on the far left is its loop space object as in prop. \ref{LoopingAndDeloopingInIntroduction}.
%  \label{TheArchetypeExtension}
%\end{proposition}
The above is the general formalization of the basic idea schematically indicated
in \ref{TraditionalPrequantumGeometryViaSlicing}.
This class of extensions is the archetype of all the $\infty$-group
extensions in higher prequantization that we will meet, namely the integrated Atiyah extension in Section \ref{HigherAtiyah} and the
quantomorphism $\infty$-group extension in Section \ref{TheCentralTheorems}.

}

\subsection{Principal $G$-bundles, associated bundles and their sections}
 \label{HigherFiberBundles}

Let $G$ be a Lie group. We have already met the stack $\mathbf{B}G$ of
principal $G$-bundles. Moreover, by the considerations in the previous
section, we have that a morphism $\nabla^0\colon X\to \mathbf{B}G$
from a smooth manifold $X$ to $\mathbf{B}G$ in $\mathbf{H}$ is
precisely the datum of a principal $G$-bundle $P\to X$ over $X$.
\footnote{Here and in the following we tend to denote modulating maps
  of principal $G$- bundles by $\nabla^0$, because in
higher geometric prequantization it is natural to regard these maps as the leftmost stage in a sequence of
analogous but richer maps whose rightmost stage is a principal $G$-connection.
} Even more precisely, we have a natural equivalence of $\infty$-groupoids between $\mathbf{H}(X,\mathbf{B}G)$ and the groupoid of principal $G$-bundles over $X$ with gauge transformations as morphisms. We are now going to describe a few classical constructions in the theory of principal bundles from the point of view of smooth stacks. A reference for all the results in this section, with complete proofs and additional details is \cite{NikolausSchreiberStevensonI}.

The stack $\mathbf{B}G$ is a pointed stack, with the distinguished point given by the trivial principal $G$-bundle. To any morphism  $\nabla^0\colon X\to \mathbf{B}G$ it is therefore naturally associated a stack over $X$, the fiber of $\nabla^0$. This turns out to be nothing but the total space $P$ of the $G$-bundle modulated by $\nabla^0$, i.e., we have a homotopy pullback diagram
$$
    \raisebox{20pt}{
    \xymatrix{
	  P \ar[r] \ar[d] & {*} \ar[d]
	  \\
	  X \ar[r]^{\nabla^0} & \mathbf{B}G
	}
	}
	\,.
  $$
Namely, by the universal property of the pullback, to give a morphism from a smooth manifold $Y$ to the fiber of $\nabla^0$ is equivalent to giving a smooth map $f\colon Y\to X$ together with a trivialization of the principal $G$-bundle $f^*P\to Y$. Since a trivialisation of a principal $G$-bundle is equivalent to the datum of a section, we are giving a pair $(f,\sigma)$ where $f\colon Y\to X$ is a smooth map and $\sigma\colon Y\to f^*P$ is a section of $f^*P\to Y$. But such a pair is precisely the datum of a smooth map $Y\to P$. Equivalently, one can identify the fiber of $\nabla^0$ with $P$ by a local data description: by replacing the stack $X$ by the equivalent stack presented by the cover groupoid associated with a good open cover $\mathcal{U}=\{U_i\}_{i\in I}$, one sees that the fiber of $\nabla^0$ is presented by the Lie groupoid whose manifold of objects is $\coprod_i U_i\times G$ and whose manifold of morphisms is $\coprod_{i,j}U_{i,j}$ with source and target maps given by $s(x)=(x,1_G)\in U_i\times G$ and $t(x)=(x,g_{ij}(x))\in U_j\times G$ for any $x\in U_{ij}$, and these are manifestly the glueing data for the smooth manifold $P$.

\begin{remark}
Conversely, every principal $G$-bundle $P$ over a smooth manifold $X$ can be realised as the fiber of a modulating morphism $X\to \mathbf{B}G$. This means that in the world of smooth stacks the
point inclusion  $*\to \mathbf{B}G$ is equivalently the
\emph{universal $G$-principal $\infty$-bundle} $\mathbf{E}G \to \mathbf{B}G$. Under geometric realization, the equivalence of stacks $\mathbf{E}G\cong *$ reproduces one of the most classical results from the homotopy theory of principal bundles: the total space of the universal $G$-bundle $EG\to BG$ on the classifying space $BG$ is contractible.
\label{ModuliForPrincipalBundles}
\end{remark}

\begin{remark}
As a particular case of the above construction we obtain the pullback diagram
$$
    \raisebox{20pt}{
    \xymatrix{
	  G \ar[r] \ar[d] & {*} \ar[d]
	  \\
	  {*} \ar[r]^{\nabla^0} & \mathbf{B}G
	}
	}
	\,.
  $$
exhibiting the group $G$ as the total space of the trivial $G$-bundle over the point. Since the above diagram equivalently presents the loop space of $\mathbf{B}G$, we see that we have a natural equivalence of stacks $G\cong \mathbf{\Omega}\mathbf{B}G$. In other words, $G$ is a delooping of $\mathbf{B}G$.
\end{remark}

Let now $V$ be a smooth manifold on which the Lie group $G$ acts smoothly. Then we can form the action groupoid $V/\!/G$, and so the smooth stack associated to it, which we will denote by the same symbol. Since, as noticed above, $\mathbf{B}G$ is the smooth stack associated with the action groupoid $*/\!/G$, the morphism of $G$-manifolds $V\to *$ induces a natural morphism of stacks $\rho\colon V/\!/G\to \mathbf{B}G$, which precisely encodes the action of $G$ on $V$ (so that we will denote this action by the same symbol). The following proposition expresses the unversality property of the morphism $\rho$. Its proof is similar to the argument given above to identify the total space of the principal bundle modulated by $\nabla^0$ with the fiber of $\nabla^0$ itself. A detailed proof can be found in \cite[prop. 4.6]{NikolausSchreiberStevensonI}.
\begin{proposition} The morphism of stacks $\rho\colon V/\!/G\to \mathbf{B}G$ is the \emph{universal $\rho$-associated $V$-fiber bundle}:
  for $P \to X$ a $G$-principal bundle modulated by $\nabla^0$ and for $(V,\rho)$ a $G$-action,
  the $\rho$-associated $V$-fiber bundle $P \times_G V \to X$
  fits into a homotopy pullback square of the form
  $$
    \raisebox{20pt}{
    \xymatrix{
	  P \times_G V \ar[r] \ar[d] & V/\!/G \ar[d]^\rho
	  \\
	  X \ar[r]^{\nabla^0} & \mathbf{B}G
	}
	}
	\,.
  $$
  \label{UniversalAssociatedBundle}
\end{proposition}
 \begin{remark}As an immediate corollary, we see one has a homotopy pullback diagram
$$
    \raisebox{20pt}{
    \xymatrix{
	  V \ar[r] \ar[d] & V/\!/G \ar[d]^\rho
	  \\
	  {*} \ar[r] & \mathbf{B}G
	}
	}
	\,.
  $$
This pullback is the starting point for a far reaching generalisation of the notion of smooth Lie group action in the context of smooth stacks. Namely, one can define an $\infty$-group object in $\mathbf{H}$ as a smooth stack $G$ equipped with a pointed delooping $\mathbf{B}G$, i.e., such that there exist a pointed connected stack $*\to\mathbf{B}G$ together with an equivalence $G\cong \mathbf{\Omega}\mathbf{B}G$, and call $\mathbf{B}G$ the stack of $G$-principal $\infty$-bundles. Next one can \emph{define} an $\infty$-action of $G$ on a smooth stack $V$ as a homotopy commutative diagram as the one above, where now $ V/\!/G$ is some given stack which is part of the data of the diagram: in other words an $\infty$-action of $G$ on $V$ is defined as a morphism $\rho$ to $\mathbf{B}G$ with fiber $V$. Once actions have been defined this way, one uses the statement of Proposition \ref{UniversalAssociatedBundle} to
define $\rho$-associated $V$-fiber bundles. See \cite{NikolausSchreiberStevensonI} for a comprehensive treatement of the theory of $\infty$-bundles.
\end{remark}

{
\begin{remark}
The requirement that $\mathbf{B}G$ be a connected object is motivated by the fact that the loop space functor $\mathbf{\Omega}$ only sees the connected component of the marked point. Once connectedness is required, the delooping is uniquely determined up to canonical equivalence, and the assignment $G\rightsquigarrow \mathbf{B}G$ becomes a functor. More precisely, we have a pair of inverse equivalences of
  $\infty$-categories
  $$
    \xymatrix{
     \{\text{group objects in }\mathbf{H}\}
	 \ar@/_2pc/[r]_-{\mathbf{B}}
	 	 &\simeq \qquad\,
	 \{\text{pointed connected objects in }\mathbf{H}\} \ar@{->}@/_2pc/[l]_-{\mathbf{\Omega}}
	}
  $$
   \label{LoopingAndDelooping}
\end{remark}
}

{
\begin{remark}
The above notion of group action has an immediate generalization to groupoid actions. Namely, let $\mathcal{G} = \left(\xymatrix{\mathcal{G}_1 \ar@<+4pt>[r]^t \ar@<-4pt>[r]_s\ar@{<-}@<+0pt>[r] & \mathcal{G}_0}\right)$ be a groupoid object in $\mathbf{H}$, and let $\iota_{\mathcal{G}}\colon \mathcal{G}_0\to \mathcal{G}$ the canonical inclusion of the identities.
  Then a \emph{groupoid action} of $\mathcal{G}$
  on the smooth stack $V$ is
%  another groupoid object
%  $$
%    (E/\!/\mathcal{G})_\bullet \in \mathrm{Grpd}(\mathbf{H})
%  $$
%  corresponding to an $\infty$-groupoid with atlas $\xymatrix{E \ar@{->>}[r] & E/\!/\mathcal{G}}$
%  and an $\infty$-pullback diagram of atlases of the form
 the datum of a homotopy pullback diagram
  $$
    \raisebox{20pt}{
    \xymatrix{
	 V \ar[r] \ar[d] & V/\!/\mathcal{G} \ar[d]
	  \\
	  \mathcal{G}_0 \ar[r]^{\iota_{\mathcal{G}}} & \mathcal{G}
	}
	}
	\,.
  $$
  \label{ActionOfGroupoidInIntroduction}
%
%To see heuristically how such a definition indeed encodes an action, it is helpful to
%think of path lifting: For an element $e \in E$ and a morphism
%$(p(e) \stackrel{g}{\to} y) \in \mathcal{G}^{(\Delta^1)}$ in
%$\mathcal{G}_\bullet$, the $\mathcal{G}$-action of $g$ on $e$
%corresponds to a lift of $g$ to a morphism
%$(e \stackrel{\hat g}{\to} \tilde e) \in (E/\!/\mathcal{G})^{\Delta^1}$
%in the action groupoid, which takes $e$ to a morphism $\tilde e$ sitting over $y$.
%Notice that for $\mathcal{G} \simeq \mathbf{B}G$ the delooping groupoid of
%an $\infty$-group, this reduces to the definition of actions of $\infty$-groups
%discussed around prop. \ref{SliceOverModuliStackOfGIsGAxtions}.
\end{remark}
}

\begin{remark}If $G$ is a Lie group and $P\to X$ is a $G$-principal bundle modulated by $\nabla^0\colon X\to \mathbf{B}G$,
then we have seen a homotopy pullback diagram
$$
    \raisebox{20pt}{
    \xymatrix{
	  P \ar[r] \ar[d] & X \ar[d]^{\nabla^0}
	  \\
	  {*} \ar[r] & \mathbf{B}G
	}
	}
	\,.
  $$
This exhibits the equivalence of stacks $X\cong P/\!/G$. This precisely encodes a well known fact from the classical theory of $G$-principal bundles: there is a free $G$-action on the total space $P$ of a principal $G$-bundle and the projection to the base induces a diffeomorphism of smooth manifolds $P/G\xrightarrow{\sim} X$.
\end{remark}

Let now $\pi\colon P\times_G V\to X$ be the $\rho$-associated $V$-fiber bundle for some action $\rho$ of a group $G$ on $V$ and for some principal $G$-bundle $P\to X$ modulated by some map $\nabla^0\colon X\to \mathbf{B}G$. By definition, a \emph{section} $\sigma$ of $P\times_G V\to X$ is a lift of the identity morphism of $X$ through $\pi$. Equivalently, this is a morphism $\sigma\colon \mathrm{id}_X\to \pi$ in the slice $\infty$-category $\mathbf{H}_{/X}$. This immediately tells us that we actually have a whole space of sections
\[
\Gamma_X\left(P \times_G V\right):=\mathbf{H}_{/X}(\mathrm{id}_X,\pi).
\]
By the universal property of the pullback and by Proposition \ref{UniversalAssociatedBundle} we therefore have the following
\begin{corollary}
 \label{cor.sections}
  The space of sections of a $V$-fiber bundle
  which is $\rho$-associated to a principal bundle modulated by $\nabla^0$
  is naturally equivalent to the space of maps $\Psi : \nabla^0 \to \rho$ in the slice $\infty$-category over
  $\mathbf{B}G$:
  $$
    \Gamma_X\left(P \times_G V\right)
	\cong
	\mathbf{H}_{/\mathbf{B}G}(\nabla^0,\rho)
	\,.
  $$
  \end{corollary}
  {
  \begin{remark}
 By the above Corollary, one has in particular that the equivalence classes of sections of  the $V$-fiber bundle $\rho$-associated to $\nabla^0\colon X\to \mathbf{B}G$ are the cohomology classes of $\nabla^0$ with coefficients in $\rho$. They can be interpreted as $\nabla^0$-twisted cohomology classes of $X$ with coefficients in $V$.
  \end{remark}
  }

\begin{example}The above considerations follow from universal properties and apply to any $\infty$-action of an $\infty$-group $G$ on a smooth stack $V$. It is however instructive to explicitly show how a morphism $\nabla^0\to\rho$ over $\mathbf{B}G$ precisely encodes a section of the associated bundle in the classical case of a Lie group action on a smooth manifold. Namely, in terms of a open cover $\mathcal{U}=\{U_i\}_{i\in I}$ of $X$, a morphism $X\to V/\!/G$ is given by
\begin{itemize}
\item smooth maps $\sigma_i\colon U_{i}\to V$ for every $i\in I$;
\item smooth maps $g_{ij}\colon U_{ij}\to G$ for every $i,j\in I$
\end{itemize}
such that
\begin{itemize}
\item $\sigma_i\bigr\vert_{U_{ij}}=g_{ij}\cdot \sigma_{j}\bigr\vert_{U_{ij}}$  for every $i,j\in I$;
\item  $g_{ij} g_{jk} g_{ki}=1_G$ on $U_{ijk}$ for every $i,j,k\in I$
\end{itemize}
These are precisely the local data for a section $\sigma$ of the $V$-fiber bundle associated with the cocycle $\{g_{ij}\}$. Requiring that the morphism $X\to V/\!/G$ is a morphism over $\mathbf{B}G$ is precisely the datum of an isomorphism between the principal $G$-bundle defined by the cocycle $\{g_{ij}\}$ and the principal $G$-bundle $P$ modulated by $\nabla^0$, and this isomorphism identifies $\sigma$ with a section of $P \times_G V$.
\label{LocalDataFor1Sections}
\end{example}
{
Following the argument in Section \ref{group-stacks}, we can do a further step and consider the smooth \emph{stack of sections} of $P \times_G V$,
\[
\mathbf{\Gamma}_X\left(P \times_G V\right)
	:=
	[\nabla^0,\rho]_{\mathbf{H}}
	\,.
\]
We then have the following.
\begin{proposition}\label{SectionsByMapsInSlice}
Let $P\to X$ be a $G$-principal bundle, modulated by a morphism $\nabla^0\colon X\to \mathbf{B}G$, and $\rho$ be an action of $G$ on some stack $V$. Then there is a natural action
\[
\mathbf{Aut}_{\mathbf{H}}(\nabla^0)\times \mathbf{\Gamma}_X\left(P \times_G V\right)\to \mathbf{\Gamma}_X\left(P \times_G V\right)
\]
of the group stack of automorphisms of $\nabla^0$ on the stack of sections of the $\rho$-associated $V$-bundle $P \times_G V\to X$.
\end{proposition}
}

\subsection{Higher deloopings, higher cohomology}
\label{GroupExtension}

As we have seen, an $\infty$-group object $G$ in $\mathbf{H}$ is an object admitting a delooping $\mathbf{B}G$. One can consider group objects $G$ in $\mathbf{H}$ admitting further deloopings
$\mathbf{B}^k G$, for $k\geq 1$. In particular, as we will see in Section \ref{DoldKanCorrespondence}, an abelian Lie group is infinitely deloopable, i.e., it is an  \emph{infinite loop space object} in $\mathbf{H}$. This means that for a $k$-deloopable group object $G$, the higher the value of $k$, the \emph{closer to abelian}
the $\infty$-group $G$ is.  For the lowest values of $k$, there is some established terminology in category theory: For instance a group object $G$ in $\mathbf{H}$ is called  a  \emph{braided group object}
if $\mathbf{B}G$ is equipped with a further delooping $\mathbf{B}^2 G$; notice how this is equivalent to saying that $\mathbf{B}G$ is itself equipped with the structure of a group object. It is called a \emph{sylleptic group object} if it is a 3-fold deloopable object, i.e., if it is endowed a 3-fold delooping $\mathbf{B}^3 G$. Equivalently, this means that $\mathbf{B}G$ is equipped with the structure of a braided group object. It is convenient to complete this picture by considering also $0$-deloopable objects: these are arbitrary stacks.

\begin{remark}
A braided $\infty$-group is in particular a group object whose multiplication is commutative up to homotopy. It is worth noticing that for a compact connected Lie group, this condition is equivalent to abelianity \cite{araki-james-thomas}, so the only braided compact connected Lie groups are abelian Lie groups.
\end{remark}

Let now $A$ be a $k$-deloopable object in $\mathbf{H}$, for some $k\geq 0$, with $\mathbf{B}^kA$ be a $k$-delooping of $A$. We set, for any object $X$ of $\mathbf{H}$,
\[
H^k(X,A):=\pi_0\mathbf{H}(X,\mathbf{B}^kA)
\]
and call this the $k$-th cohomology group of $X$ with coefficients in $A$. When $X$ is a smooth manifold and $A$ is an abelian Lie group, then $H^k(X,A)$ is precisely the $k$-th cohomology group of $X$ with coefficients in the sheaf of smooth $A$-valued functions, justifying the notation. Also, when $k=1$ and $A$ is a Lie group $G$, we find that for a smooth manifold $X$
\[
H^1(X,G)=\pi_0\mathbf{H}(X,\mathbf{B}G)=\{\text{principal $G$-bundles}\}/\sim
\]
since $\mathbf{B}G$ is the classifying stack for principal $G$-bundles. So again we see that the notation $H^k(X,A)$ is coherent with the usual notations from sheaf cohomology.

As a matter of notation, when $G$ is a group object in $\mathbf{H}$ we will write
\[
H^k_{\mathrm{grp}}(G,A):=H^k(\mathbf{B}G,A),
\]
i.e., $H^k_{\mathrm{grp}}(G,A)=\pi_0\mathbf{H}(\mathbf{B}G,\mathbf{B}^kA)$. The space $\mathbf{H}(\mathbf{B}G,\mathbf{B}^kA)$ will be called the space of group $k$-cocycles on $G$ with coefficients in $A$ and $H^k_{\mathrm{grp}}(G,A)$ will be called the $k$-th group cohomology group of $G$ with coefficients in $A$. Again, these notations are compatible with and generalize the usual notations from group cohomology.
\begin{remark}
If $c\colon \mathbf{B}G\to \mathbf{B}K$ is a group 1-cocycle, then its fiber is again deloopable, and so it determines a group object $H$:
\[
\xymatrix{
\mathbf{B}H\ar[r]^{\varphi_c}\ar[d] & \mathbf{B}G\ar[d]^{c}\\
{*}\ar[r] &\mathbf{B}K
}
\]
Looping this diagram we obtain the homotopy pullback diagram
\[
\xymatrix{
H\ar[r]^{\mathbf{\Omega}\varphi_c}\ar[d] & G\ar[d]^{\mathbf{\Omega}c}\\
{*}\ar[r] &K
}
\]
where now the morphisms are not only morphisms of stacks but are morphisms of $\infty$-group objects. In particular, if $G$ and $K$ were Lie groups, the morphism $\mathbf{\Omega}c\colon G\to K$ is a Lie group homomorphism, and $\mathbf{\Omega}\varphi_c\colon H\to G$ is the inclusion of the kernel of $\mathbf{\Omega}c$. Notice that every Lie group homomorphism between $G$ and $K$ can be obtained this way, that is, Lie group homomorphisms from $G$ to $K$ are precisely the morphisms of smooth stacks between $\mathbf{B}G$ and $\mathbf{B}K$.
\end{remark}
If $A$ is an $(n+2)$-deloopable object with $n>0$, then an $(n+2)$-cocycle $c\colon \mathbf{B}G\to \mathbf{B}^{n+2}A$ can be equivalently seen as a 1-cocycle with coefficients in $\mathbf{B}^{n+1}A$, so the above reasoning produces the homotopy pullback diagram
\[
\xymatrix{
\mathbf{B}\widehat{G}\ar[r]\ar[d]_{\varphi_c} & {*}\ar[d]\\
\mathbf{B}G\ar[r]^-{c} &\mathbf{B}^{n+2}A
}
\]
for a suitable $\infty$-group $\widehat{G}$. Taking the fiber of the vertical morphism on the left we get the homotopy pullback
\[
\xymatrix{
\mathbf{B}^{n+1}A\ar[r]\ar[d] & \mathbf{B}\widehat{G}\ar[d]^{\varphi_c}\\
{*}\ar[r] &\mathbf{B}G
}\,,
\]
looping which we finally obtain the homotopy pullback diagram
\[
\xymatrix{
\mathbf{B}^{n}A\ar[r]\ar[d] & \widehat{G}\ar[d]^{\mathbf{\Omega}\varphi_c}\\
{*}\ar[r] &G
}\,,
\]
where all the maps are morphisms of $\infty$-groups. This diagram exhibits $\widehat{G}$ as the central $\mathbf{B}^nA$-extension of $G$ classified by the $(n+2)$-cocycle $c$. In particular, for $n=0$, for $G$ a Lie group and $A$ an abelian Lie group, this construction recovers the classical construction of a central $A$-extension of $G$ from a 2-cocycle on $G$ with coefficients in $A$.

%\begin{example}
%  {\color{red}For $G \in \mathrm{Grp}(\mathbf{H})$ an $\infty$-group} and
%  $A \in \mathrm{Grp}_{n+2}(\mathbf{H})$ a sufficiently deloopable $\infty$-group,
%  a map of the form $\mathbf{c} : \mathbf{B}G \to \mathbf{B}^{n+2}A$
%  is, by remark \ref{GroupCohomology}, equivalently a cocycle representing
%  a class $c \in H^{n+2}_{\mathrm{grp}}(G,A)$ in the degree-$n$ $\infty$-group cohomology of $G$
%  with coefficients in $A$. The $\mathbf{B}^{n+1} A$-principal $\infty$-bundle
%  $$
%    \xymatrix{
%	  \mathbf{B}^{n+1} A \ar[r] & \mathbf{B}\widehat G \ar[d]^{\mathrm{fib}(\mathbf{c})}
%	  \\
%	  & \mathbf{B}G
%	}
%  $$
%  which is classified by $\mathbf{c}$ according to
%  prop. \ref{EquivalenceOfGPrincipalInfinityBundlesWithCocycles} is
%  the delooping of the \emph{$\infty$-group extension}
%  $$
%    \xymatrix{
%	  \mathbf{B}^n A \ar[r] & \widehat G \ar[d]^{\Omega \mathrm{fib}(\mathbf{c})}
%	  \\
%	  & G
%	}
%  $$
%  which is classified by $c$.
%\end{example}
\begin{remark}\label{remark-cs-wzw}In \ref{AnExtended6dSomething} we will see how higher bundles of the
  form $\varphi_c\colon \mathbf{B}\widehat{G}\to \mathbf{B}G$ appear in Chern-Simons-type field theories, while the morphisms
  $\mathbf{\Omega} \varphi_c\colon \widehat{G}\to G$ are those appearing in the corresponding
  Wess-Zumino-Witten-type field theories.
  \end{remark}

\begin{example}
 \label{TwistedBundlesAre2Sections}
Since $U(1)$ is an abelian group, it admits arbitrarily high deloopings. In particular, we have a group stack $\mathbf{B}^2U(1)$. For any $N\geq 1$, the central extension of Lie groups $U(1) \to U(N) \to PU(N)$
  deloops in $\mathbf{H}$ to a long homotopy fiber sequence of the form
  $$
    \raisebox{20pt}{
    \xymatrix{
	  \mathbf{B}U(1) \ar[r]\ar[d] & \mathbf{B}U(N) \ar[d] \ar[r]&{*}\ar[d]\\
	 {*}\ar[r]  & \mathbf{B} PU(N) \ar[r]^{\mathbf{dd}_N}
	   &  \mathbf{B}^2 U(1)
	}
	}
	\,,
  $$
  where $\mathbf{dd}_N\colon \mathbf{B} PU(N)\to  \mathbf{B}^2 U(1)$ is the \emph{Dixmier-Douady map}: the refinement at the level of smooth stacks of the  Dixmier-Douady class $dd_N$ in $H^2_{\mathrm{grp}}(PU(N),U(1))$, classifying the $U(1)$-central extension $U(N)$ of $PU(N)$.
  Notice how in the framework of smooth stacks the map $\mathbf{dd}_N$ is the map modulating the principal $\mathbf{B}U(1)$-bundle $\mathbf{B}U(N)\to \mathbf{B}PU(N)$ over the stack of principal $PU(N)$-bundles. Also, it can be equivalently seen as the datum of an action of the $\infty$-group stack $\mathbf{B}U(1)$ on the stack $\mathbf{B}U(N)$, with $\mathbf{B}U(n)/\!/\mathbf{B}U(1)\simeq \mathbf{B}PU(N)$.
  Therefore, for any map $\nabla^0 : X \to \mathbf{B}^2 U(1)$ modulating
  a $\mathbf{B}U(1)$-principal 2-bundle (or \emph{bundle gerbe}) $P$ over $X$, we have an associated $\mathbf{B}U(N)$-fiber 2-bundle $P\times_{\mathbf{B}U(1)}\mathbf{B}U(N)$.\footnote{This is a \emph{Giraud $U(n)$-gerbe} over $X$,
  see \cite[section 4.4]{NikolausSchreiberStevensonI}.} By Corollary \ref{cor.sections},
 a  section $\sigma$ of the
  associated $\mathbf{B}U(N)$-fiber 2-bundle $P\times_{\mathbf{B}U(1)}\mathbf{B}U(N)\to X$
  is a lift $\Psi$ as in the diagram
  $$
    \raisebox{20pt}{
    \xymatrix{
	  & \mathbf{B}PU(n) \ar[d]^{\mathbf{dd}_n}
	  \\
	  X \ar[r]^{\nabla^0} \ar[ur]^{\Psi} & \mathbf{B}^2 U(1)
	}
	}\,,
  $$
 and so it is
  equivalent to what is called a \emph{bundle gerbe module} or a
  rank-$n$ \emph{twisted unitary bundle}
  (see for instance \cite{CareyJohnsonMurray}). This exhibits twisted bundles as
  sections of 2-bundles and so as \emph{twisted cohomology classes}, which is a general pattern in the theory of smooth higher stacks. In particular, twisted unitary bundles define classes in \emph{twisted K-theory}.
      % in accord with the general remark \ref{TwistedCohomology}.
 \end{example}

\subsection{Principal $G$-bundles with connections}
\label{GPrincipalConnections}

Let $G$ be a Lie group and let $\mathfrak{g}$ its Lie algebra. Then principal $G$-bundles with $\mathfrak{g}$-connections form a smooth stack, which we denote by the symbol $\mathbf{B}G_{\mathrm{conn}}$. This is nothing but an elegant way of stating the fact that the datum of a principal $G$-bundle with $\mathfrak{g}$-connection $(P,\nabla)$ over a smooth manifold $X$ can be encoded into local data relative to a good open cover $\mathcal{U}=\{U_i\}_{i\in I}$ of $X$, and that these data can be pulled back along smooth maps of differential manifolds. More in detail, since the local data description of $(P,\nabla)$ consists of
\begin{itemize}
\item $\mathfrak{g}$-valued 1-forms $A_i\in \Omega^1(U_i;\mathfrak{g})$ for every $j$;
\item smooth functions $g_{ij}\colon U_{ij}\to G$, for every $i,j$;
\end{itemize}
such that
\begin{itemize}
\item $A_j=g_{ij}^{-1}A_ig_{ij} +g_{ij}^{-1}(d+A_i)g_{ij}$ on $U_{ij}$;
\item $g_{ij}g_{jk}g_{ki}=1$ on $U_{ijk}$.
\end{itemize}
This shows that $\mathbf{B}G_{\mathrm{conn}}$ is the stackification of the prestack (associated with the nerve of)
\[
U\mapsto \Omega^1(U;\mathfrak{g})/\!/C^\infty(U;G),
\]
where the action on the right is the usual gauge action of $C^\infty(U;G)$ on $\Omega^1(U;\mathfrak{g})$. Notice that we have an obvious forgetful morphism $\mathbf{B}G_{\mathrm{conn}}\to \mathbf{B}G$, which forgets the connection data.
\begin{remark}\label{curvature}
Let $\Omega^2(U;\mathfrak{g}))/\!/_{\mathrm{Ad}}C^\infty(U;G)$ denote the conjugacy action of $C^\infty(U;G)$ on $\Omega^2(U;\mathfrak{g})$, and let $\Omega^2(-;\mathrm{Ad}\mathfrak{g})$ be the smooth stack which is the stackification of $U\mapsto \Omega^2(U;\mathfrak{g})/\!/_{\mathrm{Ad}}C^\infty(U;G)$. A section of this stack over a smooth manifold $X$ is precisely the datum of a principal $G$-bundle $P$ over $X$, together with a smooth 2-form on $X$ with coefficients in the associated $\mathfrak{g}$-fiber bundle $P\times_G \mathfrak{g}$, where the action is the conjugacy action. Then taking the curvature of a principal $\mathfrak{g}$-connection defines a morphism of smooth stacks
\[
F\colon \mathbf{B}G_{\mathrm{conn}} \to \Omega^2(-;\mathrm{Ad}\mathfrak{g}).
\]
\end{remark}
 Imposing the curvature to be zero, one gets the stack $\flat\mathbf{B}G$ of \emph{flat} $G$-bundles. It is a classical fact that the datum of a flat $G$-bundle over a smooth manifold $X$ is equivalent to the datum of a $G$-local system on $X$, i.e., to a representation of the Poincar\'e groupoid of $X$ with values in the group $G$ (or, better, in the smooth groupoid with a single object and the group $G$ as group of isomorphisms). This is actually a manifestation of an $\infty$-adjunction
 \[
 \mathbf{H}(X,\flat\mathbf{B}G)\simeq  \mathbf{H}(\Pi X,\mathbf{B}G)
 \]
 between endofunctors\footnote{
 To be precise, the construction of fundamental $\infty$-groupoids (``Poincar\'e $\infty$-groupoids'') is an  $\infty$-functor $\Pi\colon \mathbf{H}\to \infty$-$\mathbf{groupoids}$, and one promotes it to an endofunctor of $\mathbf{H}$ by composing it with the natural embedding $\mathrm{LConst}\colon \infty$-$\mathbf{groupoids}\to \Pi\colon \mathbf{H}$ which looks at any $\infty$-groupoid $\mathcal{G}$ as at the smooth stack of locally constant functions to $\mathcal{G}$.}
of the topos $\mathbf{H}$ of smooth stacks.
 \begin{remark}
 From a categorical point of view, the $\Pi/\flat$-adjunction is one of the fundamental structural features of the $\infty$-topos $\mathbf{H}$. It is indeed part of those properties of $\mathbf{H}$ that make it a \emph{cohesive} $\infty$-topos, see \cite{dcct}.
 \end{remark}

%%%%%%%%%%%%%%%%%%%%%%%%%%%%%%%%%%%%%%%%%%%%%%%%%%%%%%
\subsection{The Dold-Kan correspondence}
\label{DoldKanCorrespondence}
%%%%%%%%%%%%%%%%%%%%%%%%%%%%%%%%%%%%%%%%%%%%%%%%%%%%%%

The Dold-Kan correspondence exhibits a natural equivalence between chain complexes of abelian groups concentrated in nonnegative degrees with simplicial abelian groups; see \cite{GoerssJardine}. Hence, its sheafification (or, more precisely, its $\infty$-stackification) associates with any complex of sheaves of abelian groups over the site of smooth manifolds a sheaf of simplicial abelian groups. Since this is in particular a sheaf of Kan complexes, we see that the (sheafified) Dold-Kan correspondence associates a smooth $\infty$-stack with any complex of sheaves of abelian groups (concentrated in nonnegative degrees).

The explict description of this construction is extremely simple: if
\[
\mathcal{A}_\bullet:=\left(\cdots\xrightarrow{d_{n+1}} \mathcal{A}_n\xrightarrow{d_n}\mathcal{A}_{n-1}\xrightarrow{d_{n-1}}\cdots \xrightarrow{d_2} \mathcal{A}_1\xrightarrow{d_1} \mathcal{A}_0\right)
\]
is a chain complex of abelian groups over the site of smooth manifolds, the smooth stack $\mathrm{DK}(\mathcal{A}_\bullet)$ associates with a smooth manifold $X$ endowed with a good open cover $\mathcal{U}=\{U_i\}_{i\in I}$ the following set of local data:
\begin{itemize}
\item sections $A_{k;i_1,\dots,i_k}$ of $\mathcal{A}_k$ over $U_{i_0,\dots,i_k}$ for every $k\geq 0$ and every $i_0,\dots,i_k$
\end{itemize}
such that
\begin{itemize}
\item $\displaystyle{d_kA_{k;i_0,\dots,i_k}=\sum_{j=0}^k(-1)^jA_{k-1;i_1,\dots,\hat{i_j},\dots,i_k}\big\vert_{U_{i_0,\dots,i_k}}}$ over $U_{i_0,\dots,i_k}$, with the convention that $A_{-1;\emptyset}=0$.
\end{itemize}
For instance if $\mathcal{A}_\bullet$ is a 2-term complex $(\mathcal{A}_1\xrightarrow{d} \mathcal{A}_0)$, then a section of $\mathrm{DK}(\mathcal{A}_\bullet)$ over a smooth manifold $X$ endowed with a good open cover  $\mathcal{U}$ consists of
\begin{itemize}
\item an element $A_{0;i}\in \mathcal{A}_0(U_i)$ for any $i$;
\item an element $A_{1;ij}\in \mathcal{A}_1(U_{ij})$ for any $i,j$
\end{itemize}
such that
\begin{itemize}
\item $dA_{1;ij}=A_{0;j}-A_{0;i}$ on $U_{ij}$;
\item $A_{1;jk}-A_{1;ik}+A_{1;ij}=0$ on $U_{ijk}$.
\end{itemize}

\begin{remark}
Notice that, since the Dold-Kan correspondence takes values in simplicial abelian groups, the smooth stacks of the form $\mathrm{DK}(\mathcal{A}_\bullet)$ are actually smooth $\infty$-groups. Moreover, they are infinitely deloopable.
\end{remark}

\begin{definition}
Let $A$ be an abelian Lie group. The smooth $n$-stack $\mathbf{B}^nA$ is the $\infty$-stack associated by the Dold-Kan correspondence to the chain complex
\[
\cdots\to 0\to C^\infty(-;A)\to 0\to 0\to\cdots\to 0\to 0,
\]
with $C^\infty(-;A)$ in degree $n$. It will be called the $n$-stack of principal $A$-$n$-bundles.
\end{definition}
\begin{remark}
It is immediate from the definition that a section of $\mathbf{B}^nA$ over a smooth manifold $X$ is the datum of a \v{C}ech $n$-cocycle with values in $C^\infty(-;A)$ for a good open cover of $X$. In particular we see that
\[
\check{H}^n(X;\underline{A})=\pi_0\mathbf{H}(X,\mathbf{B}^nA),
\]
where on the left we wrote $\underline{A}$ for the sheaf of smooth $A$-valued functions on $X$. Notice how for $n=1$ one recovers the local data description of the stack of principal $A$-bundles, for the abelian Lie group $A$. For $n=2$ one recovers the local data description of an $A$-bundle gerbe, see for instance \cite{BrylinskiLoop}.
\end{remark}
\begin{remark}
The notation $\mathbf{B}^n A$ used for the $n$-stack of $A$-$n$-bundles is consistent with the notation $\mathbf{B}$ to denote the delooping: one has a natural equivalence $\mathbf{B}^{n+1}A\cong \mathbf{B}(\mathbf{B}^{n}A)$. This shows in particular that an abelian Lie group $A$ as well as all the stacks $\mathbf{B}^nA$ are infinitely deloopable. More generally, one has
\[
\mathbf{B}(\mathrm{DK}(\mathcal{A}_\bullet))\simeq \mathrm{DK}(\mathcal{A}_{\bullet}[1]),
\]
where $\mathcal{A}_{\bullet}[1]$ is the chain complex obrained from $\mathcal{A}_\bullet$ by shifting it by one on the left, i.e. is the chain complex
\[
\cdots\xrightarrow{d_{n+1}} \mathcal{A}_n\xrightarrow{d_n}\mathcal{A}_{n-1}\xrightarrow{d_{n-1}}\cdots \xrightarrow{d_2} \mathcal{A}_1\xrightarrow{d_1} \mathcal{A}_0\to 0,
\]
with $\mathcal{A}_0$ in degree 1. The ``shifting by 1 on the left'' operation indeed corresponds to delooping in the $\infty$-category $\mathrm{Chains}_{\geq 0}$ of chain complexes of abelian groups concentrated in nonnegative degree.
\end{remark}
\begin{remark}\label{looping-chains}
The looping operation in $\mathrm{Chains}_{\geq 0}$ is a bit less elementary. Namely, one can not simply shift by 1 on the right (as one would do in the $\infty$-category of unbounded complexes) since shifting on the right the component $\mathcal{A}_0$ would ``fall'' into degree $-1$. One remedies this by taking a truncation: if
\[
\mathcal{A}_\bullet:=\left(\cdots\xrightarrow{d_{n+1}} \mathcal{A}_n\xrightarrow{d_n}\mathcal{A}_{n-1}\xrightarrow{d_{n-1}}\cdots \xrightarrow{d_2} \mathcal{A}_1\xrightarrow{d_1} \mathcal{A}_0\right),
\]
then
\[
\Omega\mathcal{A}_\bullet:=\left(\cdots\xrightarrow{d_{n+1}} \mathcal{A}_n\xrightarrow{d_n}\mathcal{A}_{n-1}\xrightarrow{d_{n-1}}\cdots \xrightarrow{d_3} \mathcal{A}_2\xrightarrow{d_2} \ker(d_1)\right).
\]
The Dold-Kan correspondence preserves the looping, so one has
\[
\Omega\mathrm{DK}(\mathcal{A}_{\bullet})\simeq \mathrm{DK}(\Omega\mathcal{A}_{\bullet}).
\]
Notice that $\Omega\mathbf{B}\mathcal{A}_{\bullet}\simeq \mathcal{A}_{\bullet}$ for any $\mathcal{A}_\bullet$ in $\mathrm{Chains}_{\geq 0}$, but generally $\mathbf{B}\Omega\mathcal{A}_{\bullet}\not\simeq \mathcal{A}_{\bullet}$.
\end{remark}

\begin{remark}\label{loop-space-dk}
Since $\mathrm{DK}(\mathcal{A}_\bullet)$ is a smooth $\infty$-group, for any stack $X$ the hom-space $\mathbf{H}(X,\mathrm{DK}(\mathcal{A}_\bullet))$ is an $\infty$-group, with identity element the ``zero morphism'' $0\colon X\to \mathrm{DK}(\mathcal{A}_\bullet)$. Being a group, $\mathbf{H}(X,\mathrm{DK}(\mathcal{A}_\bullet))$ is in particular a homogeneous space, so that for any $\varphi\colon X\to \mathrm{DK}(\mathcal{A}_\bullet)$ one has a natural equivalence induced by the ``translation by $\varphi$''
\[
\Omega_\varphi\mathbf{H}(X,\mathrm{DK}(\mathcal{A}_\bullet))\simeq \Omega_0\mathbf{H}(X,\mathrm{DK}(\mathcal{A}_\bullet))\simeq \mathbf{H}(X,\mathrm{DK}(\Omega\mathcal{A}_\bullet)).
\]
\end{remark}
\begin{remark}
Since the Dold-Kan correspondence is functorial, a morphism $\mathcal{A}_\bullet\to \mathcal{B}_\bullet$ of chain complexes of sheaves of abelian groups over the site of smooth manifolds induces a morphism of smooth (higher) stacks $\mathrm{DK}(\mathcal{A}_\bullet)\to \mathrm{DK}(\mathcal{B}_\bullet)$.
\end{remark}

\subsection{Principal $U(1)$-$n$-bundles with connections}
 \label{HigherConnections}

For any $n\geq 1$, we denote by $\mathbf{B}^nU(1)_{\mathrm{conn}}$ the $n$-stack which is presented, via the Dold-Kan correspondence, by the degree $n$ Deligne complex
\[
\left(	  \xymatrix{
	 C^\infty(-;U(1))
	    \ar[r]^-{\frac{1}{2\pi i}d \mathrm{log}}
	    &
	    \Omega^1
	    \ar[r]^-d
	    &
	    \cdots
	    \ar[r]^-d
	    &
	    \Omega^n
	  }
	\right),
\]
with the sheaf $\Omega^n$ of degree $n$ real valued differential forms placed in degree 0.
  \begin{remark}
For $n=1$ the stack $\mathbf{B}U(1)_{\mathrm{conn}}$ is precisely the stack of $U(1)$-principal bundles with connection,
as defined in Section \ref{GPrincipalConnections}; this shows that the notation $\mathbf{B}^nU(1)_{\mathrm{conn}}$ is coherent with the general notation $\mathbf{B}G_{\mathrm{conn}}$ we have introduced there. Similarly, one sees from the local data description, that a section of $\mathbf{B}^2U(1)_{\mathrm{conn}}$ over a smooth manifold $X$ is equivalently the datum of a \emph{$U(1)$-bundle gerbe with connection and curving} over $X$, while a section of $\mathbf{B}^2U(1)_{\mathrm{conn}}$ over $X$ is equivalently the datum of a \emph{$U(1)$-bundle 2-gerbe with connection and curving and 3-form connection}. \end{remark}
For any $n\geq 1$, the morphism of chain complexes
\[
    \raisebox{20pt}{
	  \xymatrix{
	C^\infty(-;U(1))
	    \ar[r]^-{\frac{1}{2\pi i}d \mathrm{log}}
		\ar[d]
	    &
	    \Omega^1
	    \ar[r]^-d
		\ar[d]
	    &
	    \cdots
	    \ar[r]^-d
		\ar@{}[d]|{\cdots}
	    &
	    \Omega^n
		\ar[d]		\\
		C^\infty(-;U(1))\ar[r] & 0 \ar[r] & \cdots \ar[r] & 0
	  }
	  }\,
 \]
 induces the forgetful morphism $\mathbf{B}^nU(1)_{\mathrm{conn}} \to \mathbf{B}^nU(1)$, which forgets \emph{all} of the $n$-connection data. In several occasion it is interesting to forget only a top part of the connection data, i.e., to consider the morphism of chain complexes
 \[
    \raisebox{20pt}{
	  \xymatrix{
	C^\infty(-;U(1))
	    \ar[r]^-{\frac{1}{2\pi i}d \mathrm{log}}
		\ar[d]
	    &
	    \Omega^1
	    \ar[r]^-d
		\ar[d]
	    &
	    \cdots
	    \ar[r]^-d
		\ar@{}[d]|{\cdots}
	    &
	    \Omega^k
		\ar[r]^-d
		\ar[d]
		&
	    \Omega^{k+1}
		\ar[r]^-d
		\ar[d]	     &
	    \cdots
	    \ar[r]^-d
		\ar@{}[d]|{\cdots}
	    &
	    \Omega^n
		\ar[d]
		\\
		C^\infty(-;U(1))\ar[r]^-{\frac{1}{2\pi i}d \mathrm{log}}
 & \Omega^1 \ar[r] & \cdots \ar[r]^d & \Omega^k\ar[r]&0\ar[r] & \cdots \ar[r] & 0
	  }
	  }\,.
 \]
This induces the partial forgetting of the connection data morphism $\mathbf{B}^nU(1)_{\mathrm{conn}} \to \mathbf{B}^{n-k}(\mathbf{B}^kU(1)_\mathrm{conn})$. Notice how we have a whole tower of forgetting morphisms
\[
\mathbf{B}^nU(1)_{\mathrm{conn}}\to \mathbf{B}(\mathbf{B}^{n-1}U(1)_\mathrm{conn})\to \mathbf{B}^{2}(\mathbf{B}^{n-2}U(1)_\mathrm{conn})\to \cdots \to \mathbf{B}^{n-1}(\mathbf{B}U(1)_\mathrm{conn})\to \mathbf{B}^nU(1).
\]
\begin{remark}
At least one of the stacks $\mathbf{B}^{n-k}(\mathbf{B}^kU(1)_\mathrm{conn})$ mentioned above is commonly encountered in the literature on bundle gerbes: $\mathbf{B}(\mathbf{B}U(1)_\mathrm{conn})$ is the 2-stack whose sections are ``$U(1)$-bundle gerbes with connective structure but without curving''.
\end{remark}
Finally, for any $n\geq 1$, the morphism of chain complexes
\[
    \raisebox{20pt}{
	  \xymatrix{
	C^\infty(-;U(1))
	    \ar[r]^-{\frac{1}{2\pi i}d \mathrm{log}}
		\ar[d]
	    &
	    \Omega^1
	    \ar[r]^-d
		\ar[d]
	    &
	    \cdots
	    \ar[r]^-d
		\ar@{}[d]|{\cdots}
	    &
	    \Omega^n
		\ar[d]^{d}
		\\
		0 \ar[r] & 0 \ar[r] & \cdots \ar[r]^0 & \Omega^{n+1}_{\mathrm{cl}}
	  }
	  }\,,
 \]
 where $\Omega^{n+1}_{\mathrm{cl}}$ denotes the sheaf of closed $(n+1)$-forms, induces the \emph{curvature} morphism
 \[
 F\colon \mathbf{B}^nU(1)_{\mathrm{conn}}\to  \Omega^{n+1}_{\mathrm{cl}}.
 \]
 \begin{remark}
 Notice that for $n=1$ this notion of curvature does not exactly coincide with the one introduced in remark \ref{curvature}. However, since $U(1)$ is an abelian Lie group, the adjoint action of $U(1)$ on 2-forms is trivial and so we have a canonical morphism $\Omega^2(-;\mathrm{Ad}\mathfrak{u}_1)\to \Omega^2$ relating (and essentially identifying) the two notions of curvature one has for $\mathbf{B}U(1)_{\mathrm{conn}}$, by the following commutative diagram:
 \[
 \xymatrix{\mathbf{B}U(1)_{\mathrm{conn}} \ar[r]^F\ar[d]_F & \Omega^2(-;\mathrm{Ad}\mathfrak{u}_1)\ar[d]\\
 \Omega^2_{\mathrm{cl}}\ar[r]& \Omega^2
 }.
 \]
 \end{remark}
 Imposing the curvature to be zero, one gets the stack $\flat\mathbf{B}^nU(1)$ of \emph{flat} $U(1)$-$n$-bundles. This is presented under the Dold-Kan correspondence by the complex
 \[
\left(	  \xymatrix{
	 C^\infty(-;U(1))
	    \ar[r]^-{\frac{1}{2\pi i}d \mathrm{log}}
	    &
	    \Omega^1
	    \ar[r]^-d
	    &
	    \cdots
	    \ar[r]^-d
	    &
	    \Omega^n_{\mathrm{cl}}
	  }
	\right).
\]
By the Poincar\'e lemma, this complex is equivalent to the chain complex
 \[
\left(	  \xymatrix{
	 C^\infty(-;U(1)_{\mathrm{discr}})
	    \ar[r]
	    &
	   0
	    \ar[r]
	    &
	    \cdots
	    \ar[r]
	    &
	    0
	  }
	\right),
\]
where $U(1)_{\mathrm{discr}}$ denotes the Lie group which is obtained by endowing the abstract group $U(1)$ with the discrete topology, and where $C^\infty(-;U(1)_{\mathrm{discr}})$ is placed in degree $n$. This second description of $\flat\mathbf{B}^nU(1)$ makes it manifest that it is the stack of flat $U(1)$-$n$-bundles.
\begin{remark}
As for $\flat\mathbf{B}G$, also $\flat\mathbf{B}^nU(1)$ can be characterized in terms of the $\Pi/\flat$ $\infty$-adjunction. Namely, one has
\[
\mathbf{H}(X,\flat\mathbf{B}^nU(1))\simeq \mathbf{H}(\Pi X,\mathbf{B}^nU(1)).
\]
\end{remark}
\begin{remark}\label{rem.loop-is-flat}
The first description of the $n$-stack $\flat\mathbf{B}^nU(1)$, together with Remark \ref{looping-chains}, shows that $\flat\mathbf{B}^nU(1)\simeq \Omega\mathbf{B}^{n+1}U(1)_{\mathrm{conn}}$. Together with Remark \ref{loop-space-dk}, this implies that, for any $U(1)$-$(n+1)$-connection $\nabla\colon X\to \mathbf{B}^{n+1}U(1)_{\mathrm{conn}}$ one has a natural equivalence
\[
\Omega_{\nabla}\mathbf{H}(X,\mathbf{B}^{n+1}U(1)_{\mathrm{conn}})\simeq \mathbf{H}(X,\flat\mathbf{B}^nU(1)).
\]
\end{remark}

\subsection{Truncations and $n$-images}
\label{sec.truncations}
If $\mathbf{S}$ is an arbitrary higher stack and $X$ is an $n$-stack for some fixed $n\geq 0$, it is pretty clear that all the amount of data contained into $\mathbf{S}$ ``above level $n$'' is lost by mapping into $X$. Stated a bit more formally, this means that the hom-space $\mathbf{H}(\mathbf{S},X)$ is actually equivalent to a hom-space $\mathbf{H}(\mathbf{S}_{\leq n},X)$, where $\mathbf{S}_{\leq n}$ is the $n$-stack obtained from $\mathbf{S}$ by (coherently) forgetting all the information above level $n$. A completely rigorous version of this statement is the following. Denote by $\mathbf{H}_{\leq n}$ the $\infty$-category of $n$-stacks. Then the inclusion $\mathbf{H}_{\leq n}\hookrightarrow \mathbf{H}$ has a left adjoint, called $n$-truncation. Equivalently, one says that $\mathbf{H}_{\leq n}$ is a reflective $\infty$-subcategory of $\mathbf{H}$ \cite[{section 5.2.7}]{Lurie}.

Since the $n$-truncation is a left adjoint, for any stack $X$ we have a canonical map $X\to X_{\leq n}$, which we can read as a factorisation of the terminal morphism of $X$:
\[
X\to X_{\leq n}\to *.
\]
Moreover, since $\mathbf{H}_{\leq n}\hookrightarrow \mathbf{H}_{\leq n+1}$, we have a finer factorisation $X\to X_{\leq n+1}\to X_{\leq n}\to *$. Letting $n$ range over the nonnegative integers, we get a tower of factorisations of the terminal morphism
$$
    \xymatrix{
	  & \ar@{..}[d]
	  \\
	  & X_{\leq 2} \ar[d] \ar[ddr]
	  \\
	  &X_{\leq 1}\ar[d] \ar[dr]
	  \\
	  X
	  \ar[r] \ar[ur] \ar[uur]
	  \ar@/_1pc/[rr]
	  & X_{\leq 0} \ar[r] & {*}
	}\,,
  $$
the Postnikov tower of $X$. More generally, one has a notion of a Postnikov tower for a morphism $f\colon X\to Y$ in $\mathbf{H}$. In order to present it, we need to recall that there is a  notion of \emph{homotopy groups} $\mathbf{\pi}_n$ of objects in $\mathbf{H}$, which, for $n\geq 1$, are group objects in $\mathbf{H}_{\leq 0}$ \cite[{section 6.1.2}]{Lurie}. For $n=0$ one has $\mathbf{\pi}_0X\cong X_{\leq 0}$. In terms of the homotopy groups of stacks one can then state the following result, a detailed discussion and a proof of which can be found in  \cite[section 5.5.6 and 6.5]{Lurie}.
%\subsubsection{Higher toposes of geometric $\infty$-groupoids}
 \label{HigherToposesOfHigherStacks}
\begin{proposition}
%The full sub-$\infty$-category of $n$-truncated objects
%$\tau_{\leq n}\mathbf{H} \hookrightarrow \mathbf{H}$ in an $\infty$-topos is reflectively
%embedded, which means that there is an idempotent truncation projection
%$\tau_n : \mathbf{H} \to \mathbf{H}$ which sends an arbitrary $\infty$-stack
%$X$ to its universal approximation by an $n$-truncated object $\tau_n X$,
%the $n$th \emph{Postnikov stage} of $X$ as seen in $\mathbf{H}$.
%
 % More generally given
  Given a morphism $f : X \to Y$ in $\mathbf{H}$, there is a
  tower of factorizations
  $$
    
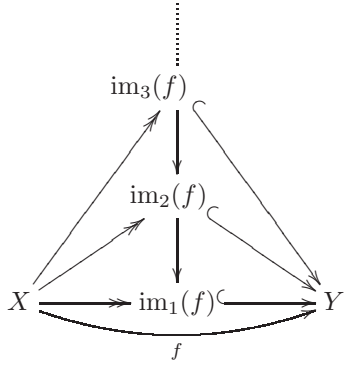
\xymatrix{
	  & \ar@{..}[d]
	  \\
	  & \mathrm{im}_3(f)\phantom{\int_cup} \ar[d] \ar@{^{(}->}[ddr]
	  \\
	  & \mathrm{im}_2(f)\phantom{\int}\ar[d] \ar@{^{(}->}[dr]
	  \\
	  X
	  \ar@{->>}[r] \ar@{->>}[ur] \ar@{->>}[uur]
	  \ar@/_1pc/[rr]_{f}
	  & \mathrm{im}_1(f) \ar@{^{(}->}[r] & Y
	}
  $$
  with the property that, for all $n \in \mathbb{N}$,
  the morphism $\xymatrix{X \ar@{->>}[r] & \mathrm{im}_n}(f)$ is an epimorphism on $\mathbf{\pi}_0$
  and an isomorphism on $\mathbf{\pi}_i$ for all $i<n-1$, and that
  $\xymatrix{\mathrm{im}_n(f) \ar@{^{(}->}[r] & Y}$ is
  a monomorphism on $\mathbf{\pi}_{n-1}$ and
  an isomorphism on all $\mathbf{\pi}_i$ for $i\geq n$.
  \label{PostnikovSystemInOntroduction}
\end{proposition}

%\begin{definition}
We call the object $\mathrm{im}_n(f)$ in Proposition \ref{PostnikovSystemInOntroduction}
the \emph{$n$-image} of $f$ and say that
morphisms of the form $\xymatrix{X \ar@{->>}[r] & \mathrm{im}_n(f)}$ are \emph{$n$-epimorphisms}
and that morphisms of the form $\xymatrix{\mathrm{im}_n(f) \ar@{^{(}->}[r] & Y}$ are
\emph{$n$-monomorphisms}. In \cite{Lurie} these are called \emph{$(n-1)$-connective}
and \emph{$(n-2)$-truncated} morphisms, respectively.
 % \label{ImagesInIntroduction}
%\end{definition}

\begin{example}
A classical example of 1-epimorphism is the following: if $\mathcal{G} = \left(\xymatrix{\mathcal{G}_1 \ar@<+4pt>[r]^t \ar@<-4pt>[r]_s\ar@{<-}@<+0pt>[r] & \mathcal{G}_0}\right)$ is a Lie groupoid, then the canonical map
$ \xymatrix{
    \mathcal{G}_0
	\ar@{->>}[r]
	&
	\mathcal{G}
  }$
from the space of objects (regarded as a groupoid with only identity morphisms) to $\mathcal{G}$ is a 1-epimorphism. In particular, if $G$ is a Lie group the morphism $*\to \mathbf{B}G$ is a 1-epimorphism. This generalises to arbitrary group objects and groupoid objects in $\mathbf{H}$.
\end{example}

{
\begin{remark}\label{differential-to-gauge}
Let $\pi\colon Z\to Y$ be a morphism in $\mathbf{H}$. Then, by the essential uniqueness of the $n$-epi/$n$-mono factorisation of morphisms, if $\tilde{f}\colon X\to Z$ is a lift of a morphism $f\colon X\to Y$, i.e., if we have a homotopy commutative diagram
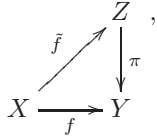
\[
\xymatrix{
& Z\ar[d]^{\pi}\\
X\ar[ru]^{\tilde{f}}\ar[r]_{f}&Y
},
\]
then there is an induced natural morphism $\mathrm{im}_n(\tilde{f})\to \mathrm{im}_n(f)$. We will encounter this phenomenon in
Section \ref{HigherCourant}, where we will exhibit a natural morphism from the differential gauge groupoid of a principal $G$-bundle with connection to the  gauge groupoid of the underlying principal $G$-bundle.
\end{remark}
}

\begin{remark}
By construction, one has the following compatibility between $n$-images and loopings:\footnote{U.S. thanks Egbert Rijke for discussion of this point.}
for $f$ a morphism of pointed objects in $\mathbf{H}$, there is
   a natural equivalence
  $$
    \mathrm{im}_n\left(\mathbf{\Omega}(f)\right)
	\simeq
	\mathbf{\Omega}\, \mathrm{im}_{n+1}(f)
  $$
  between the $n$-image of the looping of $f$ and the looping of the
  $(n+1)$-image of $f$, for each $n \in \mathbb{N}$.
  \label{LoopingOfNImage}
%\end{proposition}
\end{remark}
\begin{remark}
For $n = 1$ the $n$-image factorization has a useful  explicit characterization.
%\begin{proposition}
  \label{Cech1Image}
  For $f : X \to Y$ a morphism in $\mathbf{H}$, consider the associated augmented simplicial diagram in $\mathbf{H}$
  \[
   \xymatrix{
    \ar@{..}[r]
    &
    X \underset{Y}{\times} X \underset{Y}{\times} X
	\ar@<+8pt>[r]
	\ar@{<-}@<+4pt>[r]
	\ar@<+0pt>[r]
	\ar@{<-}@<-4pt>[r]
	\ar@<-8pt>[r]
    &
    X \underset{Y}{\times}X
	\ar@<+4pt>[r]
	\ar@{<-}@<-0pt>[r]
	\ar@<-4pt>[r]
	&
	X\ar[r]^{f}&Y
	}\,.
  \]
%
%  consider the homotopy-colimiting cocone under its {\v C}ech nerve simplicial
%  diagram as indicated in the top row of the following diagram
By the universal property of the $\infty$-colimit, one gets a canonical factorisation
$$
  \raisebox{20pt}{
  \xymatrix{
    \ar@{..}[r]
    &
    X \underset{Y}{\times} X \underset{Y}{\times} X
	\ar@<+8pt>[r]
	\ar@{<-}@<+4pt>[r]
	\ar@<+0pt>[r]
	\ar@{<-}@<-4pt>[r]
	\ar@<-8pt>[r]
    &
    X \underset{Y}{\times}X
	\ar@<+4pt>[r]
	\ar@{<-}@<-0pt>[r]
	\ar@<-4pt>[r]
	&
	X
	\ar@{->>}[rr]^-{p}
	\ar[drr]_{f}
	&&
	\left(\underset{\longrightarrow}{\lim}_n X^{\times^{n+1}_{Y}}\right)
%	\simeq
%	\mathrm{im}_1(f)
	\ar@{^{(}->}[d]^i
	\\
	&&&&& Y
  }
  }
  \,.
  $$
  which exhibits $\left(\underset{\longrightarrow}{\lim}_n X^{\times^{n+1}_{Y}}\right)$ as the 1-image of $f$.
%  Since $f : X \to Y$ canonically extends to a homotopy cocone under its
%  own {\v C}ech nerve, the universal property of the $\infty$-colimit induces a vertical
%  dashed map $i$, as indicated. The resulting factorization of $f$ is its 1-image
%  factorization, as shown.
\end{remark}

\begin{example}
Let $G$ be a Lie group and let $*\to \mathbf{B}G$ the distinguished point corresponding to the trivial $G$-bundle. Then the morphism $*\to \mathbf{B}G$ has a 1-image given by the homotopy colimit of the diagram
  \[
   \xymatrix{
    \ar@{..}[r]
    &
    {*} \underset{\mathbf{B}G}{\times} {*} \underset{\mathbf{B}G}{\times} {*}
	\ar@<+8pt>[r]
	\ar@{<-}@<+4pt>[r]
	\ar@<+0pt>[r]
	\ar@{<-}@<-4pt>[r]
	\ar@<-8pt>[r]
    &
    {*} \underset{\mathbf{B}G}{\times}{*}
	\ar@<+4pt>[r]
	\ar@{<-}@<-0pt>[r]
	\ar@<-4pt>[r]
	&
	{*}
	}\,.
  \]
Since ${*} \underset{\mathbf{B}G}{\times}{*}\cong \mathbf{\Omega}\mathbf{B}G\cong G$ and, more generally, $*^{\times^{n+1}_{\mathbf{B}G}}\cong G^n$, this is the homotopy colimit of the simplicial manifold
 \[
   \xymatrix{
    \ar@{..}[r]
    &
   G^2
	\ar@<+8pt>[r]
	\ar@{<-}@<+4pt>[r]
	\ar@<+0pt>[r]
	\ar@{<-}@<-4pt>[r]
	\ar@<-8pt>[r]
    &
    G	\ar@<+4pt>[r]
	\ar@{<-}@<-0pt>[r]
	\ar@<-4pt>[r]
	&
	{*}
	}\,,
  \]
  i.e., the simplicial manifold associated with the Lie groupoid $\xymatrix{G	\ar@<+4pt>[r]
	\ar@{<-}@<-0pt>[r]
	\ar@<-4pt>[r]
	&
	{*}}$.
On the other hand, since $\mathbf{\pi}_0(\mathbf{B}G)\cong *$, the trivial factorisation $*\to \mathbf{B}G\to \mathbf{B}G$ is a factorisation of $*\to \mathbf{B}G$ into a 1-epi and a 1-mono, so $ \mathbf{B}G\cong \mathrm{im}_1(*\to \mathbf{B}G)$ and one recovers the familiar bar construction of $\mathbf{B}G$.
\end{example}

\begin{example}
\label{ex.atiyah}
The above example has a natural generalization. Let $P\to X$ be a principal $G$-bundle modulated by a morphism $\nabla^0\colon X\to \mathbf{B}G$ from a smooth manifold $X$ to the 1-stack $\mathbf{B}G$. Then the 1-image of $\nabla^0$ is realised by the homotopy colimit  of the simplicial diagram
 \[
   \xymatrix{
    \ar@{..}[r]
    &
    X \underset{\mathbf{B}G}{\times} X \underset{\mathbf{B}G}{\times} X
	\ar@<+8pt>[r]
	\ar@{<-}@<+4pt>[r]
	\ar@<+0pt>[r]
	\ar@{<-}@<-4pt>[r]
	\ar@<-8pt>[r]
    &
    X \underset{\mathbf{B}G}{\times}X
	\ar@<+4pt>[r]
	\ar@{<-}@<-0pt>[r]
	\ar@<-4pt>[r]
	&
	X
	}\,,
  \]
   i.e., the simplicial stack associated with the stacky groupoid $\xymatrix{X \underset{\mathbf{B}G}{\times}X	\ar@<+4pt>[r]
	\ar@{<-}@<-0pt>[r]
	\ar@<-4pt>[r]
	&
	X}$.
Since $X$ is a smooth manifold, it is 0-truncated as an object in $\mathbf{H}$ and so $\mathbf{\pi}_0X\cong X$. Therefore $X\to \mathbf{B}G$ is not a 1-epimorphism unless $X$ is the 1-point manifold. This means that the 1-image of $\nabla^0$ will be some new stack, different from $\mathbf{B}G$. It turns out that the 1-image $\mathrm{im}_1(\nabla^0)$ (see Section \ref{sec.truncations}) is a well known object from gauge theory. Namely, the homotopy fibre product $X \underset{\mathbf{B}G}{\times}X$ is naturally equivalent to the smooth manifold given by the quotient $(P \times P)/G$ where $G$ acts diagonally on $P\times P$, and the stack groupoid $\xymatrix{X \underset{\mathbf{B}G}{\times}X	\ar@<+4pt>[r]
	\ar@{<-}@<-0pt>[r]
	\ar@<-4pt>[r]
	&
	X}$ is therefore equivalent to the Lie groupoid
	\[
	\mathrm{At}(P)=\left(\xymatrix{(P\times P)/G	\ar@<+4pt>[r]
	\ar@{<-}@<-0pt>[r]
	\ar@<-4pt>[r]
	&
	X}\right),
	\]
i.e., to the \emph{gauge groupoid} of $P$. Since the gauge groupoid integrates the  \emph{Atiyah Lie algebroid} of $P$, we find it convenient to call it \emph{Atiyah groupoid} of $P$ and to denote it by the symbol $\mathrm{At}(P)$. We will also use the symbol $\mathrm{At}(\nabla^0)$ to denote the gauge groupoid, when we want to make the relevance of the modulating map $\nabla^0$ more explicit.
By universality of the 1-image factorization
$$
  \nabla^0 : \xymatrix{
    X \ar@{->>}[r]^{\iota_{\mathrm{At}(P)}}
	&
	\mathrm{At}(P)
	\ar@{^{(}->}[r]
	&
	\mathbf{B}G
  }
  \,
$$
of $\nabla^0\colon X\to \mathbf{B}G$, any natural transformation
from $\nabla^0$ to itself  factors through the fully faithful 1-monomorphism $\mathrm{At}(P)\hookrightarrow \mathbf{B}G$, i.e., we have an equivalence
%\end{remark}
%In particular we have a canonical factorizing map from
%$\mathrm{At}(P)$ to $\mathbf{B}G$ which is a 1-monomorphism,
%and this implies that the components of any natural transformation
%from $\nabla^0$ to itself factor through this \emph{fully faithful} inclusion:
$$
  \begin{array}{ccccc}
    \left\{
      \raisebox{30pt}{
      \xymatrix{
        X \ar[ddr]_{\nabla^0}^{\ }="t" \ar[rr]^\phi|\simeq_{\ }="s"  && X \ar[ddl]^{\nabla^0}
	    \\
	    \\
	    & \mathbf{B}G
	    \ar@{=>} "s"; "t"
     }
    }
  \right\}
   &\simeq&
   \left\{
     \raisebox{33pt}{
	   \xymatrix{
	     X \ar[ddr]_{\nabla^0} \ar@{->>}[dr]|{\iota}^{\ }="t" \ar[rr]^\phi|\simeq_{\ }="s"
		 && X
		 \ar[ddl]^{\nabla^0} \ar@{->>}[dl]|{\iota}
		 \\
		 & \mathrm{At}(P)
		  \ar@{^{(}->}[d]
		 \\
		 & \mathbf{B}G
		 \ar@{=>} "s"; "t"
	   }
	 }
   \right\}
   &\simeq &
     \left\{
        \raisebox{20pt}{
        \xymatrix{
	      X \ar@{->>}[dr]_{\iota_{\mathrm{At}(P)}}^{\ }="t" \ar[rr]^{\phi}|\simeq_{\ }="s"
		   && X \ar@{->>}[dl]^{\iota_{\mathrm{At}(P)}}
	     \\
	     & \mathrm{At}(P)
	     \ar@{=>} "s"; "t"
	    }
	    }
      \right\}

  \end{array}
  \,.
$$
Recalling Example \ref{Bisections}, this gives a canonical equivalence $\mathbf{BiSect}(\mathrm{At}(P))
	\simeq
	\mathbf{Aut}_{\mathbf{H}}(\nabla^0)$ between the group stack of bisections of the Atiyah groupoid and the group stack of automorphisms of the modulating map $\nabla^0$.
\end{example}

\subsection{The stack of $G$-connections on a given stack}
 \label{ModuliOfConnections}

 Let $X$ be a smooth manifold, with a given good open cover
 $\mathcal{U}$, and let $G$ be a Lie group. Then a principal
 $G$-bundle on $X$ is encoded in local data given by a set of smooth
 functions $g_{ij}\colon U_{ij}\to G$ such that $g_{ij}g_{jk}g_{ki}=1$
 over $U_{ijk}$. Therefore, if $U$ is some smooth manifold of
 parameters, e.g., a Cartesian space, there is a clear naive idea of
 what a family of principal $G$-bundles on $X$, parametrized by $U$:
 the datum of a collection of smooth functions $g_{ij}(u)\colon
 U_{ij}\to G$, with $u$ ranging in $U$, such that
 $g_{ij}(u)g_{jk}(u)g_{ki}(u)=1$ over $U_{ijk}$, for any $u\in
 U$. Moreover, one would ask the dependence on the parameter $u\in U$
 to be smooth. All this is conveniently expressed by saying that one
 considers smooth functions $g_{ij}\colon U_i\times U\to G$ such that
 $g_{ij}g_{jk}g_{ki}=1$ over $U_{ijk}\times U$. A brief reflection
 shows that the naive notion of a family of principal $G$-bundles on
 $X$ parametrized by $U$ is exactly encoded in a section of the
 internal hom stack $[X,\mathbf{B}G]$ over $U$.

However, when we move to principal $G$-bundles with connection, we see the things go differently. Namely, in the naive version of a family of $G$-connections on $X$ parametrized by $U$, one would have families of $\mathfrak{g}$-valued 1-forms on $U_i$, depending on a parameter $u$ in $U$,  by hence expressions of the form
\[
A_{i;\alpha}^j(x_i^1,\dots,x_i^n;u)dx^\alpha_i\otimes \gamma_j,
\]
where $\{\gamma_j\}_{j\in J}$ is a linear basis of the Lie algebra $\mathfrak{g}$ of $G$, $A_{i;\alpha}$ is a smooth real valued function over $U_i\times U$, and the $x^\alpha_i$'s are coordinates on $U_i$. This is a $\mathfrak{g}$-valued 1-form on $U_i\times U$, but it is not the most general element in $\Omega^1(U_i\times U;\mathfrak{g})$; rather, it is an element in the linear subspace $A^{1,0}(U_i\times U;\mathfrak{g})$ of the $\mathfrak{g}$-valued 1-forms that are of type (1,0) with respect to the product structure of $U_i\times U$. In other words, the naive notion of a $U$-family of $G$-connections on $X$ is not encoded into a section of the stack $[X,\mathbf{B}G_{\mathrm{conn}}]$ over $U$, but rather in the section of another stack, which we will call $G\mathbf{Conn}(X)$, whose local data are precisely
\begin{itemize}
\item $\mathfrak{g}$-valued 1-forms $A_i\in \Omega^{1,0}(U_i\times U;\mathfrak{g})$ for every $i$;
\item smooth functions $g_{ij}\colon U_{ij}\times U\to G$, for every $i,j$;
\end{itemize}
such that
\begin{itemize}
\item $A_j=g_{ij}^{-1}A_ig_{ij} +g_{ij}^{-1}d_{U_{ij}}g_{ij}$ on $U_{ij}\times U$;
%\item $A_j=g_{ij}^{-1}A_ig_{ij} +g_{ij}^{-1}(d_{U_{ij}}+A_i)g_{ij}$ on $U_{ij}\times U$;
\item $g_{ij}g_{jk}g_{ki}=1$ on $U_{ijk}\times U$,
\end{itemize}
where $d_{U_{ij}}$ is the (1,0)-part of the de Rham differential decomposed as $d=d_{U_{ij}}+d_U$ according to the product structure of $U_{ij}\times U$. In a similar way one defines the stack $\mathbf{Flat}$-$G$-$\mathbf{Conn}(X)$ of flat $G$-connections on $X$, the stack $U(1)$-$n$-$\mathbf{Conn}(X)$ of $U(1)$-$n$-connections on $X$ and $\mathbf{Flat}\text{-}U(1)\text{-}n\text{-}\mathbf{Conn}(X)$ of flat $U(1)$-$n$-connections on $X$.
\begin{remark}
The projection $\Omega^{1}(U_i\times U;\mathfrak{g})\to \Omega^{1,0}(U_i\times U;\mathfrak{g})$ induces a morphism of stacks $[X,\mathbf{B}G_{\mathrm{conn}}]\to G\mathbf{Conn}(X)$. We will refer to this morphism as the \emph{concretification} map.
\end{remark}

\begin{remark}
One may wonder whether the stack $G\mathbf{Conn}(X)$ has a more intrinsic definition. It is indeed so. Namely, the ``swithching off'' of the contribution from (0,1)-forms from a section of $[X,\mathbf{B}G_{\mathrm{conn}}]$ over a Cartesian space $U$ can be obtained by thinking of $U$ as as endowed with the discrete topology. However, declaring the hom-space $\mathbf{H}(U,G\mathbf{Conn}(X))$ to be $\mathbf{H}(U_{\mathrm{discr}},[X,\mathbf{B}G_{\mathrm{conn}}])$ would not be quite correct, since by discretizing $U$ one looses the condition that the functions $g_{ij}\colon U_{ij}\times U\to G$ are smooth with respect to the smooth structure of $U$, as well as the regularity in the $U$-direction of the 1-forms $A_i$. What one really wants, is to consider only those morphisms in $\mathbf{H}(U_{\mathrm{discr}},[X,\mathbf{B}G_{\mathrm{conn}}])$ which come from $\mathbf{H}(U,[X,\mathbf{B}G_{\mathrm{conn}}])$ via the natural morphism $U_{\mathrm{discr}}\to U$. To correctly formalize this idea, notice that a morphism from a smooth manifold $M$ to $U_{\mathrm{discr}}$ is the same thing as a locally constant $U$-valued morphism from $M$ to $U$. This identifies $U_{\mathrm{discr}}$ with the smooth stack $\flat U$. Namely, smooth maps to $U_{\mathrm{discr}}$ are \emph{flat} (i.e.: constant) maps to $U$. With this in mind, the hom-space $\mathbf{H}(U_{\mathrm{discr}},[X,\mathbf{B}G_{\mathrm{conn}}])$ mentioned above becomes $\mathbf{H}(\flat U,[X,\mathbf{B}G_{\mathrm{conn}}])$. Now, a crucial feature of the $\infty$-topos $\mathbf{H}$, which is part of its structure of \emph{cohesive} topos, is that the endomorphism $\flat$ does not only have a left adjoint $\Pi$ but also a right adjoint $\sharp$. Therefore we have
\[
\mathbf{H}(\flat U,[X,\mathbf{B}G_{\mathrm{conn}}])\simeq \mathbf{H}(U,\sharp[X,\mathbf{B}G_{\mathrm{conn}}]).
\]

This means that the first Postnikov stage $\sharp_1 $
of the canonical map $[X,\mathbf{B}G_{\mathrm{conn}}] \to \sharp [X,\mathbf{B}G_{\mathrm{conn}}]$
is getting closer to being the right moduli
stack $G$-$\mathbf{Conn}(X)$. But it still differs: while this now has the right moduli structure on the $G$-connections
itself, it does not know anymore about smooth collections of gauge transformations.
But since gauge transformations of $G$-connections are just gauge transformations of the underlying bundle
subject to the condition that the connection is respected, we may finally correct this by
forming a homotopy fiber product like so:
$$
  \mbox{$G$-$\mathbf{Conn}(X)$} \simeq \sharp_1[X,\mathbf{B}G_{\mathrm{conn}}] \underset{\sharp_1 [X,\mathbf{B}G]}{\times} [X,\mathbf{B}G]
  \,.
$$
In the generalization of this procedure, one may abstractly construct $U(1)$-$n$-$\mathbf{Conn}(X)$
and $\mathbf{Flat}$-$U(1)$-$n$-$\mathbf{Conn}(X)$ as an iterated fiber products over
the Postnikov stages of the $\sharp$-unit on the Hodge filtration of the Deligne complex.
See \cite{dcct} for more details on this differential concretification map  and its image.
\end{remark}

\section{Higher prequantum bundles}
\label{higherprequantumbundles}

This section gives the key construction of the article in subsection \ref{HigherCourantQuantomorphimsGroupoids},
and the central theorem in subsection \ref{TheCentralTheorems}.

\subsection{Higher gauge groupoids}
 \label{HigherAtiyah}

Notice how, in Example \ref{ex.atiyah}, defining the gauge groupoid as the 1-image of the modulating morphism for a principal $G$-bundle we have an immediate generalisation to gauge groupoids for $\infty$-bundles.
Let $\mathbf{H}$ be an $\infty$-topos, let $G \in \mathrm{Grp}(\mathbf{H})$
be an $\infty$-group and let $P \to X$ be a $G$-principal $\infty$-bundle in $\mathbf{H}$,
as discussed above in \ref{HigherFiberBundles}.
\begin{definition} Let $G$ be an $\infty$-group object in $\mathbf{H}$ and let $\nabla^0\colon X\to \mathbf{B}G$ a morphism in $\mathbf{H}$ modulating a principal $G$-bundle $P\to X$.
  The \emph{higher gauge groupoid} (or higher Atiyah groupoid) $\mathrm{At}(P)$
  of $P$ is the groupoid object in $\mathbf{H}$
 % which under prop. \ref{1EpisAreEquivalentToGroupoidObjects}
  %corresponds to the 1-image projection $p_{\mathrm{At}(P)}$
  given by the 1-image of $\nabla^0$.
%  $$
%    \nabla^0 :
%	\xymatrix{
%	  X \ar@{->>}[rr]^{p_{\mathrm{At}(P)}}
%	  &&
%	  \mathrm{At}(P)
%	  \ar@{^{(}->}[rr]
%	  &&
%	  \mathbf{B}G
%	}
%  $$
%  of the map $\nabla^0$ which modulates $P \to X$ via
%  prop. \ref{EquivalenceOfGPrincipalInfinityBundlesWithCocycles}.
%  \label{HigherAtiyahGroupoidInIntroduction}
\end{definition}
We will also write $\mathrm{At}(\nabla^0)$ for the higher gauge groupoid of $P$. Reasoning as in Example \ref{ex.atiyah} we find a a canonical equivalence
\[
\mathbf{BiSect}(\mathrm{At}(P))
	\simeq
	\mathbf{Aut}_{\mathbf{H}}(\nabla^0).
	\]
 Moreover, by
 Corollary \ref{cor.central-extension}, we have the fiber sequence
 \[
\xymatrix{
\mathbf{\Omega}_{\nabla^0}[X,\mathbf{B}G]\ar[r]\ar[d] &\mathbf{Aut}_{\mathbf{H}}(\nabla^0)\ar[d]\\
{*}\ar[r] &\mathbf{Aut}(X)
}\,,
\]
thus presenting the group stack of bisections of the Atiyah groupoid as an $\infty$-group extension of the group stack of automorphisms of $X$ by the loop group of $[X,\mathbf{B}G]$  based at the point $\nabla^0$.
%
%\begin{theorem}
%  In the situation of def. \ref{HigherAtiyahGroupoidInIntroduction},
%  there is a canonical equivalence
%  $$
%    \mathbf{BiSect}(\mathrm{At}(P)_\bullet)
%	\simeq
%	\mathbf{Aut}_{\mathbf{H}}(\nabla^0)
%  $$
%  between the $\infty$-group of bisections, def. \ref{BisectionInIntroduction},
%  of the higher Atiyah groupoid of
%  a $G$-principal $\infty$-bundle $P$ and the $\mathbf{H}$-valued automorphism $\infty$-group of
%  its modulating map $\nabla^0$, according to
%  prop. \ref{EquivalenceOfGPrincipalInfinityBundlesWithCocycles}.
%  Moreover, the $\infty$-group of bisections of the higher Atiyah $\infty$-groupoid
%  is an $\infty$-group extension, example \ref{GroupExtension}, of the form
%$$
%  \xymatrix{
%    \Omega_{\nabla^0} [X, \mathbf{B}G]
%	\ar[r]
%	&
%	\mathbf{BiSect}(\mathbf{At}(P)_\bullet)
%	\ar[r]
%	\ar@{}[d]|\simeq
%	&
%	\mathbf{Aut}(X)
%	\\
%	& \mathbf{Aut}_{\mathbf{H}}(\nabla^0)
%  }
%  \,,
%$$
%where on the right we have the canonical forgetful map.
%\label{TheAutomorphismInSliceExtension}
%\end{theorem}
Also, by Proposition \ref{SectionsByMapsInSlice}, we get a canonical
action of the group stack $\mathbf{BiSect}(\mathrm{At}(P))$ on the
stack of sections of any associated $V$-fiber bundle $P \times_G V\to
X$.
%$$
%  \mathbf{BiSect}(\mathrm{At}(P)_\bullet) \times \mathbf{\Gamma}_X(P \times_G V)
%  \to
%  \mathbf{\Gamma}_X(P \times_G V)
%  \,.
%$$
%\label{ActionOfAtiyahBisectionsOnPrequantumStates}
%\end{corollary}
\begin{remark}
As an illustration for the use of higher Atiyah groupoids in higher geometry,
notice how we can immediately rederive and generalize to higher geometry the classical statement in
Lie groupoid theory, which says that every $G$-principal bundle $P\to X$ arises as the source fiber
of its Atiyah groupoid. Namely, for any point $x$ of $X$, we have
a homotopy commutative diagram
$$
   \raisebox{20pt}{
    \xymatrix{
	   {*} \ar@{=}[r] \ar@{->>}[d]^{x} & {*} \ar@{->}[d]
	  \\
	  \mathrm{At}(P)
	  \ar@{^{(}->}[r]
	  &
	  \mathbf{B}G
	}
	}
  $$
  which, by the universal property of the pullback factors as
$$
   \raisebox{20pt}{
    \xymatrix{
    {*}\ar@/^1pc/@{=}[rrd]\ar@/_1pc/@{->}[rdd]_{x}\ar[dr]\\
	&   Q \ar@{^{(}->}[r] \ar@{->>}[d] & {*} \ar@{->>}[d]
	  \\
	 & \mathrm{At}(P)
	  \ar@{^{(}->}[r]
	  &
	  \mathbf{B}G
	}
	}
  $$
where we have used the $\infty$-pullback stability
  of 1-epimorphisms and 1-monomorphisms \cite{Lurie}. Now, $Q\hookrightarrow *$ is a  1-monomorphism from
  $Q$ to the terminal object, and $Q$ is not the empty stack since it receives a morphism from $*$, so $Q\hookrightarrow *$ is an equivalence. Therefore, if we denote by $\nabla^0\colon X\to \mathbf{B}G$ a modulating map for $P$, we find the homotopy commutative diagram
%\begin{proposition}
%  For $G \in \mathrm{Grp}(\mathbf{H})$ an $\infty$-group, every $G$-principal $\infty$-bundle
%  $P \to X$ in $\mathbf{H}$ over an inhabited (= (-1)-connected) object $X$
%  is equivalently the source-fiber of a transitive
%  higher groupoid $\mathcal{G}_\bullet \in \mathrm{Grpd}(\mathbf{H})$ with
%  vertex $\infty$-group $G$ (automorphism $\infty$-group of any point).
%  Here in particular we can set $\mathcal{G}_\bullet = \mathrm{At}(P)_\bullet$.
%\end{proposition}
%\proof
  $$
   \raisebox{20pt}{
    \xymatrix{
	  P \ar@{->>}[r] \ar[d] & {*} \ar[r]^\simeq \ar@{->>}[d]^{x} & {*} \ar@{->>}[d]
	  \\
	  X
	  \ar@{->>}[r]
	  \ar@/_1pc/[rr]_{\nabla^0}
	  &
	  \mathrm{At}(P)
	  \ar@{^{(}->}[r]
	  &
	  \mathbf{B}G
	}
	}
  $$
  where both the outer rectangle and the right sub-square are $\infty$-pullbacks.
   By the 2-out-of-3 law for $\infty$-pullbacks also the left sub-square is an $\infty$-pullback
  and this exhibits $P$ as the source fiber of $\mathrm{At}(P)$ over $x \in X$.
%\endofproof
\end{remark}
%Here we are interested in the following generalization to higher Atiyah groupoids
%of the classical facts reviewed at the beginning of this section. While this is
%a fairly elementary result in higher topos theory, we highlight it as a theorem
%since it serves as the blueprint for the differential refinement in
%theorem \ref{TheLongHomotopyFiberSequenceOfTheQuantomorphimsGroup} below.
%\proof
%  By the defining property of 1-monomorphisms and by prop. \ref{TheArchetypeExtension}.
%\endofproof
%\begin{remark}
%Together with prop. \ref{SliceOverModuliStackOfGIsGAxtions},
%this theorem says that higher Atiyah groupoids are related to
%$G$-equivariant maps between the fibers of their principal $\infty$-bundles
%in just the way that one expects from the traditional situation.
%\end{remark}

\begin{remark}\label{action.bisections}
%Looking back through the discussion in
%\ref{TraditionalPrequantumGeometryViaSlicing}, we see
%that the main reason why one passes to groups of bisections is because these canonically
%\emph{act}. For instance we saw that a prequantum operator
%is a tangent to a global bisection of the quantomorphism groupoid,
%and its action on prequantum states is
%inherited from the canonical action of that group of bisections.
%
%But in fact there is a natural notion of actions of higher groupoids themselves,
%which refines the notion of action of their $\infty$-groups of bisections:
%
  A perfectly similar argument exhibits the canonical action of the
  Atiyah groupoid on the total space of an associated $V$-bundle.
  Namely, if $\rho\colon V/\!/G \to \mathbf{B}G$ denotes the
  $G$-action on $V$, then we have the following pasting of homotopy
  pullback diagrams
$$
  \raisebox{20pt}{
  \xymatrix{
    P \times_G V
	\ar[r]
	\ar[d]
	&
	(P \times_G V)/\!/\mathrm{At}(P)
	\ar[r]
	\ar[d]
	&
	V/\!/G
	\ar[d]^{\rho}
	\\
	X \ar@{->>}[r]
	\ar@/_1pc/[rr]_{\nabla^0}
	& \mathrm{At}(P)
	\ar@{^{(}->}[r]
	&
	\mathbf{B}G
  }
  }
  \,,
$$
and the left square exhibits the canonical action of $\mathrm{At}(P)$
on $P \times_G V$. Passing to global sections, this returns the action
of the group of bisections of the Atiyah groupoid on the space of
sections of $P \times_G V\to X$.
\label{CaonicalGroupoidActionOfAtiyahGroupoid}
\end{remark}

 In view of the discussion in Section \ref{GroupExtension} we may ask for a $\infty$-group cocycle that classifies the
higher Atiyah extension. This will not exist on all of $\mathbf{Aut}(X)$, in general,
but just on the 1-image of the morphism $\mathbf{Aut}_{\mathbf{H}}(\nabla^0)\to \mathbf{Aut}(X)$, i.e., in more colloquial terms, on the subgroup of $\mathbf{Aut}(X)$ consisting of automorphisms of $X$ that admit a lift to an auto equivalence of the principal $G$-bundle $P$. We denote by $ \mathbf{Aut}_P(X)$ this subgroup, i.e., we write
  $$
    \xymatrix{
	  \mathbf{Aut}_{\mathbf{H}}(\nabla^0)
	  \ar@{->>}[r]
	  &
	  \mathbf{Aut}_P(X)
	  \ar@{^{(}->}[r]
	  &
	  \mathbf{Aut}(X)
	}
	\,.
  $$
for the 1-epi/1-mono factorisation of $\mathbf{Aut}_{\mathbf{H}}(\nabla^0)\to \mathbf{Aut}(X)$.
\begin{theorem}
  We have a pasting of homotopy pullback diagrams in $\mathbf{H}$ of the form
$$
    \xymatrix{
      \mathbf{\Omega}_{\nabla^0} [X,\mathbf{B}G]
	  \ar[r]\ar[d]
	  &
	  \mathbf{Aut}_{\mathbf{H}}(\nabla^0)
	  \ar@{->>}[d]\ar[r] &{*}\ar[d]
	  \\
	 {*}\ar[r] &
	  \mathbf{Aut}_P(X)
	  \ar[r]^-{\nabla^0_*}
	  &
	  \mathbf{B}\left(\mathbf{\Omega}_{\nabla^0} [X,\mathbf{B}G] \right)
		}
	\,.
  $$
%
%  $$
%    \xymatrix{
%      \Omega_{\nabla^0} [X,\mathbf{B}G]
%	  \ar[r]
%	  &
%	  \mathbf{BiSect}(\mathrm{At}(P)_\bullet)
%	  \ar@{->>}[r]
%	  &
%	  \mathbf{Aut}_P(X)
%	  \ar[r]^-{\nabla^0 \circ (-)}
%	  &
%	  \mathbf{B}\left(\Omega_{\nabla^0} [X,\mathbf{B}G] \right)
%		}
%	\,.
%  $$
  Moreover, if $G$ is a %sylleptic
braided  group,\footnote{I.e., if $G$ admits a %triple
double delooping $\mathbf{B}^2G$, which implies that
  %$ \mathbf{B}\left(\Omega_{\nabla^0} [X,\mathbf{B}G] \right)$
  $[X,\mathbf{B}G]\cong \mathbf{\Omega}[X,\mathbf{B}^2G]$
  is
   naturally an $\infty$-group object. This hypothesis is clearly satisfied when $G$ is an abelian group.} then the above is actually a long homotopy fiber
  sequence of group objects in $\mathbf{H}$, i.e., it can be delooped.
  In this case the delooping
  \[
  \mathbf{B}(\nabla^0_*)\colon \mathbf{B}\mathbf{Aut}_P(X)\to \mathbf{B}^2\left(\mathbf{\Omega}_{\nabla^0} [X,\mathbf{B}G] \right)
  \]
   is the $\infty$-group cocycle
  that classifies $\mathbf{Aut}_{\mathbf{H}}(\nabla^0)$
  as an $\Omega_{\nabla^0}[X, \mathbf{B}G]$-extension of
  the $\infty$-group $\mathbf{Aut}_P(X)$.
  \label{TheLongSequenceForHigherAtiyahBisections}
\end{theorem}
\proof
  % First consider the underlying morphisms in $\mathbf{H}$.
  By Proposition \ref{prop.aut-in-slice}
  we have a homotopy pullback.
   $$
    \xymatrix{
	   \mathbf{Aut}_{\mathbf{H}}(\nabla^0)
	  \ar[r]
	  \ar[d]
	 	  &
	  \mathbf{Aut}(X)
	  \ar[d]^{\nabla^0_*
	  }
	  \\
	  {*}
	  \ar[r]_{
	  \nabla^0}
	  	  &
	  [X, \mathbf{B}G]
	}
  $$
  Form the 1-image factorization
 \[
 * \twoheadrightarrow\mathbf{B}\left(\mathbf{\Omega}_{\nabla^0} [X,\mathbf{B}G] \right) \hookrightarrow  [X, \mathbf{B}G]
 \]
  of the bottom
  map and by the homotopy pullback stability of
  1-monomorphisms and 1-epimorphisms in $\mathbf{H}$
 we get the pasting of homotopy pullback diagrams
   $$
    \xymatrix{
	  %\mathbf{BiSect}(\mathrm{At}(P)_\bullet)
	   \mathbf{Aut}_{\mathbf{H}}(\nabla^0)
	  \ar@{->>}[r]
	  \ar[d]
	  &
	  \mathbf{Aut}_P(X)
	  \ar@{^{(}->}[r]
	  \ar[d]
	  &
	  \mathbf{Aut}(X)
	  \ar[d]^{\nabla^0_* %\circ (-)
	  }
	  \\
	  {*}
	  \ar@{->>}[r]
%	  \ar@/_1pc/[rr]_{%\vdash
%	  \nabla^0}
	  &
	  \mathbf{B}\left(\mathbf{\Omega}_{\nabla^0} [X,\mathbf{B}G] \right)
	  \ar@{^{(}->}[r]
	  &
	  [X, \mathbf{B}G]
	}\,.
  $$
%  We form the 1-image factorization of the bottom
%  map as indicated and observe that by homotopy pullback stability of
%  1-monomorphisms and 1-epimorphisms in $\mathbf{H}$ also the right and
%  in particular also the left sub-square are then homotopy pullbacks.
\par
  Now if $G$ is equipped with the structure of a braided
  group,
  it remains to see that the vertical map in the middle
  is a homomorphism of $\infty$-groups and the left
  square is a homotopy pullback of group objects in $\mathbf{H}$. In other words, we want to show that the left square deloops.
To this end, notice that if a double deleting $\mathbf{B}^2G$ of $G$ exists, then one can deloop the homotopy pullback diagram we started with. Then, taking the 2-image factorization of $\mathbf{B}(\nabla_0)\colon *\to [X,\mathbf{B}^2G]$ and recalling from Remark \ref{LoopingOfNImage} that
$
    \mathrm{im}_1\left(\nabla_0\right)
	\simeq
	\mathbf{\Omega}\, \mathrm{im}_{2}(\mathbf{B}(\nabla_0))
  $
  one sees that the 2-image factorisation gives the desired delooping of the left square.
\endofproof

\subsection{Higher differential gauge groupoids}
 \label{HigherCourantQuantomorphimsGroupoids}

In Section \ref{HigherAtiyah} we have seen that the higher Atiyah
groupoid (or higher gauge groupoid) of a $G$-principal $\infty$-bundle $P \to X$
modulated by a map
$\nabla^0 : \xymatrix{X \ar[r] & \mathbf{B}G}$
is equivalently
%just
the 1-image projection of $\nabla^0$.
As the notion of 1-image projection does not depend on $\mathbf{B}G$ being the moduli stack of $G$-principal
bundles, this construction of the Atiyah groupoid immediately generalizes to the case of an arbitrary target $\infty$-stack $\mathbf{S}$.
In particular we will consider as targets the stacks $\mathbf{B}^nU(1)_{\mathrm{conn}}$ of $U(1)$-$n$-bundles with connection and will
discuss how the general construction of higher Atiyah groupoids
leads in this case to \emph{differential} higher gauge groupoids which refine the traditional notion of quantomorphism group
and Poisson bracket Lie algebra of a symplectic manifold
to higher geometry. Similarly, by considering the stacks  $\mathbf{B}(\mathbf{B}^{n-1}U(1)_{\mathrm{conn}})$ one obtains a natural construction and generalization of Courant groupoids and algebroids.
%
%In the following section \ref{TheCentralTheorems} we then state the corresponding
%differential and higher analogs of the Atiyah sequence.

\subsubsection{Higher quantomorphism- and Heisenberg-groupoids}
 \label{QuantomorphismAndHeisenbergGroup}

Let us begin by considering the classical case, i.e. a $U(1)$-principal bundle with connection over a smooth manifold $X$. Denoting $\nabla\colon X\to \mathbf{B}U(1)_{\mathrm{conn}}$ the modulating map for this bundle, we can consider the 1-image of $\nabla$ and call it the Atiyah groupoid (or gauge groupoid) of $\nabla$, i.e., we can set $\mathrm{At}(\nabla):= \mathrm{im}_1(\nabla)$. By Example \ref{ex.atiyah} we have a canonical equivalence
\[
\mathbf{BiSect}(\mathrm{At}(\nabla))
	\simeq
	\mathbf{Aut}_{\mathbf{H}}(\nabla).
	\]
and  Proposition \ref{prop.aut-in-slice} gives us the homotopy pullback.
   $$
    \xymatrix{
	   \mathbf{Aut}_{\mathbf{H}}(\nabla)
	  \ar[r]
	  \ar[d]
	 	  &
	  \mathbf{Aut}(X)
	  \ar[d]^{\nabla_*
	  }
	  \\
	  {*}
	  \ar[r]_-{
	  \nabla}
	  	  &
	  [X, \mathbf{B}U(1)_{\mathrm{conn}}]
	}
  $$
  However, as observed in \ref{ModuliOfConnections}, the moduli stack
  $ [X, \mathbf{B}U(1)_{\mathrm{conn}}]$ does not parameterize
  smooth families of $U(1)$-connections on $X$ in the correct way, and so
  when one is interested in such families, it has to be replaced by its
  concretified version $U(1)\mathbf{Conn}(X)$. When one does this, in
  the above fiber sequence, the smooth group
  $\mathbf{Aut}_{\mathbf{H}}(\nabla)$ is replaced by a smooth group
  having the same global points, i.e., whose ``elements'' are still
  the automorphisms of the map $\nabla$. More explicitly, these are
  pairs $(\phi,\eta)$ consisting of a diffeomorphism $\phi \colon X
  \to X$ and a gauge transformation $\eta \colon \phi^* \nabla \to
  \nabla$, but whose smooth structure is such that a morphism from a
  Cartesian space $U$ to this smooth group are precisely smooth
  families $(\phi_u,\eta_u)$ of such automorphisms, parametrized by
  $U$. The group parametrizing such smooth families is known in
  geometric quantization as the \emph{quantomorphism group} of the
  $U(1)$-bundle with connection modulated by $\nabla$. In other words,
  we see that the quantomorphism group $\mathbf{QuantMorph}(\nabla)$
  of $\nabla$ is defined by the homotopy fiber sequence
  $$
    \xymatrix{
	  \mathbf{QuantMorph}(\nabla)
	  \ar[r]
	  \ar[d]
	 	  &
	  \mathbf{Aut}(X)
	  \ar[d]^{\nabla_*
	  }
	  \\
	  {*}
	  \ar[r]_-{\nabla}
	  	  &
	 U(1)\mathbf{Conn}(X)
	}
  $$
where the morphism on the right is actually the composite of $\nabla_*$ with the
differential concretification projection
$\xymatrix{[X, \mathbf{B}U(1)_{\mathrm{conn}}] \ar[r] & U(1)\mathbf{Conn}(X)}$
of Section \ref{ModuliOfConnections}.

All this suggests the following immediate generalization to the case of arbitrary $U(1)$-$n$-bundles with connection.
\begin{definition}\label{def.quantomorphism.group}
Let $X$ be a smooth manifold and let $\nabla\colon X\to \mathbf{B}^nU(1)_{\mathrm{conn}}$ be a map modulating a $U(1)$-$n$-bundle with connection on $X$. The \emph{higher differential gauge groupoid} (or higher differential Atiyah groupoid) of $\nabla$ is defined as the 1-image of $\nabla$, i.e.,
$\mathrm{At}(\nabla):= \mathrm{im}_1(\nabla)$. The \emph{quantomorphism $\infty$-group} of $\nabla$ is defined by the homotopy fiber sequence
 $$
    \xymatrix{
	  \mathbf{QuantMorph}(\nabla)
	  \ar[r]
	  \ar[d]
	 	  &
	  \mathbf{Aut}(X)
	  \ar[d]^{\nabla_*
	  }
	  \\
	  {*}
	  \ar[r]_-{\nabla}
	  	  &
	 U(1)\text{-}n\text{-}\mathbf{Conn}(X)
	}
  $$
\end{definition}

\begin{remark}
Let $\nabla\colon X\to \mathbf{B}U(1)_{\mathrm{conn}}$ be a map modulating a principal $U(1)$-bundle with connection on a smooth manifold $X$, and let $P\to X$ be the underlying principal $U(1)$-bundle. By Remark \ref{differential-to-gauge}, the canonical forgetful map
$\mathbf{B}U(1)_{\mathrm{conn}} \to \mathbf{B}U(1)$, mapping $\nabla$ to the morphism $\nabla^0\colon X\to \mathbf{B}U(1)$ modulating $P$, induces a canonical morphism $\mathrm{At}(\nabla)\to \mathrm{At}(P)$ from the differential gauge groupoid of $\nabla$ to the gauge groupoid of the underlying
principal bundle $P$. In turn, this induces a canonical morphism
$\mathbf{QuantMorph}(\nabla) \to  \mathbf{BiSect}(\mathrm{At}(P))\simeq
	\mathbf{Aut}_{\mathbf{H}}(\nabla^0)$ from the quantomorphism group of $\nabla$ to the group of bisections of $P$.
Thereby, by Remark \ref{action.bisections}, we have a canonical action of the quantomorphism group
%via prop. \ref{ActionOfAtiyahBisectionsOnPrequantumStates},
%the quantomorphism $\infty$-group acts
on the space of sections of any associated $V$-fiber bundle $E\to X$ for $P$. This is the
\emph{prequantum operator} action we have discussed in the introduction in the case $E$ is the $\mathbb{C}$-fiber bundle associated with $P$ by the canonical action of $U(1)$ on $\mathbb{C}$. All this immediately generalizes to higher $U(1)$-bundles with connection, giving the $\infty$-action of the quantomorphism $\infty$-groups on the space of sections of  $V$-fiber $\infty$-bundles associated to the underlying principal $U(1)$-$n$-bundles.
\end{remark}

\begin{example}\label{prequantum-operator} Let us work out in detail the example of the prequantum operator action described in the introduction. Globally, a quantomorphism of $\nabla\colon  X\to \mathbf{B}U(1)_{\mathrm{conn}}$ is a pair $(\phi,\eta)$, where $\phi$ is a diffeomorphism of $X$ and $\eta\colon \phi^*\nabla\to \nabla$ is an isomorphism of principal $U(1)$-connections on $X$. In particular $\eta$ induces an isomorphism (which we will denote by the same symbol) $\eta\colon \phi^*P\to P$ between the principal $U(1)$-bundles underlying $\phi^*\nabla$ and $\nabla$, respectively. The action of $(\phi,\eta)$ on a section $\sigma$ of the $\mathbb{C}$-fiber bundle associated with $P$ by the canonical $U(1)$-action on $\mathbb{C}$ is then given by
\[
(\phi,\eta)\colon \sigma \to \eta(\phi^*\sigma).
\]
That is, we first pull back the section $\sigma$ via $\phi$ to get a
section $\phi^*\sigma$ of the pullback-bundle
$\phi^*(P\times_{U(1)}\mathbb{C})$, and then identify this section
with a section of $P\times_{U(1)}\mathbb{C}$ by the isomorphism
$\eta\colon \phi^*(P\times_{U(1)}\mathbb{C}) \to
P\times_{U(1)}\mathbb{C}$.  A local data description of this action is
easily given. To do so, let us fix a good open cover
$\mathcal{U}=\{U_i\}_{i\in I}$ of $X$ and write $\tilde{\mathcal{U}}$
for the good open cover of $X$ defined by
$\tilde{U}_i=\phi^{-1}(U_i)$, so that for any $i$ in $I$, the
diffeomorphism $\phi$ of $X$ restricts to a diffeomorphism $\phi\colon
\tilde{U}_i\to U_i$. Next, denote by $g_{ij}\colon U_{ij}\to U(1)$ and
by $\tilde{g}_{ij}\colon \tilde{U}_{ij}\to U(1)$ the transition
functions for the $U(1)$-bundle $P$ relative to the covers
$\mathcal{U}$ and $\tilde{\mathcal{U}}$, respectively. Then the
isomorphism $\eta\colon \phi^*P\to P$ is encoded into a set of smooth
functions $\eta_i\colon \tilde{U}_i\to U(1)$ such that
\[
\phi^*g_{ij}=\eta_i^{-1}\tilde{g}_{ij}\eta_j
\]
on $\tilde{U}_{ij}$, for any $i,j$.\footnote{Since $\eta$ is actually an isomorphism of principal $U(1)$-connections, there are other constrains for the $\eta_i$'s, expressing the compatibility with the local connection 1-forms. Here we are not writing these explicitly since, as seen above, $\eta$ will come into play only as an isomorphism of principal $U(1)$-bundles.}  A section $\sigma$ of the associated $\mathbb{C}$-fiber bundle is given by local functions $\sigma_i\colon U_i\to \mathbb{C}$ such that $\sigma_i=g_{ij}\sigma_j$ on $U_{ij}$, for every $i,j$. Pulling back these functions via $\phi$ we get local functions $\phi^*\sigma_i\colon \tilde{U}_i\to \mathbb{C}$ which satisfy
\[
\phi^*\sigma_i=\phi^*g_{ij}\phi^*\sigma_j=\eta_i^{-1}\tilde{g}_{ij}\eta_j \phi^*\sigma_j.
\]
Hence,
\[
\eta_i\phi^*\sigma_i=\tilde{g}_{ij}\eta_j \phi^*\sigma_j,
\]
and we see that $\{\eta_i\phi^*\sigma_i\}_{i\in I}$ are the local data for a section of $P\times_{U(1)}\mathbb{C}$. Indeed these are precisely the local data for the section $\eta(\phi^*\sigma)$ described above.
\end{example}

 If $\nabla\colon X\to \mathbf{B}U(1)_{\mathrm{conn}}$ is a geometric prequantization of a symplectic manifold $(X,\omega)$, then it is well known that the quantomorphism group of $\nabla$ integrates the Poisson algebra of $(X,\omega)$, i.e., the vector field components of ``infinitesimal quantomorphisms'' are Hamiltonian vector fields.
  Moreover, given a Hamiltonian action for a Lie group $G$ on $X$, a momentum map for this action is equivalently a lift
\[
\xymatrix{
& (C^\infty(X),\{\,,\,\})\ar[d]\\
\mathfrak{g}\ar[r]\ar[ru]&\mathfrak{X}_{\mathrm{Ham}}(X)
}
\]
of the Lie algebra morphism describing the infinitesimal Hamiltonian action of $G$. Finally, if the symplectic manifold is $\mathbb{R}^{2n}$ endowed with its canonical symplectic structure, and we consider the Hamiltonian action of $\mathbb{R}^{2n}$ over itself by translations, then the Heisenberg algebra on $2n$ variables is realized as the pullback of Lie algebra morphisms
\[
\xymatrix{
\mathrm{Heis}_{2n}\ar[r]\ar[d]& (C^\infty(\mathbb{R}^{2n}),\{\,,\,\})\ar[d]\\
\mathbb{R}^{2n}\ar[r]&\mathfrak{X}_{\mathrm{Ham}}(\mathbb{R}^{2n})
}
\]
Given all this, we now have the following list of evident generalizations of traditional notions in geometric prequantization to the context of higher prequantum bundles $\nabla\colon X\to \mathbf{B}^nU(1)_{\mathrm{conn}}$ which we are considering here.
\begin{definition}
 \label{def.hamiltonian.group}
  Let $\nabla\colon X \to \mathbf{B}^nU(1)_{\mathrm{conn}}$ be a prequantum $n$-bundle on a smooth manifold, or more generally on a smooth stack, $X$.
 Then
  \begin{itemize}
    \item the \emph{Hamiltonian symplectomorphism group}
	  $\mathbf{HamSympl}(\nabla)$ is the sub-$\infty$-group
	  of the smooth automorphisms of $X$ which is the 1-image
	  of the quantomorphism group in $ \mathbf{Aut}(X)$:
	  $$\xymatrix{
		   \mathbf{QuantMorph}(\nabla)
		   \ar@{->>}[r]
		   &
		   \mathbf{HamSympl}(\nabla)
		   \ar@^{^{(}->}[r]
		   &
		   \mathbf{Aut}(X)\,
		}
	  $$
	 \item
       for $G$ a smooth $\infty$-group,
       a \emph{Hamiltonian $G$-action} on $X$ is an $\infty$-group homomorphism
	   $G \xrightarrow{\rho} \mathbf{HamSympl}(\nabla)$;
	 \item
	   an \emph{integrated $G$-momentum map} is a lift
	\[
\xymatrix{
& \mathbf{QuantMorph}(\nabla)
\ar[d]\\
G\ar[r]_-{\rho}\ar[ru]^{\hat{\rho}}& \mathbf{HamSympl}(\nabla)
}
\]
	   	 \item
	   given a Hamiltonian $G$-action $\rho$, the corresponding
	   \emph{Heisenberg $\infty$-group} $\mathbf{Heis}_\rho(\nabla)$ is the
	   homotopy fiber product in
	   $$
	     \raisebox{20pt}{
	     \xymatrix{
		    \mathbf{Heis}_\rho(\nabla) \ar[r] \ar[d] & \mathbf{QuantMorph}(\nabla) \ar[d]
			\\
			G \ar[r]^-\rho & \mathbf{HamSympl}(\nabla)
		 }
		 }\,.
	   $$
  \end{itemize}
  \label{HamiltonianActionMomentumMapAndHeisenbergGroup}
\end{definition}

\begin{example}
A remarkable example of a higher Heisenberg group is the String group of a simply connected simple compact Lie group $G$. We describe this example in detail in Section \ref{ExtendedWZW} below.

\end{example}

\subsubsection{Higher Courant groupoids}
 \label{HigherCourant}

 Given a  $U(1)$-$n$-bundle with connection refining a given principal $U(1)$-$n$-bundle,
$$
  \xymatrix{
    & \mathbf{B}^nU(1)_{\mathrm{conn}}
	\ar[d]
    \\
    X \ar[ur]^{\nabla} \ar[r]^{\nabla^0} & \mathbf{B}^nU(1)
  },
$$
we have considered in the previous section the corresponding higher
gauge groupoid $\mathrm{At}(\nabla^0)$ and the higher differential
gauge groupoid $\mathrm{At}(\nabla)$, and have seen that they come
with a canonical ``forgetful'' map $ \mathrm{At}(\nabla)\to
\mathrm{At}(\nabla^0)$.
Furthermore, in Section \ref{HigherConnections},
we considered towers of stacks of higher
$U(1)$-bundles with connection ``without higher degree connection
data'':
$$
  \xymatrix{
	&& \mathbf{B}^nU(1)_{\mathrm{conn}} \ar[d]
	\\
	&& \mathbf{B}(\mathbf{B}^{n-1}U(1)_{\mathrm{conn}})
	   \ar[d]
	\\
	&& \vdots \ar[d]
	\\
    X \ar[rr]|{\nabla^0}^{\ }="t"
	 \ar[uurr]|{\nabla^{n-1}}_{\ }="s"
	 \ar@/^1pc/[uuurr]^{\nabla}
	&& \mathbf{B}^nU(1)	%
	\ar@{..} "s"; "t"
  }
$$
Hence, % we have a whole
% tower of higher gauge groupoids
% then
there is a corresponding tower of higher gauge groupoids:
$$
  \xymatrix{
%    \mbox{\begin{tabular}{c}higher \\ Quantomorphism \\ groupoid \end{tabular}}
%	&
%	\mbox{\begin{tabular}{c} higher \\ Courant \\ groupoid \end{tabular}}
%	& \cdots &
%	\mbox{\begin{tabular}{c} intermediate \\ differential \\ higher \\ Atiyah \\ groupoid \end{tabular}}
%	& \cdots
%	&
%	\mbox{\begin{tabular}{c} higher \\ Atiyah \\ groupoid \end{tabular}}
%    \\
    \mathrm{At}(\nabla)
	\ar[r]
	&
	\mathrm{At}(\nabla^{n-1})
	\ar[r]
	&
	\cdots
	\ar[r]
	&
	\mathrm{At}(\nabla^k)
	\ar[r]
	&
	\cdots
	\ar[r]
	&
	\mathrm{At}(\nabla^0)
  }
  \,
$$
interpolating between $\mathrm{At}(\nabla^0)$ and $\mathrm{At}(\nabla)$.

\begin{example}
  Let us have a close look at the $\mathrm{At}(\nabla^{n-1})$
  groupoid, for $n=2$. If $X$ is a smooth manifold, a map
  $\nabla^1\colon X\to \mathbf{B}(\mathbf{B}U(1)_{\mathrm{conn}})$
  modules a geometric structure known in the literature as a
  ``$U(1)$-bundle gerbe with connective structure but without
  curving'' on $X$. The points of the smooth 2-group
  $\mathbf{BiSect}(\mathrm{At}(\nabla))\simeq
  \mathbf{Aut}_{\mathbf{H}}(\nabla)$ are pairs $(\phi, \eta)$
  consisting of a diffeomorphism $\phi\colon X \xrightarrow{\sim} X$
  and an equivalence of bundle gerbes with connective structure but
  without curving $\eta \colon \phi^* \nabla^{1} \xrightarrow{\sim}
  \nabla^{1}$.  A homotopy between two such pairs $(\phi_1, \eta_1)
  \to (\phi_2, \eta_2)$ exists only if $\phi_1 = \phi_2$ and is then
  given by a higher gauge equivalence $\kappa \colon \eta_1
  \xrightarrow{\sim}\eta_2$. By inducing on this 2-group of bisections
  the ``concretified'' smooth structure, as we did for the
  quantomorphism group in Section
  \ref{QuantomorphismAndHeisenbergGroup}, we get a smooth 2-group
  whose $U$-plots are precisely smooth $U$-parameterized collections
  of pairs consisting of diffeomorphisms of $X$ and of compatible
  bundle gerbe gauge transformations.  It is precisely this smooth 2-group
  that was studied in \cite{BC}. There it was shown that the Lie
  2-algebra corresponding to it via Lie differentiation is the Lie
  2-algebra of sections of the \emph{Courant Lie 2-algebroid}, which
  is traditionally associated with the bundle gerbe with connective
  structure modulated by $\nabla^1$. (See \cite{BC} and the references
  therein for details on Courant Lie 2-algebroids.) Therefore, we see
  that the smooth higher groupoid $\mathrm{At}(\nabla^{1})$ is indeed
  a Lie integration of the traditional Courant Lie 2-algebroid
  assigned to $\nabla^{1}$: the \emph{smooth Courant 2-groupoid} of
  $\nabla^1$.
  \end{example}

The above example suggests the following general definition.

\begin{definition}
  For $\nabla^{n-1} \colon X \to \mathbf{B}(\mathbf{B}^{n-1}U(1)_{\mathrm{conn}})$
  a $U(1)$-$n$-principal connection without top-degree connection data,
   the  \emph{higher Courant groupoid} of $\nabla^{n-1}$ is defined as the
higher Atiyah groupoid
  $\mathrm{At}(\nabla^{n-1})$, i.e., as the smooth $\infty$-groupoid given by the 1-image
  factorization of $\nabla^{n-1}$.
%  $$
%    \xymatrix{
%	  X \ar@{->>}[r]
%	  &
%	  \mathrm{At}(\nabla^{n-1})
%	  :=
%	  \mathrm{im}_1(\nabla^{n-1})
%	}
%	\,.
%  $$
  \label{HigherCourantGroupoid}
\end{definition}
\begin{example}
%  More generally, in the situation of example \ref{TheTraditionalCourant2Groupoid}
%  consider now for some $n \geq 1$ the smooth circle $n$-group
%  $\mathbb{G} = \mathbf{B}^{n-1}U(1)$ as in example \ref{GroupsInSmoothInfinityGroupoids}.
%  Then by example \ref{DeligneComplexInExample} a map
%  $$
%    \nabla^{n-1} :
%	\xymatrix{X \ar[r] & \mathbf{B}(\mathbf{B}^{n-1}U(1)_{\mathrm{conn}})}
%  $$
%  is equivalently a Deligne cocycle on $X$ in degree $(n+1)$ without $n$-form data.
%
  To see what the smooth higher Courant groupoid
  $\mathrm{At}(\nabla^{n-1})$ is like, consider first the topologically trivial case, in which
  the underlying $U(1)$-$n$-bundle is trivial. In this case a bisection of $\mathrm{At}(\nabla^{n-1})$
  is readily seen to be a pair consisting of a diffeomorphism
  $\phi$ of $X$ together with an $(n-1)$-form $H \in \Omega^{n-1}(X)$,
  satisfying no further compatibility condition. This suggests that, in the general case, there should be a model for the Lie differentiation of
  the higher Courant groupoid $\mathrm{At}(\nabla^{n-1})$ which is given by an $L_\infty$-algebra
  having in the lowest degree the space of sections
  of a vector bundle on $X$ locally isomorphic to the direct sum $T X \oplus \wedge^{n-1} T^* X$.
  This is precisely what the sections of higher Courant Lie $n$-algebroids
  are supposed to be like, see for instance \cite{Zambon}.
\end{example}

%Finally, if we are given a tower of differential refinements of $\mathbb{G}$-principal
%bundles as discussed in \ref{DifferentiaCoefficients}
%
%$$
%  \xymatrix{
%	&& \mathbf{B}\mathbb{G}_{\mathrm{conn}} \ar[d]
%	\\
%	&& \mathbf{B}\mathbb{G}_{\mathrm{conn}^{n-1}}
%	   \ar[d]
%	\\
%	&& \vdots \ar[d]
%	\\
%    X \ar[rr]|{\nabla^0}^{\ }="t"
%	 \ar[uurr]|{\nabla^{n-1}}_{\ }="s"
%	 \ar[uuurr]^{\nabla}
%	&& \mathbf{B}\mathbb{G}
%	%
%	\ar@{..} "s"; "t"
%  }
%$$
%then there is correspondingly a tower of higher gauge groupoids:
%$$
%  \xymatrix{
%    \mbox{\begin{tabular}{c}higher \\ Quantomorphism \\ groupoid \end{tabular}}
%	&
%	\mbox{\begin{tabular}{c} higher \\ Courant \\ groupoid \end{tabular}}
%	& \cdots &
%	\mbox{\begin{tabular}{c} intermediate \\ differential \\ higher \\ Atiyah \\ groupoid \end{tabular}}
%	& \cdots
%	&
%	\mbox{\begin{tabular}{c} higher \\ Atiyah \\ groupoid \end{tabular}}
%    \\
%    \mathrm{At}(\nabla)_\bullet
%	\ar[r]
%	&
%	\mathrm{At}(\nabla^{n-1})_\bullet
%	\ar[r]
%	&
%	\cdots
%	\ar[r]
%	&
%	\mathrm{At}(\nabla^k)
%	\ar[r]
%	&
%	\cdots
%	\ar[r]
%	&
%	\mathrm{At}(\nabla^0)
%  }
%  \,.
%$$
\begin{remark}The further intermediate stages $\mathrm{At}(\nabla^k)$ with $0<k<n-1$ seem not to correspond to
anything that has already been given a name in traditional literature.
We might call them \emph{intermediate higher differential gauge groupoids}.
%These structures are an integral part of higher prequantum geometry.
\end{remark}

\begin{remark}
More generally, for any sylleptic
$\infty$-group $G$ one can define the differential refinement $\mathbf{B}^2 \mathbb{G}_{\mathrm{conn}}$ of the double delooping $\mathbf{B}^2G$ and one has the sequence of morphisms of stacks
$$
  \xymatrix{
    \mathbf{B}^2 G_{\mathrm{conn}}
	\ar[r]
	&
	\mathbf{B}(\mathbf{B}G_{\mathrm{conn}})
	\ar[r]
	&
	\mathbf{B}^2G
  }
$$
and one gets a notion of higher Courant groupoid for a ``$G$-principal connection without top-degree connection data'', $\nabla^{\bullet-1} : X \to \mathbf{B}(\mathbf{B}G_{\mathrm{conn}})$.
\end{remark}

\subsection{Higher integrated Kostant-Souriau extensions}
 \label{TheCentralTheorems}

Conceptually, a key aspect of the traditional notion of the \emph{Poisson bracket} Lie
algebra of observables is that (over a connected manifold) it constitutes a central $\mathbb{R}$-extension
 of the Lie algebra
of Hamiltonian vector fields. This is
called the \emph{Kostant-Souriau extension}, see, e.g., \cite[section 2.3]{BrylinskiLoop}.
At level of (infinite dimensional) Lie groups, this is reflected by the fact that the quantomorphism group of a prequantized symplectic manifold $(X,\omega)$ is a $U(1)$-central
%This aspect is strenghtened by the key aspect of the
%\emph{quantomorphism group} extension: the corresponding $U(1)$-extension
of the Lie group of Hamiltonian symplectomorphisms of $X$, see \cite[section II.4]{BrylinskiLoop}.
%-- instead of the $\mathbb{R}$-extension,
%which also is a possible Lie integration.
The $U(1)$-\emph{phases} appearing this way
on top of classical Hamiltonian structures are the hallmark of (pre-)quantum geometry.

Here we discuss the refinement of these extensions to higher prequantum geometry.
First we consider general quantomorphism $\infty$-group extensions in Section \ref{TheQuantomorphismExtension},
and then the corresponding infinitesimal Poisson bracket $L_\infty$-algebra extensions
%over smooth manifolds
in Section \ref{TheHigherPoissonBracketExtension}. The latter are discussed in more detail in
\cite{nextArticle}.

\subsubsection{The quantomorphism $\infty$-group extensions}
\label{TheQuantomorphismExtension}

The main result in this section is a description of the quantomorphism
group of a $U(1)$-principal connection $\nabla\colon X\to
\mathbf{B}^nU(1)_{\mathrm{conn}}$ as a (higher) central extension of
the smooth $\infty$ group $\mathbf{HamSympl}(\nabla)$ introduced in
Definition \ref{def.hamiltonian.group}. The result is basically an
exercise in the definitions and in the properties of homotopy
pullbacks and 1-image factorizations.
\begin{theorem}
\label{TheLongHomotopyFiberSequenceOfTheQuantomorphimsGroup}
Let $\nabla\colon X\to \mathbf{B}^nU(1)_{\mathrm{conn}}$ a principal $U(1)$-$n$-connection over $X$. Then we have a long exact fiber sequence of smooth $\infty$-groups
$$
  \xymatrix{
     \mathbf{Flat}\text{-}U(1)\text{-}(n-1)\text{-}\mathbf{Conn}(X)
	 \ar[r]\ar[d]
	 &
	 \mathbf{QuantMorph}(\nabla)
	 \ar[r]\ar[d]
	 & {*}\ar[d]\\
	 {*}\ar[r]&\mathbf{HamSympl}(\nabla)
	 \ar[r]^-{\nabla_*}\
	 &
	 \mathbf{B}
	 \left(\mathbf{Flat}\text{-}U(1)\text{-}(n-1)\text{-}\mathbf{Conn}(X)
	 \right)
  }
  \,,
$$
\end{theorem}
\proof
  By Definition \ref{def.quantomorphism.group}, Definition \ref{def.hamiltonian.group}, and reasoning as in the proof of Theorem \ref{TheLongSequenceForHigherAtiyahBisections}, one has the following diagram, where the horizontal rows are 1-image factorizations:
    $$
    \raisebox{20pt}{
    \xymatrix{
      \mathbf{QuantMorph}(\nabla)
	  \ar@{->>}[r]
	  \ar[d]
	  &
	  \mathbf{HamSympl}(\nabla)
	  \ar@{^{(}->}[r]
	  \ar[d]^{\nabla_*}
	  &
	  \mathbf{Aut}(X)
	  \ar[d]^{\nabla_*}
	  \\
	  {*}
	  \ar@{->>}[r]
	  \ar@/_1pc/[rr]_{\nabla}
	  &
	  \mathbf{B}\left( \Omega_{\nabla} \left( U(1)\text{-}n\text{-}\mathbf{Conn}(X)\right) \right)
	  \ar@{^{(}->}[r]
	  &
	 U(1)\text{-}n\text{-}\mathbf{Conn}(X)
	}
   }
   \,.
  $$
  By homotopy pullback stability of both 1-epimorphisms and 1-monomorphisms and by
  essential uniqueness of 1-image factorizations this is a pasting diagram of homotopy pullback
  squares. In particular, the square diagram on the left is a homotopy pullback. The result then follows by noticing that
  \[
  \Omega_{\nabla} \left( U(1)\text{-}n\text{-}\mathbf{Conn}(X)\right)\simeq \mathbf{Flat}\text{-}U(1)\text{-}(n-1)\text{-}\mathbf{Conn}(X).
  \]
  Namely, this is nothing but the concretified version of Remark \ref{rem.loop-is-flat}.
%  The claim then follows with prop. \ref{LoopingOfDifferentialModuliIsFlatModuli} as in the proof of theorem
%  \ref{TheLongSequenceForHigherAtiyahBisections}.
\endofproof

By the pasting law for homotopy pullbacks, the analogous statement also holds for Heisenberg $\infty$-groups:
\begin{corollary}
If $\rho\colon G \to \mathbf{HamSympl}(\nabla)$
is any Hamiltonian $G$-action on $X$,
then the corresponding Heisenberg $\infty$-group sits in
the long homotopy fiber sequence
$$
  \xymatrix{
     \mathbf{Flat}\text{-}U(1)\text{-}(n-1)\text{-}\mathbf{Conn}(X)
	 \ar[r]\ar[d]
	 &
	 \mathbf{Heis}_\rho(\nabla)
	 \ar[r]\ar[d]
	 & {*}\ar[d]\\
	 {*}\ar[r]&G
	 \ar[r]^-{\nabla_*\circ \rho}\
	 &
	 \mathbf{B}
	 \left(\mathbf{Flat}\text{-}U(1)\text{-}(n-1)\text{-}\mathbf{Conn}(X)
	 \right)
  }
  \,,
$$
of smooth $\infty$-groups.
%
%
%$$
%  \xymatrix{
%     (\Omega \mathbb{G})\mathbf{FlatConn}(X)
%	 \ar[r]
%	 &
%	 \mathbf{Heis}_\phi(\nabla)
%	 \ar[r]
%	 &
%	 G
%	 \ar[r]^-{\nabla \circ (-)}
%	 &
%	 \mathbf{B}
%	 \left(
%	   \left(\Omega\mathbb{G}\right)
%	   \mathbf{FlatConn}(\nabla)
%	 \right)
%  }
%  \,,
%$$
\end{corollary}
The rather mysterious groups $\mathbf{Flat}\text{-}U(1)\text{-}(n-1)\text{-}\mathbf{Conn}(X)$ of flat $U(1)$-$(n-1)$-connections appearing in the above statements actually admit a much more familiar description if the base manifold $X$ is sufficiently connected.
%In order to put the higher generalizations of the quantomorphism extensions
%into this context, we notice the following basic fact.
\begin{proposition}
%  For $\mathbb{G} = \mathbf{B}U(1) \in \mathrm{Grp}(\mathrm{Smooth}\infty\mathrm{Grpd})$
%  the smooth circle 2-group
%  as in example \ref{deRhamCoefficientsForCircleNGroup} for $n = 2$,
%  consider $X \in \mathrm{SmoothMfd} \hookrightarrow \mathrm{Smooth}\infty\mathrm{Grpd}$
If $X$ is a connected and simply connected smooth manifold, %. Then from
 % prop. \ref{TheCorrectU1DifferentialModuli} and example \ref{TheCorrectU1DifferentialModuliFlat}
  %one obtains
 then one has  an equivalence of smooth $\infty$-groups
  $$
    \mathbf{Flat}\text{-}U(1)\text{-}\mathbf{Conn}(X) \simeq  \mathbf{B}U(1)
	\,.
  $$
  More generally, for any $n \geq 1$, if
%  and for $\mathbb{G} = \mathbf{B}^n U(1) \in \mathrm{Grp}(\mathrm{Smooth}\infty\mathrm{Grpd})$
%  the smooth circle $(n+1)$-group
%  as in example \ref{deRhamCoefficientsForCircleNGroup}, there is for $X$
 $X$ is an $(n-1)$-connected smooth manifold, then one has an equivalence of smooth $\infty$-groups
  $$
    \mathbf{Flat}\text{-}U(1)\text{-}(n-1)\text{-}\mathbf{Conn}(X) \simeq \mathbf{B}^{n-1} U(1)
	\,.
  $$
  \label{FlatConnectionsOnSimplyConnectedManifolds}
\end{proposition}
\proof
It is a classical fact that a flat $U(1)$-bundle over a manifold $X$ is equivalently the datum of a $U(1)$-local system on $X$, i.e., of a $U(1)$-valued representation of the Poincar\'e groupoid of $X$. Translated into the language of smooth stacks, this becomes an instance of the $\flat/\Pi$-adjunction:\footnote{See \cite{dcct} for a general treatment of this $\infty$-adjunction.} the space $\mathbf{H}(X,\flat\mathbf{B}U(1))$ of flat $U(1)$-connections on $X$ is equivalently the space $\mathbf{H}(\Pi X,\mathbf{B}U(1))$ of maps from the Poincar\'e groupoid of $X$ to $\mathbf{B}U(1)$. Since $\mathbf{B}U(1)$ is a 1-stack, we have
\[
\mathbf{H}(\Pi X,\mathbf{B}U(1))\cong \mathbf{H}((\Pi X)_{\leq 1},\mathbf{B}U(1))
\]
see Section \ref{sec.truncations}. By definition of $\mathbf{Flat}\text{-}U(1)\text{-}\mathbf{Conn}(X)$, and using the fact that a Cartesian space $U$ is a 0-stack, we have
\[
\mathbf{H}(U,\mathbf{Flat}\text{-}U(1)\text{-}\mathrm{Conn}(X))\cong \mathbf{H}(U\times \Pi X,\mathbf{B}U(1))\cong \mathbf{H}(U\times (\Pi X)_{\leq 1},\mathbf{B}U(1))
\]
If $X$ is connected and simply connected, then $(\Pi X)_{\leq 1}\simeq *$, and so
\[
\mathbf{H}(U,\mathbf{Flat}\text{-}U(1)\text{-}\mathbf{Conn}(X))\cong  \mathbf{H}(U,\mathbf{B}U(1)).
\]
This shows that the stacks $\mathbf{Flat}\text{-}U(1)\text{-}\mathbf{Conn}(X)$ and $\mathbf{B}U(1)$ coincide (or, better, are naturally equivalent) on every Cartesian space $U$, and so they are equivalent.
%The conclusion then follows from the Yoneda lemma.
%  We use the description of $U(1)\mathbf{FlatConn}(X)$ given
%  by prop. \ref{TheCorrectU1DifferentialModuli} and example \ref{TheCorrectU1DifferentialModuliFlat}.
%
%  A $U(1)$-
%  First notice then that on a simply connected manifold there is up to equivalence just a single
%   flat connection, hence $U(1)\mathbf{FlatConn}(X)$ is pointed connected.
%  Moreover, an auto-gauge transformation from that single flat connection (any one)
%to itself is a $U(1)$-valued function which is \emph{constant on X}.
%But therefore by prop. \ref{TheCorrectU1DifferentialModuli} the $U$-plots of
%the first homotopy sheaf of $U(1)\mathbf{FlatConn}(X)$ are
%smoothly $U$-parameterized collections of constant $U(1)$-valued functions on $X$,
%  hence are smoothly $U$-parameterized collections of elements in $U(1)$, hence are
%  smooth $U(1)$-valued functions on $U$. These are, by definition,
%  equivalently the $U$-plots of automorphisms of the point in $\mathbf{B}U(1)$.
%
 % The other cases work analogously.
The proof for an arbitrary $n$ is perfectly analogous.
\endofproof

\begin{corollary}\label{cor.extension}
Let $\nabla\colon X\to \mathbf{B}^nU(1)_{\mathrm{conn}}$ a principal $U(1)$-$n$-connection over an $(n-1)$-connected smooth manifold $X$. Then we have a long exact fiber sequence of smooth $\infty$-groups
$$
  \xymatrix{
     \mathbf{B}^{n-1}
	 U(1)	 \ar[r]\ar[d]
	 &
	 \mathbf{QuantMorph}(\nabla)
	 \ar[r]\ar[d]
	 & {*}\ar[d]\\
	 {*}\ar[r]&\mathbf{HamSympl}(\nabla)
	 \ar[r]^-{\nabla_*}\
	 &
	 \mathbf{B}^{n}
	 U(1)
  }
  \,.
$$
Moreover, for every Hamiltonian $G$-action $\rho\colon G \to \mathbf{HamSympl}(\nabla)$, we have a long exact fiber sequence of smooth $\infty$-groups
$$
  \xymatrix{
     \mathbf{B}^{n-1}
	 U(1)	 \ar[r]\ar[d]
	 &
	 \mathbf{Heis}_\rho(\nabla)
	 \ar[r]\ar[d]
	 & {*}\ar[d]\\
	 {*}\ar[r]&G
	 \ar[r]^-{\nabla_*\circ \rho}\
	 &
	 \mathbf{B}^{n}
	 U(1)
  }
  \,.
$$

\end{corollary}

\begin{example}
  For $n = 1$, Corollary \ref{cor.extension}
  reproduces the fact that the quantomorphism group of a prequantized connected symplectic manifold is a $U(1)$-central extension of the
  infinite dimensional Lie group of Hamiltonian symplectomorphisms, as discussed for instance in
  \cite{RS, Viz, Viz2}.
\end{example}

\begin{remark}
  For $n=2$, Corollary \ref{cor.extension} shows how in dealing with
  smooth automorphism groups of prequantized simply connected
  2-plectic manifolds one naturally meets smooth 2-group extensions by
  the 2-group $\mathbf{B}U(1)$. The archetypical example of
  $\mathbf{B}U(1)$-extensions is probably the smooth \emph{String
    2-group}, see the appendix of \cite{FiorenzaSatiSchreiber5Brane}
  for review.  Indeed, this occurs as the {Heisenberg 2-group
    extension} of the {WZW sigma-model} regarded as a local
  2-dimensional quantum field theory.  We come back to this in
  \ref{ExtendedWZW} below.
\end{remark}

\subsubsection{The higher Poisson bracket $L_\infty$-extensions}
 \label{TheHigherPoissonBracketExtension}

In Section \ref{QuantomorphismAndHeisenbergGroup} we have introduced the smooth quantomorphism $\infty$-group associated with a map $\nabla\colon X\to \mathbf{B}^nU(1)_{\mathrm{conn}}$. Taking ``infinitesimal quantomorphisms'' we therefore get an $L_\infty$-algebra $\mathfrak{quantmorph}(\nabla)$, which is well defined up to homotopy. A remarkable result in \cite{nextArticle} is that the homotopy type of $\mathfrak{quantmorph}(\nabla)$ only depends on the curvature $n+1$-form $\omega$ of $\nabla$. Even more remarkably, this homotopy type turns out to be that of the higher Poisson bracket $L_\infty$-algebras introduced in \cite{Rogers}. This generalises the classical result that the Lie algebra of the quantomorphism group of a prequantized symplectic manifold $(X,\omega)$ is isomorphic to the Poisson algebra $(C^\infty(X),\{\,,\,\})$.

To state more precisely the above results, let us follow \cite{Rogers} and introduce the following terminology.

\begin{definition}
  Let $n \geq 1$ be a positive integer. A \emph{pre-$n$-plectic manifold} is a pair $(X,\omega)$ consisting of a smooth manifold $X$ and a closed $(n+1)$-form
  $\omega \in \Omega^{n+1}_{\mathrm{cl}}(X)$. If the map $\iota_{(-)}\omega : T_x X \to \wedge^n T^*_x X$
  is injective for all $x \in X$, then we say that the pre-$n$-plectic manifold is an \emph{$n$-plectic manifold}.
\end{definition}
Denote by $\mathfrak{X}(X)$ the Lie algebra of vector fields on a smooth manifold $X$.
\begin{definition}
Let $(X,\omega)$ be a pre-$n$-plectic manifold, and let $v \in \mathfrak{X}(X)$ and $h \in \Omega^{n-1}(X)$ be such that $\iota_v \omega + d h=0$. Then
  we say that $v$ is a \emph{Hamiltonian vector field} (with respect to the pre-$n$-plectic
  structure $\omega$) and that
  $h$ is a \emph{Hamiltonian form} for $v$. We write
	$$
	  \Omega^{n-1}_{\mathrm{Ham}}(X)
	    :=
	  \left\{
	     (v,h) \in \mathfrak{X}(X)\oplus \Omega^{n-1}(X) \;|\; \iota_v \omega + d h =0
	  \right\}
	  \,.
	$$
\end{definition}
  It is immediate to check that Hamiltonian vector fields form a sub-Lie algebra $\mathfrak{X}_{\mathrm{Ham}}(X)$ of $\mathfrak{X}(X)$.
 \begin{definition}
 For $(X, \omega)$ a pre-$n$-plectic manifold, write
  $
    \mathfrak{Poisson}(X,\omega)
  $
  for the $L_\infty$-algebra
  \begin{itemize}
    \item whose underlying chain complex is the modified de Rham complex
	$$
	  \xymatrix{
	    \Omega^0(X) \ar[r]^-{d} & \Omega^1(X) \ar[r]^-d & \cdots \ar[r]^-d & \Omega^{n-2}(X)
		\ar[r]^-{(0,d)} & \Omega^{n-1}_{\mathrm{Ham}}(X)
	  }
	  \,,
	$$
	with $\Omega^{n-1}_{\mathrm{Ham}}(X)$ placed in degree zero;
		\item whose only non-vanishing brackets are on tuples of elements
	$(v_i,h_i) \in \Omega^{n-1}_{\mathrm{Ham}}(X,\omega)$,
	where the binary bracket is
	$$
	  [(v_1, h_1), (v_2,h_2)] := ([v_1,v_2], \iota_{v_1 \wedge v_2} \omega)
	$$
	and where the brackets of arity $k \geq 3$ are
	$$
	  [(v_1, h_1), \cdots, (v_k, h_k)] :=
	  (-1)^{\lfloor\frac{k-1}{2}\rfloor}
	  \iota_{v_1 \wedge \cdots \wedge v_n} \omega
	  \,.
	$$
  \end{itemize}
  We call $\mathfrak{Poisson}(X,\omega)$
 % the \emph{de Rham model} for the higher Poisson bracket of local observables.
  the higher Poisson brackets $L_\infty$ algebra of local observables.
  \label{PoissonBracketLInfinityAlgebra}
\end{definition}

For $n$-plectic manifolds this $L_\infty$-algebra
is isomorphic to the one given in \cite[theorem \ 5.2]{Rogers}.
For the more general
pre-$n$-plectic case this appears as \cite[Theorem\ 4.7]{MomentMap}.
\begin{example}\label{example.poisson}
  For $n = 1$ and non-degenerate $\omega$, hence for $(X,\omega)$
  an ordinary symplectic manifold,
  $\mathfrak{Poisson}(X,\omega)$ is the traditional Poisson bracket Lie algebra
  on the space of smooth functions $C^\infty(X)$ on $X$, which justifies its name. If $X$ is
  thought of as the
  phase space of a physical system, then each point of it corresponds
  to a configuration or trajectory of the system, and a function on $X$
  hence assigns a value to each such configuration. One thinks of this
  value as a physically observable property of the given configuration,
  for instance its energy. In this way functions on $X$ are ``observables''
  for $n= 1$.
  \label{ObservablesIn1}
\end{example}
\begin{remark}
  For $n>1$, as the name suggests, the $L_{\infty}$-algebra given in
    definition\ \ref{PoissonBracketLInfinityAlgebra} has an interpretation
    as a higher Lie algebra of ``local observables'' within the context of
    an $n$-dimensional local field theory. This was
    demonstrated, for example, in \cite{BHR} for the $n=2$ case.
    But notice that the usual rules of homotopy theory imply a nuanced
    notion of what ``a local observable'' is. {In particular, we
      have the following observations:}
  \begin{enumerate}
    \item On the one hand there are the principal
  homotopy-invariants of the chain complex of observables $\mathfrak{Poisson}(X,\omega)$,
  namely its homology groups $H_\bullet(\mathfrak{Poisson}(X,\omega))$.
  The traditional notion of a ``local observable'' is an element
  in one of these homology groups. {This means
  that a degree 0 local observable in the ``strict sense''  is a Hamiltonian $(n-1)$-form $j$ on $X$
  modulo exact forms. This is a familiar structure in quantum field theory: such a
  \emph{local current up to gauge transformation} is something that when integrated over a (closed) spatial slice
  of spacetime, i.e., when transgressed to codimension 1, produces a 0-form
  observable as in example \ref{ObservablesIn1}. This is called the \emph{total charge}
  of the current.}
   \item On the other hand, it is a crucial fact in homotopy theory
     that a homotopy type is not, in general, faithfully encoded by its homotopy groups (here: homology groups).
   {Therefore a ``local observable'', in a more accurate and less restricted sense,
   is \emph{any element} of $\mathfrak{Poisson}(X,\omega)$.
      }

  \item In $L_\infty(X, \omega)$ one also has a relevant notion of homotopy between local observables.
   For instance, the de Donder-Weyl Hamiltonian $H_{\mathrm{DW}}$ of multisymplectic geometry
   is a smooth function on the $n$-plectic manifold $(X,\omega)$
   which characterizes $n$-tuples of Hamiltonian vector fields $(v_1, \cdots, v_n)$
   tangent to the classical solution hypersurfaces by the equation
   $$
     d H_{\mathrm{DW}} = [v_1, \cdots, v_n]_{L_\infty(X,\omega)}
	 \,;
   $$
   see \cite%[around (4)]
   {Helein} for a review.
   This is the ``localized equation of motion'' which de-transgresses
   the traditional equation of motion in symplectic geometry.  In view
   of definition \ref{PoissonBracketLInfinityAlgebra} this equation exhibits
   $H_{\mathrm{DW}}$ as a homotopy in the $\infty$-groupoid of local
   observables that connects the $n$-ary $L_\infty$-bracket of the
   $n$-tuple of Hamiltonian vector fields to the origin. So the
   deDonder-Weyl Hamiltonian is {\emph{not a local observable in
       the strict sense}. Indeed, it crucially is not closed, in
     general, and hence does not represent an element in a homology
     group of $\mathfrak{Poisson}(X,\omega)$. It is, however, an observable in a
     homotopy-theoretic sense, since it gives a \emph{homotopy} between
     ``strict'' observables.}
  \end{enumerate}
\end{remark}
Within the framework of smooth stacks, a pre-$n$-plectic manifold $(X,\omega)$ is equivalently a smooth manifold $X$ equipped with a map $\omega\colon X\to \Omega^{n+1}_{\mathrm{cl}}$ to the smooth stack of closed $n+1$-forms. One can therefore give the following
\begin{definition}
A (geometric) prequantization of a pre-$n$-plectic manifold $(X,\omega)$ is a lift
\[
\xymatrix{
&\mathbf{B}^nU(1)_{\mathrm{conn}}\ar[d]^{\mathrm{curv}}\\
X\ar[r]_{\omega}\ar[ru]^{\nabla}& \Omega^{n+1}_{\mathrm{cl}}.
}
\]
\end{definition}
In more colloquial terms, this is the choice of a $U(1)$-principal connection on $X$, whose curvature $n+1$-form is $\omega$.
\begin{remark}
  It is easy to see that a prequantization in the above sense exists
  if and only if the closed $n+1$-form $\omega$ represents an
  \emph{integral} cohomology class. More generally one can consider
  prequantizations as lifts of $\omega$ to a morphism $\nabla$ from $X$
  to a stack of the form
  $\mathbf{B}^n(\mathbb{R}/\!/\Gamma)_{\mathrm{conn}}$, where $\Gamma$
  is any (non necessarily discrete) subgroup of $\mathbb{R}$. In this
  more general setting, every pre-$n$-plectic manifold admits a
  prequantizazion: it suffices to take $\Gamma$ to be the subgroup of
  periods of the $n+1$-form $\omega$. Everything we are saying in this
  section immediately generalizes to these more general notion of
  prequantization.
\end{remark}
We can now state the main result from \cite{nextArticle}.
\begin{proposition}
Let $\nabla$ be a prequantization of the pre-$n$-plectic manifold $(X,\omega)$, and let $\mathfrak{quantmorph}(\nabla)$ be the $L_\infty$-algebra of infinitesimal quantomorphisms of $\nabla$. Then $\mathfrak{quantmorph}(\nabla)$ has the same homotopy type of $\mathfrak{Poisson}(X,\omega)$. In particular, the homotopy type of $\mathfrak{quantmorph}(\nabla)$ is independent of the prequantization $\nabla$.
\end{proposition}
\begin{remark}
The above result is proven in \cite{nextArticle} by using an explicit model for the smooth $\infty$-group $\mathbf{Quantmorph}(\nabla)$, expressed in terms of local data for the $n$-connection $\nabla$. Differentiating this model one obtains a differential graded Lie algebra which is explicitly seen to be quasi-isomorphic (as an $L_\infty$-algebra) to  $\mathfrak{Poisson}(X,\omega)$. A more intrinsic construction of the $L_\infty$-algebra $\mathfrak{quantmorph}(\nabla)$ can be found in \cite{dcct}.
\end{remark}

\begin{remark}
 Differentiating the action of prequantum operators described in Example \ref{prequantum-operator}, one gets the action of a Hamiltonian $H$ on a section $\psi$ of the associated line bundle (i.e., on a ``prequantum state'' of the prequantized theory). To see this, write $(\{2\pi \sqrt{-1}A_i\}_{i\in I},\{g_{ij}\}_{i,j\in I}$ for the local connection 1-forms and transition functions for $\nabla$. Then in the differential graded Lie algebra model for  $\mathfrak{quantmorph}(\nabla)$ an infinitesimal quantomorphism is a pair $(v,\{\theta_i\}_{i\in I})$, where $v$ is a vector field on $X$ and the $\theta_i$ are smooth $\mathbb{R}$-valued functions on the $U_i$ such that
\[
\begin{cases}
d\theta_i=\mathcal{L}_vA_i\qquad\text{on $U_i$};
\\
\theta_i-\theta_j=\frac{1}{2\pi \sqrt{-1}}\mathcal{L}_v \log g_{ij}\qquad\text{on $U_{ij}$},
\end{cases}
\]
see \cite{nextArticle}. Differentiating the local data expression of the action of a quantomorphism $(\phi,\eta)$ on a section $\sigma$,
 \[
 (\phi,\{\eta_i\}_{i\in I})\colon \{\sigma_i\}_{i\in I}\mapsto  \{\eta_i\phi^*\sigma_i\}_{i\in I},
 \]
 we see that an infinitesimal quantomorphism $(v,\{\theta_i\}_{i\in I})$ acts on $\sigma$ by
 \[
 (v,\{\theta_i\}_{i\in I})\colon \{\sigma_i\}_{i\in I}\mapsto  \{2\pi \sqrt{-1} \theta_i\sigma_i+ \mathcal{L}_v\sigma_i\}_{i\in I}.
 \]
 The infinitesimal quantomorphism corresponding to a Hamiltonian $H$ via the isomorphism $C^\infty(X),\{\,,\,\})\cong \mathfrak{quantmorph}(\nabla)$ is $(v_H, \{-H\bigr\vert_{U_i}+\iota_{v_H}A_i\}_{i\in I})$, where $v_H$ is the Hamiltonian vector field of $H$ and $\iota$ is the contraction operator, see \cite{nextArticle}. Therefore $H$ acts on $\sigma$ as
 \[
O_H\colon \{\sigma_i\}_{i\in I}\mapsto  \{ (-2\pi \sqrt{-1}H\bigr\vert_{U_i}\sigma_i)+\iota_{v_H}(d+2\pi \sqrt{-1}A_i)\sigma_i\}_{i\in I},
 \]
 i.e., as
%\ccomment{added a missing $H$ to first summand below}
 \[
 O_H\colon \sigma \mapsto -2\pi \sqrt{-1}\, H \sigma+ \nabla_{v_H}\sigma.
 \]
This is precisely, up to an overall $\frac{\sqrt{-1}}{2\pi}$ factor, the classical prequantum action of Hamiltonians on prequantum states,
see \cite[p. 94]{BatesWeinstein}.
 \end{remark}

A second result in \cite{nextArticle} exhibits $\mathfrak{Poisson}(X,\omega)$ as a central $L_\infty$-algebra extension of the Lie algebra of Hamiltonian vector fields by the de Rham cohomology of $X$ up to degree $n-1$. To correctly state this result, let us denote by $\mathbf{B}\mathbf{H}(X,\flat\mathbf{B}^{n-1}\mathbb{R})$
the abelian $L_\infty$-algebra given by the chain complex
$
\Omega^0(X)\xrightarrow{d}\Omega^1(X)\xrightarrow{d}\cdots \xrightarrow{d}\Omega^{n-1}(X)\xrightarrow{d}d\Omega^{n-1}(X),
$
with $d\Omega^{n-1}(X)$ in degree zero. This complex
  serves as a resolution of the cocycle complex
$
\Omega^0(X)\xrightarrow{d}\Omega^1(X)\xrightarrow{d}\cdots \xrightarrow{d}\Omega_{\mathrm{cl}}^{n-1}(X)\xrightarrow{} 0
\,,
$
which is more recognizable as the cocycle complex for the de Rham cohomology of $X$ up to degree $n-1$,
once delooped (i.e., shifted on the left by one degree). Then we have the following result \cite[Theorem 3.3.1]{nextArticle}.
\begin{theorem}[higher Kostant-Souriau extension]
For $(X,\omega)$ a pre-$n$-plectic manifold there is a long homotopy fiber sequence
of $L_\infty$-algebras of the form
$$
  \xymatrix{
    \mathbf{H}(X,\flat \mathbf{B}^{n-1}\mathbb{R}  )
	\ar[r]\ar[d]
	&
	\mathfrak{Poisson}(X,\omega)\ar[d]
	\ar[r]&0\ar[d]
	\\
	0\ar[r]&\mathfrak{X}_{\mathrm{Ham}}(X,\omega)
	\ar[r]^-{\iota_{(-)} \omega}
	&
	\mathbf{B} \mathbf{H}(X,\flat \mathbf{B}^{n-1}\mathbb{R}  )
  }
  \,,
$$
where $\iota_{(-)} \omega$ is
  the $L_\infty$-homomorphism whose $k$-ary components are
  given by contracting skew tensor products of $k$ vector fields with $\omega$.
 \label{TheLInfinityExtension}
\end{theorem}

 Theorem \ref{TheLInfinityExtension} says that $\iota_{(-)}\omega$ is the $L_\infty$-cocycle
 which classifies the Poisson bracket $L_\infty$-algebra of local observables as an $L_\infty$-extension
 of the Lie algebra of Hamiltonian vector fields. This cocycle is an $L_\infty$-generalization
 of the traditional \emph{Heisenberg cocycle} classifying the traditional Heisenberg group extension. This suggests the following
%Finally, the image under Lie differentiation of the Heisenbger $\infty$-groups
%as in def. \ref{HamiltonianActionMomentumMapAndHeisenbergGroup} is
\begin{definition}
  For $\mathfrak{g}$ an $L_\infty$-algebra, a Hamiltonian action of $\mathfrak{g}$ on $X$ is an $L_\infty$-algebra morphism
  $\rho \colon \mathfrak{g} \to \mathfrak{X}_{\mathrm{Ham}}(X)$. The homotopy fiber
  of $\iota_{(-)} \omega \circ \rho$,
  i.e., the homotopy pullback of $\mathfrak{Poisson}(X,\omega)$ along $\rho$, is called the
  \emph{Heisenberg $L_\infty$-extension}
  $\mathfrak{Heis}_\rho(\mathfrak{g})$ of $\mathfrak{g}$:
  $$
    \raisebox{20pt}{
    \xymatrix{
	  \mathfrak{Heis}_\rho(\mathfrak{g})
	  \ar[d]
	  \ar[r]
	  &
	  \mathfrak{Poisson}(X,\omega)
	  \ar[d]
	  \ar[r]
	  &
	  0
	  \ar[d]
	  \\
	  \mathfrak{g}
	  \ar[r]^-{\rho}
	  &
	  \mathfrak{X}_{\mathrm{Ham}}(X)
	  \ar[r]^{\iota_{(-)}\omega}
	  &
	  \mathbf{B}\mathbf{H}(X, \flat \mathbf{B}^{n-1}\mathbb{R})
	}
	}
	\,.
  $$
  \label{HeisenbergLieExtension}
\end{definition}
We discuss examples of this below in Section \ref{ExtendedWZW}; see also \cite{nextArticle}.

\begin{remark}
Together with Example \ref{example.poisson},
Theorem \ref{TheLInfinityExtension} identifies
the $L_\infty$ algebra of def. \ref{PoissonBracketLInfinityAlgebra}
as a natural higher analogue of the Poisson bracket Lie bracket of ordinary symplectic geometry.
Other proposals in the literature for what a higher analog of the Poisson bracket Lie algebra should
be as one generalizes from symplectic forms to higher differential forms can be found
in \cite{OtherAttempts,OtherAttempts2}. These other definition are not manifestly equivalent to
definition \ref{PoissonBracketLInfinityAlgebra} and it seems unlikely that they may be equivalent to
it in a more subtle way.
\label{OtherDefinitionsInLiterature}
\end{remark}

\section{Examples}
 \label{Examples}
 We discuss some examples and applications of the theory of higher
 prequantum bundles that we have developed here. In particular, we
 will be dealing with higher prequantum bundles
 $\nabla_{\mathrm{CS}}\colon\mathbf{Fields}\to \mathbf{B}^nU(1)$, or,
 more generally, $\nabla_{\mathrm{CS}}\colon\mathbf{Fields}\to
 \mathbf{B}^n(\mathbb{R}/\!/\Gamma)$ for some subgroup
 $\Gamma\hookrightarrow \mathbb{R}$, which we will look at as higher
 Chern-Simons-type theories
 \cite{FiorenzaSatiSchreiberCup,FiorenzaSatiSchreiberCS}. The datum of
 the Chern-Simons connection $\nabla_{\mathrm{CS}}$ can be seen as a
 ``fully de-transgressed'' Lagrangian, with the classical action of
 the theory recovered by $n$-transgression:
 \[
 \int_{\Sigma_n}[\Sigma_n,\nabla_{\mathrm{CS}}]\colon [\Sigma_n,\mathbf{Fields}]\to U(1),
 \]
of any closed oriented $n$-dimesnional manifold $\Sigma_n$.
%Since in higher prequantum theory
% local Lagrangians are ``fully de-transgressed'' to higher prequantum bundles,
% conversely every example induces its corresponding transgressions.
% In the following we always start with a higher extended Chern-Simons-type theory
% in the sense of \cite{FiorenzaSatiSchreiberCup} and consider then its first transgression.
% As in the discussion in \cite{FiorenzaSatiSchreiberCS} this first transgression
% is the higher prequantum bundle of the topological sector
% of a higher extended
If, instead of transgressing to dimension $n$, one transgresses to dimension 1, and so, in particular, if one transgresses to the stack $[S^1,\mathbf{Fields}]$, then one obtains a
 Wess-Zumino-Witten type theory. In this way our examples appear
 at least in
 pairs as shown in the following table:

 \begin{center}
 \begin{tabular}{|c||c|c|}
   \hline
   & Higher CS-type theory & higher WZW-type theory
   \\
   \hline \hline
   \ref{ExtendedWZW} & 3d $G$-Chern-Simons theory & 2d WZW-model on $G$
   \\
   \hline
   \ref{HigherCSFromLieIntegration} & $\infty$-CS theory from $L_\infty$-integration &
   \\
   \hline
   \ref{DeformationQuantizationInQuantumMechanics} & 2d Poisson Chern-Simons theory & 1d quantum mechanics
   \\
   \hline
   \ref{AnExtended6dSomething} &  7d $\mathrm{String}$-Chern-Simons theory & 6d theory related to M5-brane
   \\
   \hline
 \end{tabular}
 \end{center}

\subsection{Higher prequantum 2d WZW model and the smooth $\mathrm{string}$ 2-group}
 \label{ExtendedWZW}

In the introduction in \ref{TheNeedForHigherPrequantumGeometryFromExtendedLocalGeometricQuantization}
we remarked that an old motivation for what we call higher prequantum geometry here
is the desire to ``de-transgress'' the traditional construction of
positive energy loop group representations
of simply connected compact Lie groups $G$ by, in our terminology, regarding the
canonical $\mathbf{B}U(1)$-2-bundle on $G$ (the ``WZW gerbe'') as a prequantum 2-bundle.
%Here we indicate what the extension
%theorem \ref{TheLongHomotopyFiberSequenceOfTheQuantomorphimsGroup}
%can say about this question.
Here we discuss how prequantum 2-states for the WZW sigma-model provide
at least a partial answer to this question. Then we analyze the quantomorphism
2-group of this model.

\medskip

For $G$ a connected and simply connected compact Lie group such as $G
= \mathrm{Spin}(N)$ for $N \geq 3$ or $G = \mathrm{SU}(N)$ for $N\geq
2$, the first nontrivial cohomology class of the classifying space
$BG$ is in degree 4. More precisely, $H^4(B G, \mathbb{Z}) \simeq
\mathbb{Z}$. For $\mathrm{Spin}(N)$ the generator here is known as the
\emph{first fractional Pontryagin class} $\tfrac{1}{2}p_1$, while for
$\mathrm{SU}(N)$ it is the second Chern class $c_2$. In \cite{FScSt},
a smooth and differential lift of this class was constructed,
namely a diagram of smooth stacks of the form

\begin{tabular}{c|c|c}
  \xymatrix{
    \mathbf{B}\mathrm{Spin}_{\mathrm{conn}}
	\ar[rr]^{\tfrac{1}{2}\widehat{\mathbf{p}}_1}
	\ar[d]^{u_{\mathbf{B}\mathrm{Spin}}}
	&&
	\mathbf{B}^3 U(1)_{\mathrm{conn}}
	\ar[d]^{u_{\mathbf{B}^2 U(1)}}
    \\
    \mathbf{B}\mathrm{Spin}
	\ar[rr]^{\tfrac{1}{2}\mathbf{p}_1}
	\ar[d]^\int
	&&
	\mathbf{B}^3 U(1)
	\ar[d]^\int
    \\
    B \mathrm{Spin}
	\ar[rr]^{\tfrac{1}{2}p_1}
	&&
	K(\mathbb{Z},4)
  }
  & \;\;\; \xymatrix{ \nabla_{\mathrm{CS}} \\ \nabla^0_{\mathrm{CS}} \\  \int \nabla^0_{\mathrm{CS}}} \;\;\;&
  \xymatrix{
    \mathbf{B}\mathrm{SU}_{\mathrm{conn}}
	\ar[rr]^{\widehat{\mathbf{c}}_2}
	\ar[d]^{u_{\mathbf{B}\mathrm{SU}}}
	&&
	\mathbf{B}^3 U(1)_{\mathrm{conn}}
	\ar[d]^{u_{\mathbf{B}^3 U(1)}}
    \\
    \mathbf{B}\mathrm{SU}
	\ar[rr]^{\mathbf{c}_2}
	\ar[d]^\int
	&&
	\mathbf{B}^3 U(1)
	\ar[d]^\int
    \\
    B \mathrm{SU}
	\ar[rr]^{c_2}
	&&
	K(\mathbb{Z},4)
  }\,
\end{tabular}

where  $\int$ denotes the geometric realization map
and $u_{(-)}$ is the ``forgetful the connection map''.
In \cite{FiorenzaSatiSchreiberCup, FiorenzaSatiSchreiberCS} we discussed how
the 3-connection $\nabla_{\mathrm{CS}}$ on the smooth stack of $\mathbf{B}G_{\mathrm{conn}}$ completely encodes the (higher) geometry of 3d $G$-Chern-Simons theory.

In particular $\nabla_{\mathrm{CS}}$ is a prequantization of the pre-3-pectic form (i.e., of the closed differential 4-form) on $\mathbf{B}G_{\mathrm{conn}}$ given by the Killing form $\langle -,-\rangle$ of $\mathfrak{g}$ evaluated on the curvature 2-form $F$ of the $G$-connection:
\[
  \raisebox{20pt}{
  \xymatrix{
    && \mathbf{B}^3 U(1)_{\mathrm{conn}} \ar[d]^{F_{(-)}}
    \\
    \mathbf{B}G_{\mathrm{conn}}
	\ar[rr]_-{\langle F_{(-)}, F_{(-)}\rangle}
	\ar[urr]^{\nabla_{\mathrm{CS}}}
	&&
	\Omega^4_{\mathrm{cl}}
  }
  }
  \,.
\]
This 3-connection on the stack of $G$-principal connections does
not descend to the stack $\mathbf{B}G$ of just $G$-principal bundles;
it does however descend \cite{Waldorf08} as a ``3-connection without
degree-3 and without degree-2 connection forms'' (as in Section \ref{HigherConnections}):
\[
  \raisebox{20pt}{
  \xymatrix{
    && \mathbf{B}^2(\mathbf{B} U(1)_{\mathrm{conn}})
	\ar[d]^{\mathbf{B}^2 F_{(-)}}
    \\
    \mathbf{B}G
	\ar[rr]^{\mathbf{}}
	\ar[urr]^{\nabla^2_{\mathrm{CS}}}
	&&
	\mathbf{B}^2\Omega^2_{\mathrm{cl}}
	}
	}
	\,.
	\label{nabla1CS}
\]
{Namely, as it is immediate to check from the the explicit formulas for $\nabla_{\mathrm{CS}}$ given in \cite{FScSt}, the morphism of stacks $\nabla_{\mathrm{CS}}\colon  \mathbf{B}G_{\mathrm{conn}}\to  \mathbf{B}^3 U(1)_{\mathrm{conn}}$ maps the translation action of $\Omega^1(-;\mathfrak{g})$ in  $\mathbf{B}G_{\mathrm{conn}}$ to the translation action of $(\Omega^2\xrightarrow{d}\Omega^3)$ in $ \mathbf{B}^3 U(1)_{\mathrm{conn}}$. Therefore we have an induced morphism between the quotient stacks
\[
\nabla^2_{\mathrm{CS}}\colon  \mathbf{B}G\cong \mathbf{B}G_{\mathrm{conn}}/\!/\Omega^1(-;\mathfrak{g})\xrightarrow{\nabla_{\mathrm{CS}}} \mathbf{B}^3 U(1)_{\mathrm{conn}}/\!/(\Omega^2\xrightarrow{d}\Omega^3)\cong \mathbf{B}^2(\mathbf{B} U(1)_{\mathrm{conn}}).
\]
}
In physics' parlance, the stack $\mathbf{B}G_{\mathrm{conn}}$ is the stack of Chern-Simons fields, while the underlying stack $\mathbf{B}G$ is the stack of the ``instanton sectors'' of fields. Therefore we see that over the stack of Chern-Simons fields
we canonically have the quantomorphism 3-groupoid $\mathrm{At}(\nabla_{\mathrm{CS}})$, while over the
stack of instanton sectors of the fields we have  the
Courant 3-groupoid $\mathrm{At}(\nabla^2_{\mathrm{CS}})$.
This kind of phenomenon we re-encounter below in Section \ref{DeformationQuantizationInQuantumMechanics}.

The transgression of $\nabla_{\mathrm{CS}}$ to loop space
is found to be the ``WZW gerbe'': the canonical $U(1)$-2-bundle with connection on  the Lie group $G$ itself, see \cite{CJMSW,FiorenzaSatiSchreiberCup, FiorenzaSatiSchreiberCS}. One obtains the WZW 2-connection $\nabla_{\mathrm{WZW}}$ on $G$
as the fiber integration of $\nabla_{\mathrm{CS}}$ restricted to $G$ (by identifying realising the elements of the simply connected Lie group $G$ as holonomies of
constant $\mathfrak{g}$-connections on $S^1$):
\[
  \nabla_{\mathrm{WZW}}
  :
  \xymatrix{
    G
	\ar[r]
	&
	[S^1, \mathbf{B}G_{\mathrm{conn}}]
	\ar[rr]^-{[S^1, \nabla_{\mathrm{CS}}]}
	&&
	[S^1, \mathbf{B}^3 U(1)_{\mathrm{conn}}]
	\ar[rr]^-{\exp(2 \pi i \int_{S^1}(-))}
	&&
	\mathbf{B}^2 U(1)_{\mathrm{conn}}
  }
\]
and this refines the looping of $\nabla_{\mathrm{CS}}^2$:
\[
%  \nabla_{\mathrm{WZW}}
%  :
  \xymatrix{
    G
	\simeq
	\mathbf{\Omega} \mathbf{B}G
	\ar[rr]^-{\mathbf{\Omega} \nabla_{\mathrm{CS}}^2}
	&&
	\mathbf{\Omega} \mathbf{B}^2(\mathbf{B} U(1)_{\mathrm{conn}})
	\simeq
	\mathbf{B}(\mathbf{B} U(1)_{\mathrm{conn}})  }
  \,.
\]
%The second description is apparently more immediate than the first one, but actually what one is doing is hiding fiber integration and constant connections on $S^1$ in the proof that $\nabla_{\mathrm{CS}}$ descends to $\nabla_{\mathrm{CS}}^2$.
This $\nabla_{\mathrm{WZW}}$ is the background gauge field of the 2d Wess-Zumino-Witten
sigma-model. In string theory, this is the Kalb-Ramon B-field under which the string propagating on $G$ is charged. { The refinement of $\nabla_{\mathrm{WZW}}$ to $\mathbf{\Omega} \nabla_{\mathrm{CS}}^2$ can be seen as a rephrasing into the language of smooth stacks of Konrad Waldorf's result on the existence of a canonical \emph{multiplicative} bundle gerbe with connection on the compact simply connected Lie group $G$, refining the WZW gerbe \cite{Waldorf08,Waldorf12}. }

In the language of geometric prequantization, we can regard  $\nabla_{\mathrm{WZW}}$ as the prequantization
of the canonical 2-plectic form on $G$ given by the canonical closed 3-form corresponding to the 3-cocycle $\langle -,[-,-]\rangle$ on $\mathfrak{g}$:
$$
  \raisebox{20pt}{
  \xymatrix{
    && \mathbf{B}^2 U(1)_{\mathrm{conn}}
	\ar[d]^{F_{(-)}}
    \\
    G
	\ar[urr]^{\nabla_{\mathrm{WZW}}}
	\ar[rr]_-{\omega_{\langle -,[-,-]\rangle}}
	&&
	\Omega^3_{\mathrm{cl}}
  }
  }
  \,.
$$

By Example \ref{TwistedBundlesAre2Sections},  there is an associated 2-bundle for $\nabla_{\mathrm{WZW}}$ whose sections
are  twisted unitary bundles with connection. This tells us that twisted differential K-theory cocycles on $G$ are naturally interpreted as prequantum 2-states
associated with $\nabla_{\mathrm{WZW}}$; these are the \emph{Chan-Paton gauge fields} of the Physics literature.
More explicitly,
with the notation of Corollary \ref{cor.sections} and Example \ref{TwistedBundlesAre2Sections}, a prequantum 2-state $\Psi$
of the WZW model supported over a D-brane submanifold $Q \hookrightarrow G$
is a map $\Psi : \nabla_{\mathrm{WZW}}|_Q \to \mathbf{dd}_{\mathrm{conn}}$
in the slice over $\mathbf{B}^2 U(1)_{\mathrm{conn}}$. That is, it is a homotopy commutative diagram as in the upper part of the diagram
\[
  \raisebox{20pt}{
  \xymatrix{
  &
	&&
	\coprod_N (\mathbf{B}U(N)/\!/\mathbf{B}U(1))_{\mathrm{conn}}
	\ar[d]_{\mathbf{dd}_{\mathrm{conn}}}
	\\
Q\ar[d]\ar@{^{(}->}[r] \ar@/^4pc/[rrru]^{\Psi}&	G
	\ar@/^1pc/[rr]^-{\nabla_{\mathrm{WZW}}}
	\ar[r]
	\ar[d]
	&
	[S^1, \mathbf{B}G_{\mathrm{conn}}]
	\ar[r]_-{\exp(\int_{S^1}[S^1, \nabla])}
	\ar[d]^{\mathrm{conc.}}
	&
	\mathbf{B}^2 U(1)_{\mathrm{conn}}
	\\
{*}\ar[r]&	G/\!/_{\!\mathrm{ad}}G
	\ar[r]^-\simeq
	&
	G \mathbf{Conn}(S^1)
  }
  }
  \,.
\]
In the lower part of the above diagram, on the left we are displaying the case of a ``symmetric D-brane'' for the WZW model: the homotopy fiber over a point of the projection map $G\to G/\!/_{\mathrm{ad}}G$, i.e., in more classical terms, the inclusion of a conjugacy class in $G$. On the right lower part of the diagram we have the identification of the quotient stack
$G\to G/\!/_{\mathrm{ad}}G$ with the stack of $G$-connections on $S^1$ and the way this fits the fiber integration description of the WZW connection.
% the map to the differential concretification
%of the transgressed moduli stack of fields,
%according to example \ref{TheModuliOfCircleConnections}.
%As indicated, this exhibits $G$ as fibered over its homotopy quotient
%by its adjoint action. The D-brane inclusion $Q \to G$ in the diagram is the homotopy fiber
%over a full point of $G/\!/_{\mathrm{ad}}G$ precisely if it is a conjugacy class
%of $G$, hence a ``symmetric D-brane'' for the WZW model.
In summary this means that this single diagram exhibiting WZW prequantum-2-states
as slice maps encodes all the WZW D-brane data as discussed in the literature
\cite{alekseev-schomerus,Gawedzki05,GaRe}.

\begin{remark} In \cite{Meinrenken} it is shown that the
ring of positive energy representations of the loop group of
$G$ is generated by the push-forward in K-theory of the twisted bundles $\Psi$ over
conjugacy classes of $G$, while in \cite{FiorenzaSatiSchreiberCS} we showed that the transgression of the
prequantum 2-states $\Psi$ to prequantum 1-states over the loop group $L G$
naturally encodes the anomaly cancellation of the open bosonic string
in the presence of D-branes (the Kapustin-part of the Freed-Witten-Kapustin quantum anomaly cancellation).
 Taken together this provides at least some aspects of
an answer to the question
in \ref{TheNeedForHigherPrequantumGeometryFromExtendedLocalGeometricQuantization},
concerning a higher stacky refinement of the geometric construction of
loop group representation theory.
\end{remark}

\medskip
We can now look at the quantomorphism 2-group
of $\nabla_{\mathrm{WZW}}$, or, %, def. \ref{HigherQuantomorphismGroupInIntroduction}, on these 2-states, hence,
in the language of twisted
K-theory,
%cohomology,
to the 2-group of twist automorphism. By the explicit description of local data for twisted unitary bundles with connections, one immediately sees that the infinitesimal action of quantomorphisms of WZW on the pre quantum 2-states are precisely the D-brane gauge transformations
which are familiar from the string theory literature. Namely, the infinitesimal action is locally given by a real valued 1-form $\lambda$ acting on the local connection 1-form $A$ on the
twisted bundle and on the  local connection 2-form on the WZW prequantum bundle as
$$
  A \mapsto A + \lambda
  \,,
  \;\;\;\;
  B \mapsto B + d \lambda
  \,,
$$
where in the first equation $\lambda$ is seen as a $\mathfrak{u}_N$-valued 1-form via the diagonal embedding of $U(1)$ into $U(N)$, inducing
\[
\Omega^1(-;\mathbb{R})\cong \Omega^1(-;\mathfrak{u}_1)\hookrightarrow \Omega^1(-;\mathfrak{u}_N).
\]
Next, notice that  since the 2-plectic form $\omega_{\langle -,[-,-]\rangle} \in \Omega^3_{\mathrm{cl}}(G)$
is a left invariant form (by definition),
the left translation action of $G$ on itself is Hamiltonian. Indeed, the infinitesimal action is given by left invariant vector fields and $\iota_v\omega_{\langle -,[-,-]\rangle}$ is left-invariant, and therefore closed, for any left-invariant vector field $X$. Since $H^2_{\mathrm{dR}}(G)=0$, there exists a 1-form $H$ such that  $\iota_v\omega_{\langle -,[-,-]\rangle}+H=0$, and so $v$ is Hamiltonian.
By the general theory of Section \ref{QuantomorphismAndHeisenbergGroup} we have therefore the corresponding Heisenberg 2-group $\mathbf{Heis}(G,\nabla_{\mathrm{WZW}})$
``inside'' the quantomorphism 2-group.
By Theorem \ref{TheLongHomotopyFiberSequenceOfTheQuantomorphimsGroup}
this is a 2-group extension of $G$ of the form
$$
  \xymatrix{
    \mathbf{Flat}\text{-}U(1)\text{-}\mathbf{Conn}(G)
	\ar[r]
	&
	\mathbf{Heis}(\nabla_{\mathrm{WZW}})
	\ar[r]
	&
	G	
  }
  \,.
$$
Since $G$ is connected and simply connected,
 by Proposition \ref{FlatConnectionsOnSimplyConnectedManifolds} there is an equivalence
of smooth 2-groups
$
  \mathbf{Flat}\text{-}U(1)\text{-}\mathbf{Conn}(G) \simeq \mathbf{B}U(1)
$
and so the WZW Heisenberg 2-group is in fact a smooth $\mathbf{B}U(1)$-2-group extension
$$
  \xymatrix{
    \mathbf{B}U(1)
	\ar[r]
	&
	\mathbf{Heis}(\nabla_{\mathrm{WZW}})
	\ar[r]
	&
	G\,.	
  }
$$
No surprise, this is the 2-group extension of $G$ classified by the cocycle $\nabla^0_{\mathrm{CS}} \colon\mathbf{B}G \to \mathbf{B}^3 U(1)$. Moreover, since
$G$ is compact, connected and simply connected, we have
$\pi_0 \mathbf{H}(\mathbf{B}G, \mathbf{B}^3 U(1)) \simeq H^4(B G, \mathbb{Z})
\simeq \mathbb{Z}$. This integer is the \emph{level} of the cocycle $\nabla^0_{\mathrm{CS}}$. In particular, the cocycle corresponding
to the generator $1$ of $\mathbb{Z}$ is $\tfrac{1}{2}\mathbf{p}_1$ for $G = \mathrm{Spin}$
and $\mathbf{c}_2$ for $G = SU$. Since the 2-group extension of $G$ corresponding to the generator of $H^4(B G, \mathbb{Z})$ is the
String 2-group of $G$ %extension of example \ref{LieGroupAsInfinityGroup}
we see that
$$
\mathbf{Heis}(\nabla_{\mathrm{WZW}})\cong \mathrm{String}_G.
$$
%
%$$
%  \xymatrix{
%    \mathbf{B}U(1)
%	\ar[r]
%	&
%	\mathrm{String}_G
%	\ar[r]
%	&
%	G	
%  }
%  \,.
%$$
At the infinitesimal level, this amounts to saying that the Heisenberg Lie 2-algebra extension of $\mathfrak{g}$ is a model for the \emph{string Lie 2-algebra} extension $\mathfrak{string}_{\mathfrak{g}}$. This was originally observed in \cite{RogersString}; a re-derivation
in the context of Section \ref{TheHigherPoissonBracketExtension} is in \cite{nextArticle}.
%
%Accordingly, under Lie differentiation, one finds
%(this was originally observed in \cite{RogersString}, a re-derivation
%in the context of def. \ref{HeisenbergLieExtension}
%is in \cite{nextArticle}) that the
%Heisenberg Lie 2-algebra extension
%of theorem \ref{TheLInfinityExtension}
%combined with def. \ref{HeisenbergLieExtension} is the
%\emph{string Lie 2-algebra} extension (see example \ref{LieGroupAsInfinityGroup})
%$$
%  \xymatrix@R=5pt{
%    \mathbf{B}\mathbb{R}
%	\ar[r]
%	&
%	\mathfrak{Heis}_{\langle -,[-,-]\rangle}(\mathfrak{g})
%	\ar[r]
%	\ar@{}[d]|\simeq
%	&
%	\mathfrak{g}
%	\\
%	& \mathfrak{string}_{\mathfrak{g}}
%  }
%  \,.
%$$

\subsection{Higher %prequantum $n$d
Chern-Simons-type theories and $L_\infty$-algebra cohomology}
 \label{HigherCSFromLieIntegration}

 The construction of the higher prequantum bundle $\nabla_{\mathrm{CS}}$
 for Chern-Simons field theory in
 \ref{ExtendedWZW} above follows a general procedure, which might be called \emph{differential Lie integration of $L_\infty$-cocycles},
 that produces
 a whole class of examples
 \cite{FScSt, frs}, see \cite{SchreiberMPI} for survey.

 Namely, just as ordinary $G$-Chern-Simons theory for a simply
 connected simple Lie group $G$ is encoded by the Killing form of
 $\mathfrak{g}$, one can canonically associate a Chern-Simons-type
 field theory to an invariant polynomial on an
 $L_\infty$-algebra.\footnote{Not all Chern-Simons-type field theories
   are of this kind. An example of another class of Chern-Simons type
   theories is given, for instance, by the the cup product higher
   $U(1)$-Chern-Simons theories considered in
   \cite{FiorenzaSatiSchreiberCup}.}
% Since also the following two examples in  \ref{DeformationQuantizationInQuantumMechanics}
% and \ref{AnExtended6dSomething} are naturally obtained this way, we here briefly
% recall this construction, due to \cite{FScSt}, with an eye towards its interpretation in higher prequantum
% geometry.
Let us briefly recall how this construction works. Two specific examples of interest for Physics will be detailed in Sections  \ref{DeformationQuantizationInQuantumMechanics}
and \ref{AnExtended6dSomething}.
 \medskip

 To begin with, given a finite-dimensional $L_\infty$-algebra
% \ccomment{finite-dim assumption added}
 $\mathfrak{g}$, there is a natural notion  of sheaves of (flat) $\mathfrak{g}$-valued smooth differential forms
 $$
   \Omega^\bullet_{\mathrm{flat}}(-,\mathfrak{g}) \hookrightarrow \Omega^\bullet(-,\mathfrak{g})
  % \in \mathrm{Sh}(\mathrm{SmthMfd})
   \,,
 $$
 and this is functorial in $\mathfrak{g}$ (with respect to $L_\infty$-morphisms).
 %(for the correct (``weak'') homomorphisms of $L_\infty$-algebras).
 Therefore there is a functor,
 denoted $\exp(-)$ in \cite{FScSt}, which assigns to an $L_\infty$-algebra
 $\mathfrak{g}$ the presheaf of Kan complexes given by
 \[
 \exp(\mathfrak{g})(U)\colon [k]\mapsto \Omega^\bullet_{\mathrm{flat}}(U\times\Delta^k,\mathfrak{g})_{\mathrm{vert}}
 \]
 for any Cartesian space $U$. Here the subscript ``vert'' means that
 one is considering only differential forms which are vertical with
 respect to the projection $U\times \Delta^k\to U$. As remarked in
 Section \ref{ModuliOfConnections}, this corresponds to considering
 smoothly $U$-parameterized collections of flat $\mathfrak{g}$-valued
 differential forms on the $k$-simplex $\Delta^k$.\footnote{As a
   technical point, in order to ensure that the differential forms in
   $ \exp(\mathfrak{g})_k$ glue smoothly when two $k$-simplices join
   along a $(k-1)$-simplex, one requires these differential forms to
   be sufficiently well behaved towards the boundary of the simplex
   (one says the differential forms have ``sitting instants'').} By
 sheafifying, one gets the Lie integration functor
 \[
 \exp\colon L_\infty\text{-algebras} \to \mathbf{H}\,.
 \]
% Under the presentation
% $L_{\mathrm{lhe}}[\mathrm{SmoothMfd}^{\mathrm{op}}] \simeq \mathrm{Smooth}\infty\mathrm{Grpd}$
% of the $\infty$-topos of smooth $\infty$-groupoids in example \ref{ToposOfSmoothInfinityGroupoids}
% this yields a Lie integration construction from $L_\infty$-algebras to smooth
% $\infty$-groupoids.
 \begin{remark}The functor $\exp$ is the fairly immediate stacky and smooth refinement of a standard
 construction in rational homotopy theory and deformation theory, see the
 references in \cite{FScSt} for a list of predecessors of this construction.
\end{remark}

 In analogy to ordinary Lie integration,
 one finds that $\exp(\mathfrak{g})$ is the ``geometrically $\infty$-connected'' Lie integration
 of $\mathfrak{g}$: its geometric realization $\int \exp(\mathfrak{g})$ is always contractible.
 %,
 %example \ref{SmoothGroupoidsAreCohesive},  of
 %$\exp(\mathfrak{g}) \in L_{\mathrm{lhe}}[\mathrm{SmoothMfd}^{\mathrm{op}}, \mathrm{KanCplx}]
 %\simeq \mathrm{Smooth}\infty\mathrm{Grpd}$
%
\begin{example}
The chain complex $\mathbb{R}[n-1]$ consisting of $\mathbb{R}$ concentrated in degree $n-1$ (we are using homological degree conventions) can naturally be seen as an abelian $L_\infty$-algebra. One finds
 $$
   \exp(\mathbb{R}[n-1]) \simeq \mathbf{B}^n \mathbb{R}
 $$
 and so
\[
\int\exp(\mathbb{R}[n-1])  \simeq \int \mathbf{B}^n \mathbb{R}\simeq B^n \mathbb{R} \simeq {*},
\]
since the abelian Lie group $\mathbb{R}$ is contractible.
\end{example}
Geometrically non-$\infty$-connected Lie integrations of
$\mathfrak{g}$ arise notably as truncations $\exp(\mathfrak{g})$. In
particular, the $n$-truncation $\exp(\mathfrak{g})_{\leq n}$ will be
the stack of $n$-groupoids integrating $\mathfrak{g}$. A remarkable
example is the following.
 \begin{example}
  If $\mathfrak{g}$ is an ordinary Lie algebra, then
 there is a natural equivalence
 \[
 \exp(\mathfrak{g})_{\leq 1} \simeq \mathbf{B}G,
 \]
 where $G$ is the simply connected Lie group $G$ integrating $\mathfrak{g}$.
\end{example}
%
% $\infty$-stack $\exp(\mathfrak{g}_1)$ to a stack of 1-groupoids reproduces (the internal delooping of)
% the simply connected Lie group $G$ corresponding to $\mathfrak{g}$ by ordinary Lie theory:
% $$
%   \tau_1 \exp(\mathfrak{g}_1) \simeq \mathbf{B}G
%   \;\;\;
%   \in \mathrm{Smooth}\infty\mathrm{Grpd}
%   \,.
% $$
\begin{example}
 Similarly for $\mathfrak{string}_{\mathfrak{g}}$ the string Lie 2-algebra of the compact simple group $G$,
 %example \ref{LieGroupAsInfinityGroup},
 the 2-truncation of $\exp(\mathfrak{string}_{\mathfrak{g}})$ to a stack of 2-groupoids
 reproduces the moduli stack of $\mathrm{String}_G$-principal 2-bundles:
 $$
   \exp(\mathfrak{string}_{\mathfrak{g}})_{\leq 2} \simeq \mathbf{B}\mathrm{String}_G.
 $$
 \end{example}

 Now the simple observation leading to Lie integration of
 $L_\infty$-cocycles
 is that a degree-$n$ $L_\infty$-cocycle $\mu$
 on an $L_\infty$-algebra $\mathfrak{g}$ is nothing but a morphism of $L_\infty$-algebras
 $$
   \mu\colon \mathfrak{g} \to \mathbb{R}[n-1]
   \,.
 $$
Since $\exp(-)$ is a functor, this immediately integrates
 to a morphism
 $$
   \exp(\mu)\colon \exp(\mathfrak{g}) \to \mathbf{B}^n \mathbb{R}.
 $$
% in $\mathrm{Smooth}\infty\mathrm{Grpd}$, hence by remark \ref{CohomologyInTopos}
% to a universal cocycle on the smooth moduli $\infty$-stack $\exp(\mathfrak{g})$.
 The morphism $\exp(\mu)$ does not descend to a morphism from the $k$-truncations $\exp(\mathfrak{g})_{\leq k}$ to $\mathbf{B}^n \mathbb{R}$ in general. However, it does descend to a morphism
  $$
   \exp(\mu)\colon  \exp(\mathfrak{g})_{\leq n-1} \to \mathbf{B}^n (\mathbb{R}/\!/\Gamma)
   \,,	
 $$
 where $\Gamma \hookrightarrow \mathbb{R}$ is the subgroup of the periods of the cocycle
 $\mu$. When $\Gamma$ is a discrete subgroup of $\mathbb{R}$ (the periods lattice of $\mu$), then up to rescaling we can identify $\Gamma$ with $\mathbb{Z}$ and so $ \mathbf{B}^n (\mathbb{R}/\!/\Gamma)$ with $ \mathbf{B}^n (\mathbb{R}/\!/\mathbb{Z})\cong \mathbf{B}^nU(1)$. This way we realize $ \exp(\mu)$ as a principal $\mathbf{B}^{n-1}U(1)$-bundle over $\exp(\mathfrak{g})_{\leq n-1}$.

\begin{example}
 For
 $$
   \langle -,[-,-]\rangle\colon \mathfrak{g} \to  \mathbb{R}[2]
 $$
 the canonical 3-cocycle on a semisimple Lie algebra $\mathfrak{g}$, the periods subgroup is isomorphic to
 $\pi_3(G) \simeq \mathbb{Z}$, where $G$ is the compact and simply connected Lie group $G$ integrating $\mathfrak{g}$. Hence,
 the Lie integration of the 3-cocycle yields a map of smooth stacks
 of the form
 $$
   \exp(\langle -,[-,-]\rangle)
     \colon
   \exp(\mathfrak{g})_{\leq 2}
	 \to
	 \mathbf{B}^3 U(1).
   \,.
 $$
 Since $G$ is 2-connected, one has $\exp(\mathfrak{g})_{\leq 2}\cong \exp(\mathfrak{g})_{\leq 1}\cong \mathbf{B}G$, and so $ \exp(\langle -,[-,-]\rangle)$ is equivalently a morphism
$$
   \mathbf{c}
     \colon
   \mathbf{B}G
	 \to
	 \mathbf{B}^3 U(1)
   \,.
 $$
The morphism $\mathbf{c}$ is a refinement of the characteristic class $c$ generating $H^4(BG,\mathbb{Z})$ to a morphism of smooth stacks, a fact we used above in \ref{ExtendedWZW}. For instance for $\mathfrak{g} = \mathfrak{so}$ the Lie algebra of the Spin group,
 the Lie integration of its canonical Lie 3-cocycle produces a smooth refinement $\tfrac{1}{2}\mathbf{p}_1$
 of the first fractional Pontryagin class.
 \end{example}

It is shown in \cite{FScSt} that the $\exp(-)$-construction can be naturally refined so to include connections into the picture. This is achieved by suitably translating Ehresmann conditions in the language of simplicial presheaves. The result is summarised by the following diagram
 $$
   \raisebox{20pt}{
   \xymatrix{
      && \mathbf{B}^n (\mathbb{R}/\!/\Gamma)_{\mathrm{conn}}
	   \ar[d]^{F_{(-)}}
      \\
      \exp(\mathfrak{g})_{\mathrm{conn},\leq n-1}
	  \ar[rr]^-{\langle F_{(-)}, \cdots, F_{(-)}\rangle}
	  \ar[urr]^-{\exp(\mu)_{\mathrm{conn}}\quad}
	  &&
	  \Omega^{n+1}_{\mathrm{cl}}
   }
   }
   \,,
 $$
where $\langle -, \cdots, -\rangle$ is an invariant polynomial on the $L_\infty$-algebra $\mathfrak{g}$ in transgression with the cocycle $\mu$ by a Chern-Simons element $\mathrm{CS}_\mu$. Notice how the above diagram gives us a canonical $n$-plectic form on the smooth stack $\exp(\mathfrak{g})_{\mathrm{conn},\leq n-1}$ together with a prequantum $n$-bundle prequantizing it. Moreover, transgression to
 codimension 0
 $$
   \exp\left( \int_{\Sigma_n} [\Sigma_n, \exp(\mu)_{\mathrm{conn}}] \right)
   :
   \xymatrix{
     [\Sigma_n, \exp(\mathfrak{g})_{\mathrm{conn},\leq n-1}]
	 \ar[r]
	 &
	 \mathbb{R}/\!/\Gamma
   }
 $$
 is an action functional on the stack of $\mathfrak{g}$-gauge fields configurations on
 a given closed oriented manifold $\Sigma_n$, which is locally given by the
 integral of the Chern-Simons $(n-1)$-form $\mathrm{CS}_\mu(A)$ on the local connection form $A$
 %(with respect to the corresponding $L_\infty$-invariant polynomial)
(and globally given by a gauge consistent globalization of such integrals). This justifies the name of \emph{higher Chern-Simons theories} given to the field theories obtained by Lie integration of $L_\infty$-algebra cocycles.
\begin{example}
For a compact simple and simply connected Lie group $G$ the $\exp(-)_{\mathrm{conn}}$ applied to the canonical 3-cocycle on $\mathfrak{g}$ produces the prequantization
$$
   \raisebox{20pt}{
   \xymatrix{
      && \mathbf{B}^3 U(1)_{\mathrm{conn}}
	   \ar[d]^{F_{(-)}}
      \\
      \mathbf{B}G_{\mathrm{conn}}
	  \ar[rr]^-{\langle F_{(-)}, F_{(-)}\rangle}
	  \ar[urr]^-{\nabla_{\mathrm{CS}}}
	  &&
	  \Omega^{4}_{\mathrm{cl}}
   }
   }
   \,
 $$
 which we discussed in Section \ref{ExtendedWZW} in the context of classical Chern-Simons theory. The transgressed action
 $$
   \exp\left( \int_{\Sigma_3} [\Sigma_3, \nabla_{\mathrm{CS}}] \right)
   :
   \xymatrix{
     [\Sigma_n, \mathbf{B} G_{\mathrm{conn}}]
	 \ar[r]
	 &
	 U(1)
   }
 $$
 is the classical Chern-Simons action for $G$-connections on closed oriented 3-manifolds.
\end{example}
%
% For example the differential refinement of
% the prequantum 3-bundle of 3d $G$-Chern-Simons theory
% $\tfrac{1}{2}\mathbf{p}_1 \simeq \tau_3 \exp(\langle -,[-,-]\rangle)$
% obtained this way is the universal Chern-Simons 3-connection
% $$
%   \exp(\langle -,[-,-]\rangle_{\mathfrak{so}})_{\mathrm{conn}}
%   \simeq
%   \tfrac{1}{2}\hat{\mathbf{p}}_1
%   :
%   \mathbf{B}\mathrm{Spin}_{\mathrm{conn}}
%   \to
%   \mathbf{B}^3 U(1)_{\mathrm{conn}}
% $$
% whose transgression to codimension 0 is the standard Chern-Simons action
% functional, as discussed above in \ref{ExtendedWZW}.
 \begin{example}
 By the above example, the Lie integration of the canonical 3-cocycle on $\mathfrak{so}$ produces a differential refinement
 $\tfrac{1}{2}\hat{\mathbf{p}}_1$ of the first fractional Pontryagin class, which further refines $\tfrac{1}{2}\mathbf{p}_1$ to morphisms of stacks of principal connections.
 Analgously, the next cocycle on $\mathfrak{so}$, the
 canonical 7-cocycle, can be regarded as a cocycle on $\mathfrak{string}$, and its differential Lie integration
 yields a prequantum 7-bundle on the stack of $\mathrm{String}$-principal
 2-connections
 $$
      \tfrac{1}{6}\hat{\mathbf{p}}_2
   \;:\;
   \mathbf{B}\mathrm{String}_{\mathrm{conn}}
   \to
   \mathbf{B}^7 U(1)_{\mathrm{conn}}
   \,,
 $$
 which refines the second fractional Pontryagin class.
 This defines a 7-dimensional nonabelian Chern-Simons theory, which
 we come to below in Section \ref{AnExtended6dSomething}.
\end{example}

\begin{remark}
 All of this discussion generalizes verbatim from $L_\infty$-algebras
 to finite--rank \emph{$L_\infty$-algebroids}.  %\ccomment{finite-rank assumption added}
Hence, as was observed in\cite{frs}, all the perturbative
 field theories known as \emph{AKSZ sigma-models} have a Lie integration
 to what here we call higher prequantum bundles for higher Chern-Simons type field theories:
 these are precisely the cases where the cocycle $\mu$ transgresses to a non-degenerate
 binary invariant polynomial $\langle-,- \rangle$ on the $L_\infty$-algebroid.
 At the level of globally defined differential forms this had been established in \cite{kotov-strobl}, of which \cite{frs} can be seen as a stackification.
 In the next section \ref{DeformationQuantizationInQuantumMechanics},
 we consider one low-dimensional example in this family and observe that
 its higher geometric prequantum and quantum theory has
 secretly been studied in some detail already, but in 1-geometric disguise.
\end{remark}

\begin{remark}\label{eq-of-motion}
 For $n$-dimensional Chern-Simons action functionals $\exp(\mu)_{\mathrm{conn}}$
 as above, one finds that their variation under $A\mapsto A+\delta A$
%
% al differential at a field configuration
% $A$ given by globally defined differential form data
is proportional to
 $$
%   \delta \exp\left( \int_{\Sigma_{n-1}} [\Sigma_n, \exp(\mu)_{\mathrm{conn}}] \right)
%   \;\propto\;
   \int_{\Sigma_n}\langle F_A \wedge \dots F_A \wedge \delta A \rangle
   \,.
 $$
 Therefore the Euler-Lagrange equations of motion of the corresponding $n$-dimensional
 Chern-Simons theory are
 $$
   \langle F_A \wedge \cdots F_A , -\rangle = 0
   \,.
 $$
 Notice that in general $F_A$ is an inhomogenous differential form, so that
this equation in general consists of several independent components.
In particular, if the invariant polynomial is binary, i.e., of the form
$\langle -,-\rangle$, and non-degenerate\footnote{This is precisely the case
in which the general $\infty$-Chern-Simons theory reproduces the AKSZ $\sigma$-models.},
then the above equations of motion reduce to
$$
  F_A = 0
$$
and hence assert that the critical/on-shell field configurations are precisely
those $L_\infty$-algbroid valued connections which are flat.
\end{remark}

\begin{remark}\label{phase-space}
The equations of motion $F_A=0$ derived in Remark \ref{eq-of-motion} offer an alternative interpretation of the stack $\exp(\mathfrak{g})_{\leq n}$.
Namely, since the equation of motions $F_A=0$ are first order
differential equations, the critical fields configurations on a
cylinder $\Delta^{n-1} \times [-T,T]$ (for small values of $T$) bijectively correspond to their initial data, i.e., to flat connections on $\Delta^{n-1}$. But these are precisely the $(n-1)$-cells of $\exp(\mathfrak{g})_{\leq n}$. Moreover, the $n$-cells in $\exp(\mathfrak{g})_{\leq n}$ implement the gauge transformations between these initial value data. Therefore, $\exp(\mathfrak{g})_{\leq n}$ can be naturally interpreted as the
the %higher/extended
\emph{reduced phase space} of the Chern-Simons/AKSZ $\sigma$-model in codimension 1, with the space of $(n-1)$-cells corresponding to the \emph{covariant phase space} for
``open $n$-branes'' in the model.
%
%by definition, the $(n-1)$-cells of $\exp(\mathfrak{g})_{\leq n}$ are
%flat $\mathfrak{g}$-valued connections on the $(n-1)$-disks and its $n$-cells
%implement gauge equivalences between such data.
%
%a flat $\mathfrak{g}$-valued connections on the $(n-1)$-disk can be seen as
%
%In this case the higher moduli stack $\tau_n \exp(\mathfrak{g})$, which in general
%is the moduli stack of instanton/charge-sectors underlying the
%topologically nontrivial $\mathfrak{g}$-connections, acquires also a different
%interpretation. By the above discussion, its $(n-1)$-cells are equivalently
%flat $\mathfrak{g}$-valued connections on the $(n-1)$-disks and its $n$-cells
%implement gauge equivalences between such data. But since the equations of motion
%$F_A = 0$ are first order differential equations, flat connections on
%$D^{n-1}$ bijectively correspond to critical field configuration on the cylinder
%$D^{n-1} \times [-T,T]$. Therefore the collection of $(n-1)$-cells of
%$\tau_n \exp(\mathfrak{g})$ is the higher/extended \emph{covariant phase space} for
%``open genus-0 $(n-1)$-branes'' in the model. Moreover, the $n$-cells between
%these $(n-1)$-cells
%implement the gauge transformations on such initial value data and hence
%$\tau_n \exp(\mathfrak{g})$ is, in codimension 1,
%the higher/extended \emph{reduced phase space}
%of the model in codimension 1.
For $n = 2$ this perspective was amplified in \cite{CattaneoFelderSymplectic}.
We turn to this special case below in Section \ref{DeformationQuantizationInQuantumMechanics}.
\end{remark}

\begin{example}
Consider the classical 3d Chern-Simons theory associated with a compact simple and simply connected Lie group $G$. Then $\exp(\mathfrak{g})_{\leq 1}\cong \mathbf{B}G$, so the
reduced phase space in codimension 2 is the stack of principal $G$-bundles and the phase space of open Chern-Simons 1-branes stretching between two $G$-bundles is equivalent to $\mathbf{\Omega}\mathbf{B}G\cong G$. The on-shell prequantum 2-bundle in codimension 2 for Dirichlet boundary conditions for the open Chern-Simons 1-brane is then seen to be precisely the WZW gerbe with connection on $G$.
%
%
%
%As an example, from this perspective the construction of the WZW-gerbe
%by looping as discussed above in \ref{HigherCSFromLieIntegration}
%is equivalently the construction of the on-shell prequantum 2-bundle in
%codimension 2 for ``Dirichlet boundary conditions'' for the open Chern-Simons
%membrane. Namely $\mathbf{B}G$ is now the extended reduced phase space,
%and so the extended phase space of membranes stretching between the unique point is
%the homotopy fiber product of the two point inclusions
%$\xymatrix{Q_0 \ar[r] & \mathbf{B}G \ar@{<-}[r] & Q_1}$, with $Q_0, Q_1 = \ast$,
%hence is
%$\Omega \mathbf{B}G \simeq G$. Since the on-shell prequantum 2-bundle
%$\nabla_{\mathrm{CS}}^1$ trivializes over these inclusions, as exhibited by
%diagrams
%$$
%  \raisebox{20pt}{
%  \xymatrix{
%    Q_i
%	\ar[r]
%	\ar[d]
%	&
%	\ast
%	\ar[d]
%	\\
%	\mathbf{B}G \ar[r]^-{\nabla_{\mathrm{CS}^2}} & \mathbf{B}^3 U(1)_{\mathrm{conn}^2}
%  }
%  }
%  \,,
%$$
%the on-shell prequantum 3-bundle $\nabla_{\mathrm{CS}}^2$ extends to a diagram
%of relative cocylces of the form
%$$
%  \raisebox{20pt}{
%  \xymatrix{
%    Q_0 \ar[r] \ar[d] & \ast \ar[d]
%	\\
%	\mathbf{B}G \ar[r]^-{\nabla_{\mathrm{CS}}^2} & \mathbf{B}(\mathbf{B}^3 U(1)_{\mathrm{conn}})
%	\\
%	Q_1 \ar[r] \ar[u] & \ast \ar[u]
%  }
%  }
%  \,,
%$$
%hence, under forming homotopy fiber products, to the WZW-2-connection $\Omega \nabla^2_{\mathrm{CS}} : G \to \mathbf{B}^2 U(1)$ on the extended phase space $G$.
\end{example}

In the next section we see another example of this phenomenon.

\subsection{Higher prequantum 2d Poisson Chern-Simons theory and
deformation quantization %quantum mechanics
}
 \label{DeformationQuantizationInQuantumMechanics}

 We discuss here how the higher geometric quantization of a
 stacky refinement of the 2d Poisson $\sigma$-model
 yields, holographically, a deformation quantization of the underlying Poisson manifold.
 %hence of a 1-dimensional field theory (quantum mechanics).
 We do so
 by unwinding what higher geometric prequantization says in this case,
 % expressed in components in ordinary prequantum theory,
 and then observe that
 %in terms of this disguised form the higher prequantization
 this prequantization
 and its holographic relation to 1d quantization has already been
 worked out, secretly, in \cite{Eli}.\footnote{At least roughly, this relation had previously been voiced in the
 introduction of \cite{CattaneoFelderSymplectic},
 but at that time the geometric quantization of symplectic groupoids
 as in \cite{Eli} had yet to be fully understood.}

 \medskip

Let $(X,\pi)$ be a Poisson manifold, and let $\mathfrak{P}_{(X,\pi)}$ be the corresponding Poisson Lie algebroid,
see for instance \cite{Bongers} for review. The $\exp$ construction provides the geometrically $\infty$-connected Lie integration of $\mathfrak{P}_{(X,\pi)}$. The 1-truncation
\[
\mathcal{P}_{(X,\pi)}=\exp(\mathfrak{P}_{(X,\pi)})_{\leq 1}
\]
is the smooth groupoid integrating the Poisson Lie algebroid of $(X,\pi)$, see \cite{Tseng-Zhu}. Moreover, as for any $L_\infty$-algebroid, we can consider the differential refinement of $\mathcal{P}_{(X,\pi)}$ obtained by the $\exp(-)_\mathrm{conn}$ construction:
\[
\mathcal{P}_{(X,\pi);\mathrm{conn}}=\exp(\mathfrak{P}_{(X,\pi)})_{\mathrm{conn};\leq 1}.
\]
The Poisson structure $\pi$ can naturally be seen as a 2-cocycle $\pi\colon \mathfrak{P}_{(X,\pi)}\to \mathbb{R}[1]$,
which is in transgression with an invariant polynomial $\langle-\rangle$ on $\mathfrak{P}_{(X,\pi)}$ via a certain Chern-Simons element $P$ (see \cite{kotov-strobl} or \cite{frs} for details). Therefore, the general construction described in Section \ref{HigherCSFromLieIntegration} gives the homotopy commutative diagram of the form
%A non-degenerate and binary invariant polynomial which induces a
%pre-2-plectic structure
%on the moduli stack of a higher Chern-Simons type theory
%
%$$
%  \omega :=
%  \langle F_{(-)} F_{(-)}\rangle
%  :
%  \xymatrix{
%    \tau_1 \exp(\mathfrak{P})_{\mathrm{conn}}
%    \to
%    \Omega^3_{\mathrm{cl}}
%  }
%$$
%exists precisely on Poisson Lie algebroids $\mathfrak{P}$, induced
%from Poisson manifolds $(X,\pi)$.
%The differential Lie integration method described above yields a
%$(\mathbf{B}(\mathbb{R}/\Gamma))$-prequantization
$$
  \raisebox{20pt}{
  \xymatrix{
	&
	\mathbf{B}^2 (\mathbb{R}/\!/\Gamma)_{\mathrm{conn}}
	\ar[d]^{F_{(-)}}
    \\
    \mathcal{P}_{(X,\pi);\mathrm{conn}}
    	\ar[ur]^{\nabla_P}
	\ar[r]_-{\omega}
	&
	\Omega^3_{\mathrm{cl}}
  }
  }
  \,.
$$
The action functional of this higher prequantum field theory over a
closed oriented 2-dimensional smooth manifold $\Sigma_2$ is %, again by \cite{FiorenzaSatiSchreiberCup, frs},
the transgression
of the prequantum 2-bundle  $\nabla_P$ to codimension 0:
$$
  \exp\left(
    \int_{\Sigma_2} [\Sigma_2, \nabla_P]
  \right)
  \colon
  \xymatrix{
    [\Sigma_2, \mathcal{P}_{(X,\pi);\mathrm{conn}}]
	\ar[r]
	&
	\mathbb{R}/\!/\Gamma
  }
  \,.
$$
%\ccomment{mild rewrite of the below sentence for better
 % readability. Please check that I didn't mangle the physics :)}
We now observe that two complementary sectors of this higher
prequantum 2d Poisson Chern-Simons field theory $\nabla_P$ have a
separate life of their own in the literature.  On the one hand, there
is the sector where the bundle structures (and hence, the nontrivial
``instanton sectors'' of the field configurations) are ignored and only
the globally defined connection differential form data is retained,
and on the other hand, there is the complementary sector where only these bundle
structures/instanton sectors are considered, and the connection data is
ignored. In more detail:
\begin{enumerate}
\item
The restriction of the action functional $\exp(\int_{\Sigma_2}[\Sigma_2, \nabla_P])$
to the linearized theory, i.e.,
along the canonical inclusion
$\Omega(\Sigma, \mathfrak{P}) \hookrightarrow [\Sigma_2, \mathcal{P}_{(X,\pi);\mathrm{conn}}]$
of globally defined $\mathfrak{P}$-valued forms into all $\mathcal{P}_{(X,\pi)}$-principal connections,
is the action functional of the \emph{Poisson sigma-model} \cite{kotov-strobl}.
\item
Forgetting the connection data, and just remembering the bundle structure yields, as mentioned above, to the smooth groupoid $\mathcal{P}_{(X,\pi)}$ integrating the Poisson manifold $(X,\pi)$.
%
%  The restriction of the moduli stack of fields
%  $\tau_1\exp(\mathfrak{P})_{\mathrm{conn}}$
%  to just $\tau_1\exp(\mathfrak{P})$ obtained by forgetting the differential refinement
%  (the connection data) und just remembering the underlying $\exp(\mathfrak{P})$-principal
%  bundles, yields what is known as the \emph{symplectic groupoid} of $\mathfrak{P}$.
%
While the prequantum 2-bundle $\nabla_P$
  does not descend along the forgetful map
  $\mathcal{P}_{(X,\pi);\mathrm{conn}} \to \mathcal{P}_{(X,\pi)}$
%  from moduli of $\tau_1\exp(\mathfrak{P})$-principal connections to their
%  underlying $\tau_1(\exp(\mathfrak{P}))$-principal bundles,
  its version $\nabla_P^1$ ``without curving''
%, given by def. \ref{PrincipalConnectionWithoutTopDegreeForms},
  does descend\footnote{Compare to the analogous statement for 3d Chern-Simons theory discussed above in Section
  \ref{ExtendedWZW}} and so
  %does hence its curvature $\omega^1$, which
  %by remark \ref{FactorizationOfCurvatureTwists}
  %has  coefficients in $\mathbf{B} \Omega^2_{\mathrm{cl}}$
  %instead of $\Omega^3_{\mathrm{cl}}$:
  we get the homotopy commutative diagram
  $$
    \raisebox{20pt}{
    \xymatrix{
	  &
	  \mathbf{B}\left( \mathbf{B}(\mathbb{R}/\!/\Gamma)_{\mathrm{conn}} \right)
	  \ar[d]^{\mathbf{B}F_{(-)}}
	  \\
	  \mathcal{P}_{(X,\pi)}
	  \ar[ur]^{\nabla_P^1}
	  \ar[r]_-{\omega^1}
	  &
	  \mathbf{B} \Omega^2_{\mathrm{cl}}
	}
	}
	\,.
  $$
  \end{enumerate}
\begin{remark}
  If  the smooth groupoid $\mathcal{P}_{(X,\pi)}$
  happens to have a presentation by a Lie groupoid\footnote{This is an integrability condition
  on $\mathfrak{P}$.}
%  under the canonical inclusion of Lie groupoids into smooth $\infty$-groupoids
%  of example \ref{ToposOfSmoothInfinityGroupoids}
  then the pair $(\mathcal{P}_{(X,\pi)},\omega^1)$
  it is called a \emph{pre-quasi-symplectic groupoid} in the literature  \cite{LGXu}.
  The de Rham hypercohomology 3-cocycle $\omega^1$ is in general is given by 3-form data and 2-form data;
  %on a {\v C}ech simplicial presheaf
  %that resolves $\tau_1\exp(\mathfrak{P})$
  %(in generalization of the simple example \ref{LocalDataFor1Sections} above) --
  when it happens to be represented by
  just a globally defined 2-form on the manifold of morphisms of the Lie groupoid $\mathcal{P}_{(X,\pi)}$
  (which is then necessarily a closed and ``multiplicative'' 2-form),
  then $(\mathcal{P}_{(X,\pi)},\omega^1)$ is called a (pre-)\emph{symplectic groupoid}.
  See \cite{Eli} for a review and further pointers to the literature.
\end{remark}

\begin{remark} In case $(\mathcal{P}_{(X,\pi)},\omega^1)$ is a (pre-)symplectic groupoid,
%the descended (pre-)2-plectic form
%$\omega^1\colon \mathcal{P}_{(X,\pi)}\to \mathbf{B}\Omega^2_{\mathrm{cl}}$
%of the higher prequantum 2d Poisson Chern-Simons theory
%is represented by a multiplicative symplectic 2-form on the manifold of morphisms
%of the Lie groupoid $\mathcal{P}_{(X,\pi)}$, then
one is faced with a situation
that looks like ordinary symplectic geometry subject to a
kind of equivariance condition. This is the perspective from which
symplectic groupoids were originally introduced and from which they are mostly studied
in the literature: as a means to
re-code Poisson geometry in terms of ordinary symplectic geometry.
 (There is at least one notable exception: \cite{LGXu}, where the
higher geometric nature of the setup is made explict.)

The goal of finding a sensible geometric quantization of symplectic groupoids, and hence in some sense of Poisson manifolds,
was finally achieved in \cite{Eli}. We will come back to this point below.
\end{remark}
\medskip

In order to further understand the conceptual role of the prequantum
2-bundle $\nabla^1_{P}$,
notice that, following \cite{CattaneoFelderSymplectic} as in Remark \ref{phase-space}, we may think of
the symplectic groupoid $\mathcal{P}$ as the
extended reduced phase space of the open string Poisson-Chern-Simons theory.
More precisely, if $\mathfrak{C}_1, \mathfrak{C}_1 \hookrightarrow \mathfrak{P}$
are two sub-Lie algebroids, and we write $\mathcal{C}_i=\exp(\mathfrak{C}_i)_{\leq 1}$ for the smooth groupoids integrating them, then the homotopy fiber product
$\mathbf{Phase}_{\mathfrak{C}_0, \mathfrak{C}_1}=\mathcal{C}_0\times_{\mathcal{P}_{(X,\pi)}}\mathcal{C}_1$
%$$
%  \xymatrix@C=0pt{
%    & \mathbf{Phase}_{\mathfrak{C}_0, \mathfrak{C}_1}
%	\ar[dl]\ar[dr]
%	\\
%	\mathcal{C}_0 \ar[dr] && \mathcal{C}_1 \ar[dl]
%	\\
%	&
%	\mathcal{P}
%  }
%$$
should be the ordinary reduced phase space of open strings that stretch
between $\mathfrak{C}_0$ and $\mathfrak{C}_1$, regarded as D-branes.
Unwinding the definitions shows that this
is precisely what is shown in \cite{CattaneoFelderCoisotropic}. Namely, for
$\mathfrak{C}_0, \mathfrak{C}_1 \hookrightarrow \mathfrak{P}$ two Lagrangian
sub-Lie algebroids\footnote{This implies, in particular, that the base manifolds of $\mathfrak{C}_0$ and $\mathfrak{C}_1$ are coisotropic submanifolds of $X$.}
the homotopy fiber product stack $\mathbf{Phase}_{\mathfrak{C}_0, \mathfrak{C}_1}$
is the symplectic reduction of the open $\mathfrak{C}_0$-$\mathfrak{C}_1$-string phase space.

Notice that the condition that $\mathfrak{C}_i \hookrightarrow \mathfrak{P}$
be Lagrangian sub-Lie algebroids implies that the prequantum 2-bundle $\nabla_P^1$ restricted to the smooth groupoids $\mathcal{C}_i$ becomes flat.
%
%, hence that we have
%commuting squares
%$$
%  \raisebox{20pt}{
%  \xymatrix{
%    \tau_1 \exp(\mathfrak{C}_i)
%	\ar[r]
%	\ar[d]
%	&
%	\flat \mathbf{B}^2 (\mathbb{R}/\Gamma)
%	\ar[d]
%	\\
%	\tau_1 \exp(\mathfrak{P})
%	\ar[r]^-{\nabla_P^1}
%	&
%	\mathbf{B}(\mathbf{B}(\mathbb{R}/\Gamma)_{\mathrm{conn}})
%  }
%  }\,.
%$$
In case the the restricted 2-bundle $\nabla_P^1\bigr\vert_{\mathcal{C}_i}$ happens to be not only flat but trivializable, one can form the homotopy commutative diagram
%If the inclusions are even such $\nabla_P^1$ entirely trivializes
%over them, hence that we have diagrams
$$
  \raisebox{20pt}{
  \xymatrix{
   \mathbf{Phase}_{\mathfrak{C}_0, \mathfrak{C}_1}\ar[d]\ar[ddr]\ar[r] &\mathbf{B}(\mathbb{R}/\!/\Gamma)_{\mathrm{conn}}\ar[ddr]\ar[d]\\
   \mathcal{C}_0\ar[ddr] \ar[r]|(.5)\hole& {*}\ar[ddr]|(.5)\hole\\
   & \mathcal{C}_1
	\ar[r]
	\ar[d]
	&
	\ast
	\ar[d]
	\\
&	\mathcal{P}_{(X,\pi)}
	\ar[r]^-{\nabla_P^1}
	&
	\mathbf{B}(\mathbf{B}(\mathbb{R}/\!/\Gamma)_{\mathrm{conn}})
  }
  }
  \,,
$$
where the morphism $\mathbf{Phase}_{\mathfrak{C}_0, \mathfrak{C}_1}
	\to \mathbf{B} (\mathbb{R}/\Gamma)_{\mathrm{conn}}$ is induced by the universal property of the pullback. It exhibits a natural principal $\mathbb{R}/\!/\Gamma$-bundle with connection on the open string phase space in 2d Poisson Chern-Simons theory.
%
%$$
%  \raisebox{20pt}{
%  \xymatrix{
%    \tau_1 \exp(\mathfrak{C}_i)
%	\ar[r]^{\nabla_{\mathfrak{C}_i}}
%	\ar[d]
%	&
%	\ast
%	\ar[d]
%	\\
%	\tau_1 \exp(\mathfrak{P})
%	\ar[r]^-{\nabla_P^1}
%	&
%	\mathbf{B}(\mathbf{B}(\mathbb{R}/\Gamma)_{\mathrm{conn}})
%  }
%  }
%  \,,
%$$
%then under forming homotopy fiber products the prequantum 2-bundle
%$\nabla_P^1$ induces a prequantum 1-bundle on the open string phase space
%by the D-brane-relative looping of the on-shell prequantum 2-bundle:
%$$
%  \nabla_{\mathfrak{C}_0} \underset{\nabla_P^1}{\times}
%  \nabla_{\mathfrak{C}_1}
%  :
%  \xymatrix{
%    \mathbf{Phase}_{\mathfrak{C}_0, \mathfrak{C}_1)}
%	\ar[r]
%	&
%	\mathbf{B} (\mathbb{R}/\Gamma)_{\mathrm{conn}}
%  }
%  \,.
%$$
%
%
\medskip

We now review the steps in the geometric quantization of the symplectic groupoid
due to Hawkins \cite{Eli},
 %-- hence the full geometric quantization of the prequantization $\nabla_P^1$ --
%while
 discussing along the way the natural re-interpretation of the various
steps  from the point of view of the higher prequantization of
2d Poisson Chern-Simons theory presented above. Assume for simplicity that the integrability condition $\Gamma=\mathbb{Z}$ is satisfied, so that
$\nabla_P^1$ is a principal $(\mathbf{B}U(1))$-2-bundle over $\mathcal{P}_{(X,\pi)}$. Assume further that $\mathcal{P}_{(X,\pi)}$ is a symplectic groupoid. In the terminology of Section \ref{HigherCourant}, this means that the curvature of $\nabla_P^1$ is represented by a globally defined closed 2-form on the manifold of morphisms of $\mathcal{P}_{(X,\pi)}$. It is therefore meaningful to ask that also the prequantization $\nabla_P^1$ have a classical description, and reduces to the datum of a multiplicative principal $U(1)$-bundle with connection on the manifold of morphisms of $\mathcal{P}_{(X,\pi)}$.
%Consider therefore $\nabla_P^1$, as above, as the $(\mathbf{B}U(1))$-prequantum 2-bundle
%of 2d Poisson Chern-Simons theory according to def. \ref{DiffModuliStack} and
%example \ref{DeligneComplexInExample}.
%If we have a genuine symplectic groupoid instead of a pre-quasi-symplectic groupoid then
%it makes sense ask for this prequantization to be presented by a {\v C}ech-Deligne
%3-cocycle on $\tau_1 \exp(\mathfrak{P})$ which is given just by a multiplicative
%circle-bundle with connection on the space of morphisms of the symplectic groupoid,
%and otherwise trivial local data on the space of objects.
While this is unlikely to be the most general higher prequantization of the 2d Poisson Chern-Simons
theory, this is the choice that allows one to think of the situation as if it were
a setup in traditional symplectic geometry equipped with
an equivariance  (or ``multiplicativity'') constraint.\footnote{
Such a ``multiplicative $U(1)$-bundle'' on the space of morphisms of a
Lie groupoid is precisely the same kind of object as the transition bundle that appears in the definition of a bundle
gerbe, only that here the underlying groupoid is not a {\v C}ech groupoid resolving
a plain manifold, but is, in general, a genuine non-trivial Lie groupoid.}
%, as opposed to a setup in higher 2-plectic geometry.
And indeed, such a multiplicative prequantum bundle is the traditional notion of prequantization of a symplectic groupoid one finds in the literature. This
 is also the situation considered in \cite{Eli},
 where the central  result is the construction of the
convolution $C^\ast$-algebra
$\mathcal{A}(\nabla^1_P)_{\mathrm{pq}}$ of sections
of the multiplicative prequantum bundle on the
space of morphisms of the symplectic groupoid, and of its subalgebra
$$
  \mathcal{A}(\nabla^1_P)_{\mathrm{q}} \hookrightarrow \mathcal{A}(\nabla^1_P)_{\mathrm{pq}}
$$
of polarized sections, once a suitable kind of polarization has been chosen.
%Observe then that convolution algebras of sections of transition bundles
%of bundle gerbes
These convolution algebras
have a natural interpretation in terms of the higher geometry
of the %corresponding
higher prequantum bundle $\nabla^1_P$: they are the algebras whose modules are the modules of sections of bundle gerbe modules twisted by $\nabla^1_P$, see  \cite[section 5]{CareyJohnsonMurray}.
In the framework of higher module theory, the category of $\mathcal{A}(\nabla^1_P)_q$-modules is a \emph{2-module}. More precisely, it is a 2-module with \emph{2-basis} the linear category
$\mathbf{B}\mathcal{A}(\nabla^1_P)_q$, see \cite[appendix]{AQFT}. On the other hand
%by \cite[section 5]{CareyJohnsonMurray} one finds
%that these are the algebras whose modules are the unitary bundles which are twisted
%by $\nabla^1$: the ``'bundle gerbe modules'.
%But by remark \ref{TwistedCohomology} and the
by the discussion in Section \ref{ExtendedWZW},
%$\nabla^1_P$-twisted unitary bundles are equivalently
 bundle gerbe modules twisted by $\nabla^1_P$ are precisely
 the (pre-)quantum 2-states of the prequantum 2-bundle $\nabla^1_P$. %regarded as a prequantum 2-bundle.
 We therefore obtain:
 $$
  \left\{
    \begin{tabular}{c}
	  quantum \mbox{2}-states of \\
	  higher prequantum 2d Poisson Chern-Simons theory
	\end{tabular}
  \right\}
  \;\;\simeq\;\;
  \mathcal{A}(\nabla^1_P)_q \text{-modules}
%  \;\;\in \;\;
%  2\mathrm{Mod}
  \,,
$$
and the quantum 2-states have therefore a natural 2-module structure.
This resolves what might be a conceptual (or maybe linguistic) puzzlement concerning
the construction in \cite{Eli} in view of the usual story of geometric quantization. Namely,
ordinarily geometric quantization directly produces the
space of states of a theory, while it requires more work to obtain the algebra of quantum
observables acting on that. On the other hand, in \cite{Eli} it superficially seems
to be the other way around: an algebra
drops out as a direct result of the quantization procedure.
However, as we showed above, from the point of view of higher prequantization this algebra
is (a 2-basis for) the 2-module of 2-states;
and indeed obtaining the \emph{2-algebra} or \emph{higher quantum operators}
which would act on these 2-states does require more work (and, up to our knowledge, has not been discussed yet in the literature).
\par
Actually, \cite{Eli} amplifies a different perspective on the central
result obtained there:
that $\mathcal{A}(\nabla^1_P)_q$ is also a \emph{strict $C^\ast$-deformation
quantization} of the Poisson manifold $(X,\pi)$.
%that corresponds to the Poisson Lie algebroid $\mathfrak{P}$!
From the point of view of higher geometric prequantization this
says that the geometrically %higher-geometric
quantized 2d Poisson Chern-Simons theory
has a 2-space of quantum 2-states
%in codimension 2
that encodes the correlators %(commutators)
of a 1-dimensional quantum mechanical system. In other words, we see
that the construction in \cite{Eli} is implicitly a ``holographic''
(strict deformation-)quantization of a Poisson manifold by directly
higher-geometrically quantizing %instead
a 2-dimensional QFT. Notice how this statement is an analogue in
$C^\ast$-deformation quantization of the seminal result on
\emph{formal} deformation quantization of Poisson manifolds. The
general formula that Kontsevich provided for the formal deformation
quantization of a Poisson manifold had been shown by Cattaneo-Felder
to be the point-particle limit of the 3-point function of the
corresponding 2d Poisson sigma-model \cite{CattaneoFelder}.  A similar
result is discussed in \cite{GukovWitten}.  There, the 2d A-model
(which is a special case of the Poisson sigma-model) is shown to
holographically encode the quantization of its target %space
symplectic manifold regarded as a 1d quantum field theory.
\par
In summary, the following table indicates how the ``holographic'' formal deformation
quantization of Poisson manifolds by Kontsevich-Cattaneo-Felder is
analogous to the ``holographic'' strict deformation quantization of
Poisson manifolds by Hawkins, when reinterpreted in terms of higher prequantization
as discussed above.

\medskip

\begin{tabular}{|l||c|c|}
  \hline
  & \begin{tabular}{c} perturbative formal/algebraic \\ quantization \end{tabular} &
   \begin{tabular}{c} non-perturbative geometric \\ quantization \end{tabular}
  \\
  \hline
  \hline
  \begin{tabular}{l}
    quantization of
	\\
	Poisson manifold
  \end{tabular}  & formal deformation quantization & strict $C^\ast$-deformation quantization
  \\
  \hline
  \begin{tabular}{l}
  ``holographically'' related \\
  2d field theory
  \end{tabular}
  &
  Poisson sigma-model & 2d Poisson Chern-Simons theory
  \\
  \hline
  \begin{tabular}{l}
    %moduli
    stack of fields
	\\
	of the 2d field theory
  \end{tabular} & Poisson Lie algebroid & symplectic groupoid
  \\
  \hline
  \begin{tabular}{l}
    quantization of
	\\
	holographically related
	\\
	2d field theory
  \end{tabular}
  &
  \begin{tabular}{l}
  perturbative quantization of \\
  Poisson sigma-model
  \end{tabular} &
  \begin{tabular}{l}
    higher geometric quantization \\
	of 2d Poisson Chern-Simons theory
  \end{tabular}
  \\
  \hline
  \begin{tabular}{l}
  1d observable algebra\\
  is holographically \\ identified with...
  \end{tabular}
  &
  \begin{tabular}{l}point-particle limit
  \\
  of 3-point function
  \end{tabular}
  &
  \begin{tabular}{c}
  basis for 2-space
  \\
  of quantum 2-states
  \end{tabular}
  \\
  \hline
\end{tabular}

\medskip

More details on this higher geometric interpretation of traditional symplectic groupoid
quantization can be found in \cite{Bongers, Nuiten}.

\subsection{Higher prequantum 6d WZW-type models and the smooth fivebrane-6-group}
\label{AnExtended6dSomething}

We close the overview of examples by providing a brief outlook on %higher dimensional
%examples in general, and on
certain higher prequantum field theories in dimensions seven and
six.% in particular.

The first example we consider is
the 7-dimensional abelian
Chern-Simons type theory given by the cup-product of a 3d $\mathbf{B}^2U(1)$-Chern-Simons theory with itself.
%in 7 dimensions, f
%For this 7d theory the ``holographic'' relation to an interesting 6d
%theory is fairly well understood.
The higher prequantum $U(1)$-7-connection behind this theory is show in \cite{FiorenzaSatiSchreiberCup} to be the morphism
\[
\mathbf{B}^3 U(1)_{\mathrm{conn}} \xrightarrow{ (-)\hat \cup(-)} \mathbf{B}^7 U(1)_{\mathrm{conn}},
\]
where $(-)\hat \cup(-)$ is a suitable refinement of the cup product $\cup$ in ordinary differential cohomology to a morphism of smooth stacks:\footnote{Actually, the theory to consider for the full holographic relation
is a quadratic refinement of this cup pairing, as discussed in \cite{FiorenzaSatiSchreiberCField}. For simplicity,
 in the present discussion we will suppress this.}
%This is the theory whose de-transgression
%is given \cite{FiorenzaSatiSchreiberCup}
%by the higher prequantum 7-bundle on the universal moduli 3-stack $\mathbf{B}^3 U(1)_{\mathrm{conn}}$
%of $\mathbf{B}^2 U(1)$-principal connections that is
%modulated by the smooth and differential refinement of the cup product $\cup$ in ordinary differential cohomology:
$$
  \begin{tabular}{c|c}
  \xymatrix{
    \mathbf{B}^3 U(1)_{\mathrm{conn}}
	\ar[rr]^{(-)\hat \cup(-)}
	\ar[d]^{u_{\mathbf{B}^3 U(1)}}
	&&
	\mathbf{B}^7 U(1)_{\mathrm{conn}}
	\ar[d]^{u_{\mathbf{B}^7 U(1)}}
	\\
	\mathbf{B}^3 U(1)
	\ar[rr]^{(-)\cup(-)}
	\ar[d]^\int
	&&
	\mathbf{B}^7 U(1)
	\ar[d]^\int
	\\
	K(\mathbb{Z}_4)
	\ar[rr]^{(-)\cup(-)}
	&&
	K(\mathbb{Z}_8)
  }
  &
  \xymatrix{
    \nabla_{7\mathrm{AbCS}}
	\\
    \nabla_{7\mathrm{AbCS}}^0	
	\\
    \int \nabla_{7\mathrm{AbCS}}^0
  }
  \end{tabular}
  \,.
$$
%\ccomment{Rewrote the first sentence for readibility}
While precise and reliable statements regarding these systems
become scarce as one proceeds into the
physics literature, the following four
seminal physics articles seem to represent the present understanding
of the story by which this 7d theory is related to a 6d theory in a higher
generalization of how 3d Chern-Simons theory is related to the 2d WZW model.

\begin{enumerate}
\item In \cite{WittenEffectiveAction} it was argued that the space of states that
the (ordinary) geometric
quantization of $\nabla_{7\mathrm{AbCS}}$ assigns to a closed 6d manifold
$\Sigma$ is naturally identified with the space of conformal blocks of
a self-dual 2-form higher gauge theory on $\Sigma$.
Moreover, this 6d theory is part of the worldvolume theory of
a single M5-brane and the above 7d Chern-Simons theory is
the abelian Chern-Simons sector of the 11-dimensional supergravity Lagrangian
compactified to a 7-manifold whose boundary is the 6d M5-brane worldvolume.

\item Then, in \cite{Maldacena,vafa-witten},  a more general relation between
the 6d theory and 11-dimensional supergravity compactified on a 4-sphere
to an asymptotically anti-de Sitter space was argued for. This is what is
today called $\mathrm{AdS}_7/\mathrm{CFT}_6$-duality, a sibling of the
$\mathrm{AdS}_5/\mathrm{CFT}_4$-duality which has received a large amount
of attention since then.

\item As a kind of synthesis of the previous two items, in \cite{WittenAdSTFT} it is
argued for both $\mathrm{AdS}_5/\mathrm{CFT}_4$ and
$\mathrm{AdS}_7/\mathrm{CFT}_6$ the conformal blocks on the CFT-side
are obtained already by keeping on the supergravity side \emph{only} the Chern-Simons terms
inside the full supergravity action.

\item  At the same time it is known that the abelian Chern-Simons term
in the 11-dimensional supergravity action relevant for
$\mathrm{AdS}_7/\mathrm{CFT}_6$ is not in general just
the abelian Chern-Simons term $\nabla_{7\mathrm{AbCS}}$ considered in the above references:
more accurately it receives Green-Schwarz-type quantum corrections
that make it a \emph{nonabelian} Chern-Simons term \cite{DLMCW}.
\end{enumerate}
\par
In \cite{FiorenzaSatiSchreiber5Brane}, we observed that
these items together %, taken at face value, imply
suggest that more generally it must be the quantum-corrected nonabelian
7d Chern-Simons Lagrangian
inside 11-dimensional supergravity which is relevant for the holographic
description of the 2-form sector of the 6d worldvolume theory of M5-branes.\footnote{See \cite{Freed} for comments on this 6d theory as an extended QFT related to extended 7d
Chern-Simons theory.}
Moreover, in \cite{FiorenzaSatiSchreiberCField} we observed that the natural lift of the
``flux quantization condition'' from \cite{WittenEffectiveAction},
 which is an identity between cohomology classes of fields
in 11d-supergravity, to the moduli stacks of the corresponding fields %(hence to higher prequantum geometry)
is given by a suitable homotopy pullback of these moduli fields,
as usual in homotopy theory.
In particular, we showed that this homotopy pullback is the smooth moduli 2-stack
$\mathbf{B}\mathrm{String}^{2\mathbf{a}}_{\mathrm{conn}}$ of $2\mathbf{a}$-twisted $\mathrm{String}$-principal
2-connections, which unifies the Spin-connection (i.e., the field of gravity) and the
3-form $C$-field into a single higher gauge field. % in higher prequantum geometry.
The nonabelian 7-dimensional Chern-Simons-type Lagrangian on String-2-connections
obtained this way in \cite{FiorenzaSatiSchreiber5Brane}
is the sum of some cup product terms and one indecomposable term, which, for the trivial twist, is a refinement $\tfrac{1}{6}\hat{\mathbf{p}}_2$ of the second fractional
Pontryagin class $\tfrac{1}{6}p_2$:
%Moreover, the refinement specifically of the indecomposable term to higher prequantum geometry
%is the stacky and differential refinement $\tfrac{1}{6}\hat{\mathbf{p}}_2$ of the universal fractional second
%Pontryagin class $\tfrac{1}{2}p_2$, which was constructed in \cite{FScSt} as reviewed in
%\ref{HigherCSFromLieIntegration} above:
$$
  \begin{tabular}{c|c}
  \xymatrix{
    \mathbf{B}\mathrm{String}_{\mathrm{conn}}
	\ar[rr]^-{\tfrac{1}{6}\hat {\mathbf{p}}_2}
	\ar[d]^{u_{\mathbf{B}\mathrm{String}}}
	&&
	\mathbf{B}^7 U(1)_{\mathrm{conn}}
	\ar[d]^{u_{\mathbf{B}^7 U(1)}}
	\\
    \mathbf{B}\mathrm{String}
	\ar[rr]^-{\tfrac{1}{6}{\mathbf{p}}_2}
	\ar[d]^\int
	&&
	\mathbf{B}^7 U(1)
	\ar[d]^\int
	\\
	B O \langle 8\rangle
	\ar[rr]^{\tfrac{1}{6}p_2}
	&&
	K(\mathbb{Z}, 8)
  }
  &
  \xymatrix{
    \nabla_{7\mathrm{CS}}
    \\
    \nabla_{7\mathrm{CS}}^0
    \\
    \int \nabla_{7\mathrm{CS}}^0
  }
  \end{tabular}
  \,.
$$
Quite  independently of whatever role this extended 7d Chern-Simons theory has as a sector in
$\mathrm{AdS}_7/\mathrm{CFT}_6$ duality,
this is the natural next example in higher prequantum theory after that of
3d $\mathrm{Spin}$-Chern-Simons theory.
Moreover, in \cite{FScSt} it was shown that the $\infty$-group extension of the smooth string 2-group classified by the cocycle $\nabla^0_\mathrm{CS}$
is a smooth refinement of the \emph{Fivebrane 6-group} \cite{SatiSchreiberStasheff, fivebranes}:
%
%the prequantum 7-bundle of this nonabelian 7d Chern-Simons
%theory over the moduli stack of its instanton sectors, hence over $\mathbf{B}\mathrm{String}$,
%is the delooping of a smooth refinement of the \emph{Fivebrane group} \cite{SatiSchreiberStasheff, fivebranes}
%to the smooth Fivebrane 6-group of example \ref{LieGroupAsInfinityGroup}:
$$
  \raisebox{20pt}{
  \xymatrix{
    \mathbf{B}\mathrm{Fivebrane}\ar[rr]\ar[d]&&{*}\ar[d]
	\\
	\mathbf{B}\mathrm{String} \ar[rr]^{\nabla^0_{7\mathrm{CS}}}
	&&
	\mathbf{B}^7 U(1)
	\,.
  }
  }
$$
Notice how, by the above general discussion this induces a WZW-type 6-bundle over the
smooth String 2-group itself, whose total space is the smooth Fivebrane group
$$
  \raisebox{20pt}{
  \xymatrix{
    \mathrm{Fivebrane}\ar[rr]\ar[d]&&{*}\ar[d]
	\\
	\mathrm{String}
	\ar[rr]^{\nabla_{6\mathrm{WZW}}^0 }
	&&
	\mathbf{B}^6 U(1)
  }}
  \,.
$$
Therefore, in view of the discussion in Section \ref{ExtendedWZW}, it is natural to expect
a 6-dimensional higher analog of traditional 2d WZW theory whose underlying
higher prequantum 6-bundle is $\nabla_{6\mathrm{WZW}}$. However, the
lift of this discussion from just instanton sectors to the full moduli stack of
fields (i.e., including the connections data) is more subtle than in the 3d/2d case and deserves
a separate discussion elsewhere. This is ongoing joint work with Hisham Sati.

\medskip

\noindent{\bf Acknowledgements.} U.S. thanks Igor Khavkine for discussion of
covariant quantization and related issues; and thanks David Carchedi for discussion
of concretification.

%%%%%%%%%%%%%%%%%%%%%%%%%%%%%%%%%%%%%%%%%%%%%%%%%%%

\end{document}